\documentclass[acmsmall,screen]{acmart}

\usepackage{listings}
\usepackage{xcolor}
\usepackage{booktabs}
\usepackage{multirow}
\usepackage{hyperref}

\hypersetup{
  colorlinks=true,
  linkcolor=yellow,
  citecolor=Red,
  urlcolor=blue,
  filecolor=blue}

\definecolor{codegreen}{rgb}{0,0.6,0}
\definecolor{codegray}{rgb}{0.5,0.5,0.5}
\definecolor{codepurple}{rgb}{0.58,0,0.82}
\definecolor{backcolour}{rgb}{0.95,0.95,0.92}
\lstdefinestyle{mystyle}{
    commentstyle=\color{codegreen},
    keywordstyle=\color{magenta},
    numberstyle=\tiny\color{codegray},
    stringstyle=\color{codepurple},
    basicstyle=\ttfamily\footnotesize,
    postbreak=\mbox{\textcolor{red}{$\hookrightarrow$}\space},
    columns=fullflexible,
    breakatwhitespace=false,
    breaklines=true,
    captionpos=b,
    keepspaces=true,
    frame=single,
    numbers=left,
    numbersep=5pt,
    showspaces=false,
    showstringspaces=false,
    showtabs=false,
    tabsize=2
}
\newcommand{\fixedwidth}[1]{{\ttfamily \small #1}}

\AtBeginDocument{%
  \providecommand\BibTeX{{%
    \normalfont B\kern-0.5em{\scshape i\kern-0.25em b}\kern-0.8em\TeX}}}

\setcopyright{acmcopyright}
\copyrightyear{2021}
\acmYear{2021}
\acmDOI{10.1145/1122445.1122456}

\acmJournal{CSUR}
\acmPrice{15.00}
\acmISBN{978-1-4503-XXXX-X/18/06}

\begin{document}

\title{A Survey on Automated Log Analysis for Reliability Engineering}

\author{Shilin He}
\affiliation{%
  \institution{Microsoft Research}
}

\author{Pinjia He}
\affiliation{%
  \institution{Department of Computer Science, ETH Zurich}
}

\author{Zhuangbin Chen}
\author{Tianyi Yang}
\author{Yuxin Su}
\author{Michael R. Lyu}
\affiliation{%
  \institution{Department of Computer Science and Engineering, The Chinese University of Hong Kong}
}

\thanks{The work was supported by Key-Area Research and Development Program of Guangdong Province (No. 2020B010165002) and the Research Grants Council of the Hong Kong Special Administrative Region, China (CUHK 14210717). This work was mainly done when Shilin He was a Ph.D. student at the Chinese University of Hong Kong.}
\authorsaddresses{Authors' addresses: S. He, Microsoft Research, No. 5, Danling Street, Haidian District, Beijing, 100080, China; email: shilin.he@microsoft.com; P. He (corresponding author), Department of Computer Science, ETH Zurich, Universitätstrasse 6, 8092 Zürich, Switzerland; email: pinjia.he@inf.ethz.ch; Z. Chen, T. Yang, Y. Su, MR. Lyu, Department of Computer Science and Engineering, The Chinese University of Hong Kong, Shatin, N.T., Hong Kong; email: \{zbchen,lyu,yxsu,tyyang\}@cse.cuhk.edu.hk}








\renewcommand{\shortauthors}{S. He et al.}

\begin{abstract}
  Logs are semi-structured text generated by logging statements in software source code. In recent decades, software logs have become imperative in the reliability assurance mechanism of many software systems because they are often the only data available that record software runtime information. As modern software is evolving into a large scale, the volume of logs has increased rapidly. To enable effective and efficient usage of modern software logs in reliability engineering, a number of studies have been conducted on automated log analysis. This survey presents a detailed overview of automated log analysis research, including how to automate and assist the writing of logging statements, how to compress logs, how to parse logs into structured event templates, and how to employ logs to detect anomalies, predict failures, and facilitate diagnosis. Additionally, we survey work that releases open-source toolkits and datasets. Based on the discussion of the recent advances, we present several promising future directions toward real-world and next-generation automated log analysis.
\end{abstract}

\ccsdesc[500]{Software and its engineering~Software maintenance tools}
\ccsdesc[500]{Software and its engineering~Software creation and management}

\keywords{log, log analysis, logging, log compression, log parsing, log mining.}

\maketitle

\section{Introduction}

In the recent decades, modern software, such as search engines, instant messaging apps., and cloud systems, has been increasingly integrated into our daily lives and becomes indispensable. Most of these software systems are expected to be available on a $24\times 7$ basis. Any non-trivial downtime can lead to significant revenue loss, especially for large-scale distributed systems~\cite{downtimeAmazon,downtimeGeneral,downtimeFacebook}. For example, in 2017, a 
downtime in Amazon led to a loss of 150+ million US dollars~\cite{downtimeexample}. Thus, the reliability of modern software is of paramount importance~\cite{lyu1996handbook}.

\begin{figure}
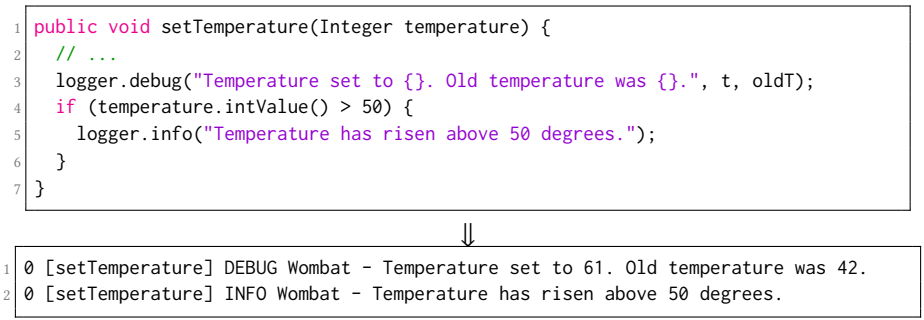

\begin{minipage}[htb]{0.83\textwidth}
\begin{lstlisting}[language=Java]
public void setTemperature(Integer temperature) {
  // ...
  logger.debug("Temperature set to {}. Old temperature was {}.", t, oldT);
  if (temperature.intValue() > 50) {
    logger.info("Temperature has risen above 50 degrees.");
  }
}
\end{lstlisting}
\end{minipage}
\\
$\Downarrow$
\\
\begin{minipage}[htb]{0.85\textwidth}
\begin{lstlisting}[]
0 [setTemperature] DEBUG Wombat - Temperature set to 61. Old temperature was 42.
0 [setTemperature] INFO Wombat - Temperature has risen above 50 degrees.
\end{lstlisting}
\end{minipage}
\caption{An example of logging statements by SLF4J and the generated logs.}
\label{fig:slf4j-example}
\vspace{-2pt}
\end{figure}

Software logs have been widely employed in a variety of reliability assurance tasks, because they are often the only data available that record software runtime information. Additionally, logs also play an indispensable role in data-driven decision making in industry~\cite{DBLP:conf/icse/PecchiaCCC15}. In general, logs are semi-structured text printed by logging statements (\textit{e.g.}, \fixedwidth{printf()}, \fixedwidth{logger.info()}) in the source code. For example, in Fig.~\ref{fig:slf4j-example}, the two log messages are printed by the two logging statements in the source code. The first few words (\textit{e.g.}, "Wombat") of the log messages are decided by the corresponding logging framework (\textit{e.g.}, SLF4J) and they are in structured form. On the contrary, the remaining words (\textit{e.g.}, "50 degrees") are unstructured because they are written by developers to describe specific system runtime events. 

\begin{figure}[thb]
\centerline{\includegraphics[width=0.83\columnwidth]{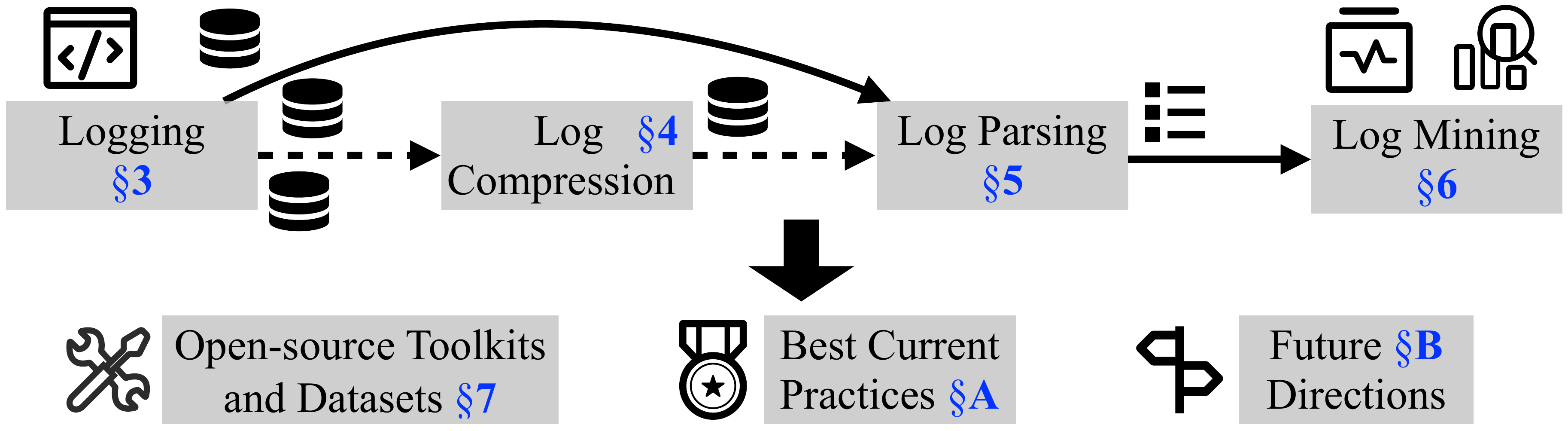}}
\caption{An overall framework for automated log analysis.}
\label{fig:framework}
\vspace{-6pt}
\end{figure}

A typical log analysis management framework is illustrated by the upper part of Fig.~\ref{fig:framework}. Particularly, traditional logging practice (\textit{e.g.}, which variables to print) mainly relies on developers' domain knowledge. During system runtime, software logs are collected and compressed as normal files using file compression toolkits~\cite{liu2019logzip} (\textit{e.g.}, WinRAR). Additionally, developers leverage the collected logs in various reliability assurance tasks (\textit{i.e.}, log mining), such as anomaly detection. Decades ago, these processes were based on specific rules specified by the developers. For example, to extract specific information related to a task (\textit{e.g.}, thread ID), developers need to design regex (\textit{i.e.}, regular expression) rules for automated log parsing~\cite{he2016experience}. The traditional anomaly detection process also relies on manually constructed rules~\cite{hansen1993automated}. These log analysis techniques were effective at the beginning because most of the widely-used software systems were small and simple.

However, as modern software has become much larger in scale and more complex in structure, traditional log analysis that is mainly based on ad-hoc domain knowledge or manually constructed and maintained rules becomes inefficient and ineffective~\cite{DBLP:journals/cacm/OlinerGX12}. This brings four major challenges to modern log analysis. (1) In many practices, while a number of senior developers in the same group share some best logging practices, most of the new developers or developers from different projects write logging statements based on domain knowledge and ad-hoc designs. As a result, the quality of the runtime logs varies to a large extent. (2) The volume of software logs has increased rapidly (\textit{e.g.}, 50 GB/h~\cite{mi2013toward}). Consequently, it is much more difficult to manually dig out the rules (\textit{e.g.}, log event templates or anomalous patterns). (3) With the prevalence of Web service and source code sharing platforms (\textit{e.g.}, Github), software could be written by hundreds of global developers. Developers who should maintain the rules often have no idea of the original logging purpose, which further increases the difficulty in manual maintenance of their rules. (4) Due to the wide adoption of the agile software development concept, a new software version often comes in a short-term manner. Thus, corresponding logging statements update frequently as well (\textit{e.g.}, hundreds of new logging statements per month~\cite{weixutheis}). However, it is hard for developers to manually update the rules.

To address these challenges, a great amount of work has been accomplished by both researchers and practitioners in recent decades. In particular, starting from 2003, a line of research efforts have been contributed to automated rule construction and critical information extraction from software logs, including the first pieces of work of log parsing~\cite{vaarandi2003data}, anomaly detection~\cite{vaarandi2003data}, and failure prediction~\cite{sahoo2003critical}. In addition, in the same year, H{\"a}t{\"o}nen \textit{et al.}~\cite{hatonen2003comprehensive} proposed the first log specific compression technique. Later, many empirical studies were conducted as the first steps towards some difficult problems in automated log analysis, including the first study on failure diagnosis by Jiang \textit{et al.}~\cite{jiang2009understanding} in 2009, the first exploration of logging practice by Yuan \textit{et al.}~\cite{DBLP:conf/icse/YuanPZ12} in 2012, and the first industrial study on logging practice by Fu \textit{et al.}~\cite{DBLP:conf/icse/FuZHLDLZX14} in 2014. Recently, machine learning  and deep learning algorithms have been widely adopted by the state-of-the-art (SOTA) papers, such as the deep learning-based "what-to-log" approach~\cite{li2020wherelog} in logging practice and Deeplog~\cite{du2017deeplog} in anomaly detection. Besides machine learning, parallelization has  been employed in various recent papers, such as Logzip~\cite{liu2019logzip} in log compression and POP~\cite{He2017TDSC} in log parsing.

These extensive studies on automated log analysis across multiple core directions have largely boosted the effectiveness and efficiency of systematic usage of software logs. However, the diversity and richness of both the research directions and recent papers could inevitably hinder the non-experts who intend to understand the SOTA and propose further improvements. To address this problem, this paper surveys 158 papers in the last 23 years across a variety of topics in log analysis. The papers under exploration are mainly from top venues in three related fields: software engineering (\textit{e.g.}, ICSE), system (\textit{e.g.}, SOSP), and networking (\textit{e.g.}, NSDI). Thus, the readers can obtain a deep understanding of the advantages and limitations of the SOTA approaches, as well as taking a glance at the existing open-source toolkits and datasets. In addition, the insights and challenges summarized in the paper can help practitioners understand the 
potential usage of the automated log analysis techniques in practice and realize the gap between academy and industry in this field. We present crucial research efforts on automated log analysis from the following seven perspectives as illustrated in Fig.~\ref{fig:framework}:
\begin{itemize}
    \item Logging: Section~\ref{sec:log_practice} introduces approaches that automate or improve logging practices, including \textit{where-to-log}, \textit{what-to-log}, and \textit{how-to-log}.
    \item Log compression: Section~\ref{sec:log_compression} presents approaches to compress software logs in runtime.
    \item Log parsing: Section~\ref{sec:log_parsing} discusses how to automatically extract event templates and key parameters from software logs. 
    \item Log mining: Section~\ref{sec:log_mining} introduces what automated log mining techniques can do to enhance system reliability. We focus on three main tasks: \textit{anomaly detection}, \textit{failure prediction}, and \textit{failure diagnosis}.
    \item Open-source toolkits and datasets: Papers providing open-source toolkits and datasets, which facilitate log analysis research, are presented in Section~\ref{sec:log_tools}. 
    \item Best current practices: Supplementary A presents some common log practices in industry, which might benefit the log analysis research and industrial deployment.
    \item Future directions: Supplementary B discusses open challenges and promising future directions that can push this field forward beyond current practices.
\end{itemize}

\section{Survey Methodology}

\begin{figure}[t!]
\centerline{\includegraphics[width=0.90\columnwidth]{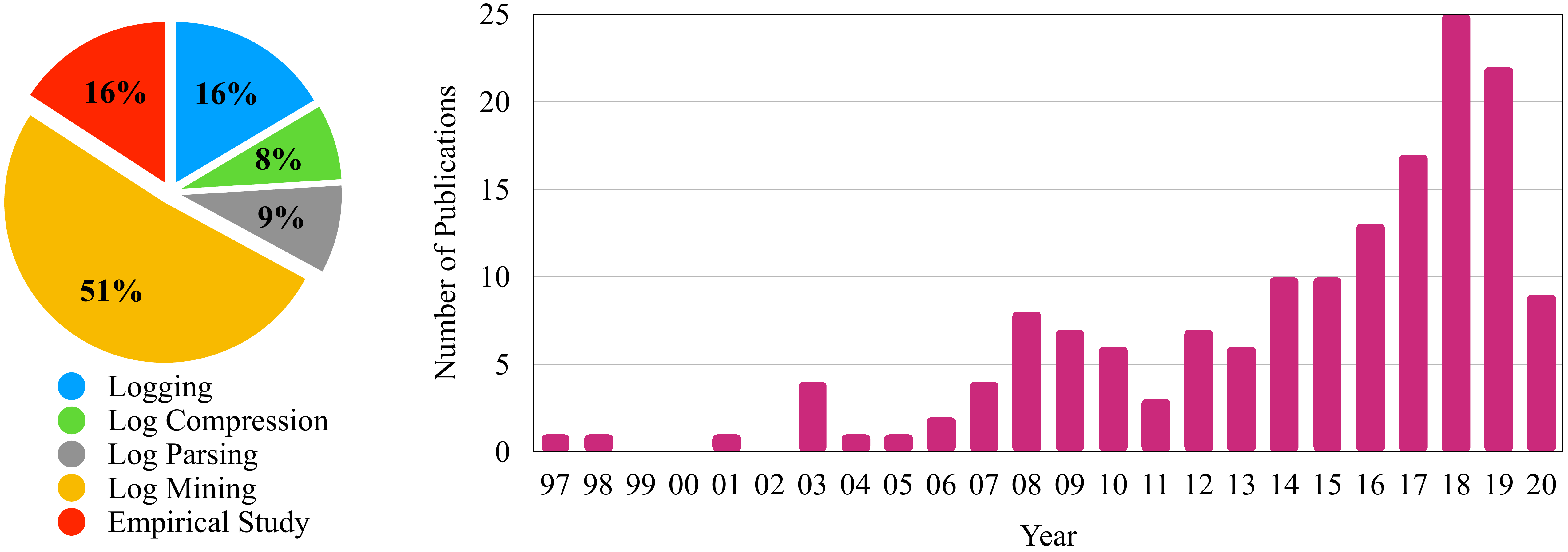}}
\caption{Paper distribution on each research topic and the associated evolution trend from 1997 to 2020.}
\label{fig:dist_year}
\vspace{-2pt}
\end{figure}

To systematically collect the publications for conducting this survey, we maintained a repository for the automated log analysis literature. We first searched relevant papers in online digital libraries and extended the repository by manually inspecting all references of these papers. The repository is now available online.\footnote{\url{https://github.com/logpai/awesome-log-analysis}}

To begin with, we searched several popular online digital libraries (\textit{e.g.}, IEEE Xplore, ACM Digital Library, Springer Online, Elsevier Online, Wiley Online, and ScienceDirect) with the following keywords: "log", "logging", "log parsing", "log compression", "log + anomaly detection", "log + failure prediction", "log + failure diagnosis". The "log" term is broadly quoted in various domains, such as the query log in database, web search log in recommendation, and the logarithm function in mathematics. Hence, to precisely collect the set of papers of our interest, we mainly focused on regular papers published in top venues (\textit{i.e.}, conferences and journals) of relevant domains, including ICSE, FSE, ASE, TSE, TOSEM, EMSE, SOSP, OSDI, ATC, NSDI, TDSC, DSN, ASPLOS, and TPDS. 
Then, we manually inspected each reference of these papers to collect additional publications that are related to the survey topics. In this paper, we focused on publications studying the reliability issues with software system logs, while publications of other topics are beyond the scope of this survey, such as logs for security and privacy~\cite{king2017log}, software change logs~\cite{chen2004open}, kernel logs~\cite{paccagnella2020custos}, and process mining logs~\cite{van2012process}.

In total, we collected 158 publications in automated log analysis area spanning from year 1997 to 2020. Fig.~\ref{fig:dist_year} (on the right part) shows the histogram of annual publications during the period. We can find that automated log analysis has been continuously and actively investigated in the past two decades. In particular, we can observe steady growth in the number of publications since 2006, indicating that the log analysis field has attracted an increasing amount of interest since then. 

Furthermore, we classify these publications into five categories by research focus: logging, log compression, log parsing, log mining, and empirical study. The distribution is presented on the left part of  Fig.~\ref{fig:dist_year}. Note that some papers may address more than one research focuses, in which case we categorized the paper based on its major contribution. As expected, Fig.~\ref{fig:dist_year} shows that a large portion (around a half) of research efforts were devoted to log mining. The reasons are two-fold: First, log mining is a comprehensive task that consists of many sub-tasks (\textit{e.g.}, anomaly detection, failure diagnosis). Each sub-task has a huge space for research exploration and thereby attracts many research studies on it; Second, research areas other than log mining are relatively mature. Besides, logging is also extensively studied in the past decades since it is more closely related to industrial practices. In addition, many log research tasks are empirical studies, and we discussed them in different sections respectively.

Furthermore, we were aware of several existing surveys on logs analysis. Oliner \textit{et al.}~\cite{DBLP:journals/cacm/OlinerGX12} briefly reviewed sixteen log analysis papers and pointed out some challenges in 2012. Since then, the log analysis area has been actively studied, which impels a more comprehensive and systematic log survey paper. Other existing surveys mainly studied logs and their use in the security area. Khan \textit{et al.}~\cite{khan2016cloud} focused on logs in the  cloud computing environment and the logs were leveraged for identifying and preventing suspicious attacks. Zeng \textit{et al.}~\cite{zeng2016computer} studied the logging mechanisms and security issues in three computer operating systems. A recent publication~\cite{landauer2020system} reviewed clustering based approaches in cyber security applications. Different from these studies, our survey mainly targets the automated log analysis for tackling the general reliability issues.


\section{Logging}
\label{sec:log_practice}

Logging is the task of constructing logging statements with proper description and necessary program variables, and inserting the logging statements to the right positions in the source code. Logging has attracted attention from both academia~\cite{DBLP:conf/icse/PecchiaCCC15, ms-event-logging, DBLP:conf/dsn/SchroederG06, xu2009detecting} and industry~\cite{DBLP:journals/ese/ChenJ17, DBLP:conf/icse/FuZHLDLZX14, DBLP:conf/icse/PecchiaCCC15, DBLP:conf/icse/YuanPZ12, DBLP:journals/ese/ZengCSC19, DBLP:conf/icse/BarikDDF16} across a variety of application domains because logging is a fundamental step for all the subsequent log mining tasks. 





\subsection{Logging Mechanism and Libraries} \label{sec:logging-mechanism}

\subsubsection{Logging Mechanism}
Logging mechanism is the set of logging statements and their activation code implemented by developers or a given software platform~\cite{DBLP:conf/icse/PecchiaCCC15}. Fig.~\ref{fig:slf4j-example} shows two example logging statements and the collected logs during the execution of the program. The logging statement at line 3 is executed every time the \fixedwidth{setTemperature} method is called, with no specific activation code. The logging statement at line 5 is controlled by the activation code \fixedwidth{if (temperature.intValue() > 50)} at line 4. According to the data collected by~\cite{DBLP:conf/icse/PecchiaCCC15}, the most widely-adopted coding pattern that is used to activate the logging statements is \fixedwidth{if (condition) then log error()}.


\subsubsection{Logging Libraries}
To improve flexibility, industrial developers often utilizes logging libraries~\cite{DBLP:conf/icse/PecchiaCCC15, chen2020studying}, which are software components that facilitate logging and provide advanced features (\textit{e.g.}, thread-safety, log archive configuration, and API separation). Toward this end, a lot of open-source logging libraries have been developed (\textit{e.g.}, Log4j~\cite{Log4j}, SLF4J~\cite{SLF4J}, AspectJ~\cite{AspectJ}, spdlog~\cite{spdlog}). 



\subsection{Challenges for Logging} \label{sec:logging-challenges}


Logging in a software system is usually decided by an empirical process in the development phase~\cite{DBLP:conf/icse/PecchiaCCC15}. In general, logging practice, \textit{i.e.}, how developers conduct the task of logging, is scarcely documented or regulated by strict standard, such as the logging mechanism and APIs~\cite{DBLP:conf/icse/BarikDDF16}. Thus, logging relies heavily on human expertise~\cite{DBLP:journals/cacm/OlinerGX12, DBLP:conf/icse/Shang12, DBLP:conf/icse/FuZHLDLZX14, DBLP:conf/icse/PecchiaCCC15, DBLP:conf/kbse/HeCHL18}. In the following, we summarize three main challenges for automated logging, which also align with the taxonomy mentioned by Chen and Jiang~\cite{DBLP:conf/icse/ChenJ17}: \emph{where-to-log}, \emph{what-to-log}, and \emph{how-to-log}. Under this categorization, every problem exhibits aspects that represent the primary concerns of the actual practice of logging. Accordingly, we summarize three major aspects, \textit{i.e.}, \emph{diagnosability}, \emph{maintenance}, and \emph{performance}.



\subsubsection{where-to-log}

Where-to-log is about determining the appropriate location of logging statements. Although logging statements provide rich information and can be inserted almost everywhere, excessive logging results in performance degradation~\cite{DBLP:conf/icse/ChenJ17} and incurs additional maintenance overhead. In addition, it is challenging to diagnose problems by analyzing a large volume of logs as most logs are unrelated to the problematic scenarios~\cite{DBLP:conf/icsm/JiangHHF08}. On the other hand, insufficient logging will also impede the logs' diagnosability. For example, an incomplete sequence of logs may hinder the reproduction of precise execution paths~\cite{DBLP:conf/sosp/ZhaoRLSYZ17}.
Therefore, developers need to be circumspect in their choices of where-to-log.

\subsubsection{what-to-log}

What-to-log is about providing sufficient and concise information within the three major components of a logging statement, \textit{i.e.}, verbosity level, static text, and dynamic content. Mis-configured verbosity level has similar consequences with inappropriate logging points. As developers typically filter logs according to the verbosity levels, under-valued verbosity levels may result in missing or ignored log messages while over-valued verbosity levels lead to overwhelming log messages~\cite{DBLP:conf/icsm/JiangHHF08}. When composing a snippet of logging code, the static text should be concise and the dynamic content should be coherent and up-to-date. Poorly written static text and inconsistent dynamic content could affect the subsequent diagnosis and maintenance activities~\cite{DBLP:conf/msr/KabinnaBSH16, liu2019variables,zhang2019robust,meng2019loganomaly}.

\subsubsection{how-to-log}

How-to-log is the "design pattern" and maintenance of logging statements systematically. Most software testing techniques focus on verifying the quality of feature code, but a few papers~\cite{DBLP:conf/icse/YuanPZ12, DBLP:journals/ese/ChenJ17, DBLP:conf/icse/ChenJ17,DBLP:journals/ese/HassaniSST18, DBLP:journals/ese/ShangNH15} pay attention to the quality and anti-patterns in the logging code. 
Most industrial and open source systems choose to scatter logging statements across the entire code base, intermixing with feature code~\cite{DBLP:conf/icse/ChenJ17}, which also hardens the maintenance of logging code.

\subsection{Logging Approaches} \label{sec:logging-solution}

\begin{table}[t]
  \centering
  \caption{Summary of logging approaches.}
  \label{tab:practice}
  \begin{tabular}{@{}l|l|p{8cm}@{}}
  \toprule
  \textbf{Problems}             & \textbf{Aspects} & \textbf{Objectives}  \\
  \midrule
  \multirow{4}{*}{\textbf{where-to-log}} & Diagnosability   & Suggest appropriate placement of logging statements into source code~\cite{DBLP:conf/dsn/CinqueCNP10,DBLP:conf/osdi/YuanPHLLTZS12,DBLP:journals/tse/CinqueCP13,DBLP:conf/icse/ZhuHFZLZ15, li2020wherelog,DBLP:conf/hotos/ZhaoRLSYZ17, DBLP:conf/sosp/ZhaoRLSYZ17, DBLP:journals/ese/YaoPSSTS20} ; Study logging practices in industry~\cite{DBLP:conf/icse/FuZHLDLZX14, DBLP:journals/ese/LiCSH18}.  \\
                                & Performance      & Minimize or reduce performance overhead~\cite{DBLP:conf/osdi/YuanPHLLTZS12, DBLP:conf/usenix/DingZLZLFZX15}.    \\
  \midrule
  \multirow{9}{*}{\textbf{what-to-log}}  & Diagnosability   & Enhance existing logging code to aid debugging~\cite{DBLP:conf/asplos/YuanZPZS11}; Suggest proper variables and text description in log~\cite{DBLP:conf/kbse/HeCHL18,liu2019variables, li2017log}.   \\
                                & Maintenance      & Determine whether a logging statement is likely to change in the future~\cite{DBLP:journals/ese/KabinnaBSSH18}; Characterize and detect duplicate logging code~\cite{DBLP:conf/icse/LiC0S19}.  \\
                                & Performance      & Study the performance overhead and energy impact of logging in mobile app~\cite{DBLP:journals/ese/ChowdhuryNHJ18, DBLP:journals/ese/ZengCSC19}; Automatically change the log level of a system in case of anomaly~\cite{DBLP:conf/iwpc/MizouchiSII19}.  \\
  \midrule
  \multirow{14}{*}{\textbf{how-to-log}}   & Diagnosability   & Characterize the anti-patterns in the logging code~\cite{DBLP:conf/icsm/ShangNHJ14}; Optimize the implementation of logging mechanism to facilitate failure diagnosis~\cite{DBLP:conf/usenix/LuoNSMC18}. \\
                                & Maintenance      & Characterize and detect the anti-patterns in the logging code~\cite{DBLP:conf/icse/YuanPZ12, DBLP:journals/ese/ChenJ17, DBLP:conf/icse/ChenJ17,DBLP:journals/ese/HassaniSST18, li2017towards, DBLP:journals/ese/LiNJLWL20}; Characterize and prioritize the maintenance of logging statements~\cite{DBLP:conf/msr/KabinnaBSH16}; Study the relationship between logging characteristics and the code quality~\cite{DBLP:journals/ese/ShangNH15, chen2019extracting}; Propose new abstraction or programming paradigm of logging~\cite{kiczales1997aspect,DBLP:conf/osdi/LockermanFKSAA018}.  \\
                                & Performance      & Optimize the compilation and execution of logging code~\cite{DBLP:conf/usenix/YangPO18}. \\
  \bottomrule
  \end{tabular}
\end{table}

Numerous solutions have been proposed to address the challenges mentioned in Section~\ref{sec:logging-challenges}. Table~\ref{tab:practice} summarizes existing studies along with corresponding problems and aspects. Each row represents a challenge. The approaches fall into three categories: \textit{static code analysis}, \textit{machine learning}, and \textit{empirical study}. A bunch of early work utilized static code analysis to analyze logging in a program without executing the source code. Machine learning-based approaches focus on learning from data. By concentrating on the inherent statistical properties of existing logging statements, learning-based approaches automatically give suggestions on improving the logging statements. In the remainder of this section, we discuss solutions by their aspects.

\subsubsection{Diagnosability}

Logs are valuable for investigating and diagnosing failures. However, a logging statement is only as helpful as the information it provides. The research on this aspect aims at (1) understanding the helpfulness of logs for failure diagnosis and (2) making logs informative for diagnosis.

\par{\textit{(1) Understanding the helpfulness of logs for failure diagnosis.}}
This is of great importance because logs are widely adopted for failure diagnosis. According to a survey~\cite{DBLP:conf/icse/FuZHLDLZX14} involving 54 experienced developers in Microsoft, almost all the participants agreed that ``logging statements are important in system development and maintenance'' and ``logs are a primary source for problem diagnosis''. Fu \textit{et al.}~\cite{DBLP:conf/icse/FuZHLDLZX14} also studied the types of logging statements in industrial software systems by source code analysis, and summarized five types of logging snippets, \textit{i.e.}, assertion-check logging, return-value-check logging, exception logging, logic-branch logging, and observing-point logging. Besides, Fu \textit{et al.}~\cite{DBLP:conf/icse/FuZHLDLZX14} further demonstrated the potential feasibility of predicting where to log.
Shang \textit{et al.}~\cite{DBLP:conf/icsm/ShangNHJ14} conducted the first empirical study to provide a taxonomy for user inquiries of logs. 
They identified five types of information that were often sought from log lines by practitioners, \textit{i.e.}, meaning, cause, context, impact, and solution. Shang \textit{et al.}~\cite{DBLP:conf/icsm/ShangNHJ14} were also the first to associate the development knowledge at present in various development repositories (\textit{e.g.}, code commits and issues reports) with the log lines and to assist practitioners in resolving real-life log inquiries. In addition, Li \textit{et al.}~\cite{DBLP:journals/ese/LiCSH18} highlighted the feasibility of guiding developers' logging practice with topic models by investigating six open source systems.

\par{\textit{(2) Making logs informative for failure diagnosis.}}
As printed logs are often the only run-time information source for debugging and analysis, the quality of log data is critically important. \emph{LogEnhancer}~\cite{DBLP:conf/asplos/YuanZPZS11} made the first attempt to systematically and automatically augment existing logging statements in order to reduce the number of possible code paths and execution states for developers to pinpoint the root cause of a failure. Zhao \textit{et al.}~\cite{DBLP:conf/hotos/ZhaoRLSYZ17, DBLP:conf/sosp/ZhaoRLSYZ17} followed the idea of \emph{LogEnhancer} and proposed an algorithm capable of completely disambiguating the call path of HDFS requests. Yuan \textit{et al.}~\cite{DBLP:conf/osdi/YuanPHLLTZS12} found that the majority of unreported failures were manifested via a generic set of error patterns (\textit{e.g.}, system call return errors) and proposed the tool \emph{Errlog} to proactively add pattern-specific logging statements by static code analysis.

As modern software becomes more complex, where-to-log has become an important but difficult decision, largely limited to the developer's domain knowledge. Around 60\% of failures due to software faults do not leave any trace in logs, and 70\% of the logging pattern aims to detect errors via a checking code placed at the end of a block of instructions~\cite{DBLP:conf/dsn/CinqueCNP10,DBLP:journals/tse/CinqueCP13}. Cinque \textit{et al.}~\cite{DBLP:journals/tse/CinqueCP13} concluded that the traditional logging mechanism has limited capacity due to the lack of a systematic error model. They further formalized the placement of the logging instruction and proposed to use system design artifacts to manually define \emph{rule-based logging} which utilizes error models about what cause errors to fail.
Zhu \textit{et al.}~\cite{DBLP:conf/icse/ZhuHFZLZ15} made an important first step towards the goal of ``learning to log''. They proposed a logging recommendation tool, \emph{LogAdvisor}, that learns the common logging rules on where-to-log from existing code via training a classifier and further leverages it for informative and viable recommendations to developers. Yao \textit{et al.}~\cite{DBLP:journals/ese/YaoPSSTS20} leveraged a statistical performance model to suggest the need for updating logging locations for performance monitoring. Li \textit{et al.}~\cite{li2020wherelog} proposed a deep learning framework to suggest where-to-log at the block level. Li \textit{et al.} also concluded that there might be similar rules regarding the implementation of logging mechanism across different systems and development teams, which agreed with the industrial survey by Pecchia \textit{et al.}~\cite{DBLP:conf/icse/PecchiaCCC15}.

The lack of strict logging guidance and domain-specific knowledge makes it difficult for developers to decide what-to-log. To address this need, Li \textit{et al.}~\cite{li2017log} employed ordinal regression model to suggest proper verbosity level based in software metrics. He \textit{et al.}~\cite{DBLP:conf/kbse/HeCHL18} conducted the first empirical study on the usage of natural language in logging statements. They showed the global (\textit{i.e.}, in a project) and local (\textit{i.e.}, in a file) repeatability of text descriptions. Furthermore, they demonstrated the potential of automated description text generation for logging statements. Liu \textit{et al.}~\cite{liu2019variables} proposed a deep learning-based approach to recommend variables in logging statements by learning embeddings of program tokens.
In order to troubleshoot transiently-recurring problems in cloud-based production systems, Luo \textit{et al.}~\cite{DBLP:conf/usenix/LuoNSMC18} put forward a new logging mechanism that assigns a blame rank to methods based on their likelihood of being relevant to the root cause of the problem. With the blame rank, logs generated by a method over a period of time are proportional to how often it is blamed for various misbehavior, thus facilitating diagnosis.


\subsubsection{Maintenance} \label{sec:practice-maintenance}

The maintenance of logging code has also attracted researchers' interest. The research on this aspect aims at (1) characterizing the maintenance and detecting anti-patterns of logging statements and (2) proposing new abstractions of logging.

\par{\textit{(1) Characterizing the maintenance and detecting anti-patterns of logging statements.}}
Anti-patterns in logging statements are bad coding patterns that undermine the quality and effectiveness of logging statements and increases the maintenance effort of projects. Many papers~\cite{DBLP:journals/ese/ShangNH15, DBLP:journals/ese/ChenJ17, DBLP:journals/ese/HassaniSST18, chen2019extracting} performed empirical studies to reveal the link between logs and defects. These papers observed a positive correlation between logging characteristics and post-release defects. Therefore, practitioners should allocate more effort to source code files with more logging statements.

Yuan \textit{et al.}~\cite{DBLP:conf/icse/YuanPZ12} made the first attempt to conduct a quantitative characteristic study of how developers log within four pieces of large open-source software. They described common anti-patterns and provided insights into where developers spend most of their efforts in modifying the log messages and how to improve logging practice. They further implemented a prototype checker to verify the feasibility of detecting unknown problematic statements using historical commit data. Chen and Jiang~\cite{DBLP:conf/icse/ChenJ17} and Hassani \textit{et al.}~\cite{DBLP:journals/ese/HassaniSST18} both studied the problem of how-to-log by characterizing and detecting the anti-patterns in the logging code. The analysis~\cite{DBLP:conf/icse/ChenJ17} of well-maintained open-source systems revealed six anti-patterns that are endorsed by developers. Chen and Jiang~\cite{DBLP:conf/icse/ChenJ17} then encoded these anti-patterns into a static code analysis tool to automatically detect anti-patterns in the source code. Li \textit{et al.}~\cite{DBLP:conf/icse/LiC0S19} developed an automated static analysis tool, \emph{DLFinder}, to detect duplicate logging statements that have the same static text description.

Just like feature code, logging code updates with time~\cite{DBLP:journals/ese/KabinnaBSSH18}. Moreover, logging statements are often changed without consideration for other stakeholders, resulting in sudden failures of log analysis tools and increased maintenance costs for such tools. Pecchia \textit{et al.}~\cite{DBLP:conf/icse/PecchiaCCC15} reviewed the industrial practice in the reengineering of logging code.
Kabinna \textit{et al.}~\cite{DBLP:conf/msr/KabinnaBSH16} empirically studied the migration of logging libraries and the main reasons for the migration. Li \textit{et al.}~\cite{li2017towards} derived and used a set of measures to predict whether a code commit requires log changes. Kabinna \textit{et al.}~\cite{DBLP:journals/ese/KabinnaBSSH18} later examined the important metrics for determining the stability of logging statements and further leveraged learning-based models (random forest classifier and Cox proportional hazards) to determine whether a logging statement is likely to remain unchanged in the future. Their findings were helpful to build robust log analysis tools by ensuring that these tools relied on logs generated by more stable logging statements. Li \textit{et al.}~\cite{DBLP:journals/ese/LiNJLWL20} designed a tool to learn log revision rules from logging context and modifications and recommend candidate log revisions.

\par{\textit{(2) Proposing new abstractions of logging.}}
Maintaining logging code along with feature code has proven to be error-prone~\cite{DBLP:journals/ese/ShangNH15, DBLP:journals/ese/ChenJ17, DBLP:journals/ese/HassaniSST18}. Hence, additional logging approaches~\cite{kiczales1997aspect, DBLP:conf/osdi/LockermanFKSAA018} have been proposed to resolve this issue. Kiczales \textit{et al.}~\cite{kiczales1997aspect} proposed a new programming paradigm that improves the modularity of the logging code. To tackle the ordering problem of logs in distributed systems, Lockerman \textit{et al.}~\cite{DBLP:conf/osdi/LockermanFKSAA018} introduced \emph{FuzzyLog} that featured strong consistency, durability, and failure atomicity.

\subsubsection{Performance}

The intermixing nature of the logging code and feature code usually incurs performance overhead, storage cost, and development and maintenance efforts~\cite{DBLP:conf/usenix/DingZLZLFZX15, DBLP:conf/asplos/YuanZPZS11, DBLP:conf/iwpc/MizouchiSII19, DBLP:journals/ese/ChowdhuryNHJ18, DBLP:journals/ese/ZengCSC19}. Tools like \emph{LogEnhancer}~\cite{DBLP:conf/asplos/YuanZPZS11}, \emph{Errlog}~\cite{DBLP:conf/osdi/YuanPHLLTZS12}, \emph{Log2}~\cite{DBLP:conf/usenix/DingZLZLFZX15}, and \emph{INFO-logging}~\cite{DBLP:conf/hotos/ZhaoRLSYZ17} all took performance into consideration while dealing with diagnosability and maintenance issues. Mizouchi \textit{et al.}~\cite{DBLP:conf/iwpc/MizouchiSII19} proposed a dynamical adjusting verbosity level to record irregular events while reducing performance overhead. Yang \textit{et al.}~\cite{DBLP:conf/usenix/YangPO18} proposed \emph{NanoLog}, a nanosecond scale logging system that achieved relatively low latency and high throughput by moving the workload of logging from the runtime hot path to the post-compilation and execution phases of the application. Chowdhury \textit{et al.}~\cite{DBLP:journals/ese/ChowdhuryNHJ18} were the first to explore the energy impact of logging in mobile apps. Zeng \textit{et al.}~\cite{DBLP:journals/ese/ZengCSC19} conducted a case study that characterized the performance impact of logging on Android apps.


\section{Log Compression}
\label{sec:log_compression}
After collecting logs by executing logging statements during runtime, logs are stored for failure diagnosis or sensitive operations auditing. Then, how to store logs efficiently becomes a challenging problem. 
Since large-scale software systems run on a 24 $\times$ 7 basis, the generated log size has been huge (\textit{e.g.}, 50 gigabytes per hour~\cite{mi2013toward}). Besides, many logs require the long-term storage for identifying duplicate problems and mining failure patterns from historical logs~\cite{yuan2019approach, amar2019mining,ding2014mining}. Auditing logs that record sensitive user operations are often stored for two years and even more to track system misuse in future. Archiving logs in such a huge volume for a long period brings inconceivably heavy burden to storage space, electrical power, and network bandwidth for transmission. To tackle these problems, a line of research has been focusing on log compression, which aims to remove the redundancy in a large log file and reduce its storage consumption. 

\subsection{Challenges for Log Compression}\label{sec:log_compress_chall}
Practically, log compression can be achieved by various approaches. The most straightforward way is to reduce the amount of logging statements in the source code or to set a less verbose log level (e.g., change from "INFO" to "ERROR")~\cite{li2017log} in runtime. Although effective, this method leads to information loss in logs, which impedes the log-based troubleshooting. Besides, it is a common practice in existing logging libraries (\textit{e.g.}, log4j) to compress log files by some general file compressors, such as gzip. However, the method is not tailored for the semi-structured log format, making it very ineffective in log compression. The structured fields (\textit{i.e.}, constant parts) are often repetitive across many raw logs, which consume a lot of storage space and should be handled specifically. To address these issues, a tailored log compression algorithm is highly in demand.

\subsection{Characteristics for Log Compression}
\label{sec:log_compress_prop}
An ideal log compressor should achieve a high compression ratio while imposing short compression and decompression time. The compression ratio denotes the ratio between the file sizes before and after the compression. A higher compression ratio indicates that less storage space is consumed and the compression algorithm is more effective. Additionally, log compression algorithm takes a certain amount of time to compress and decompress the file respectively, which is supposed to be as short as possible.
In this paper, we mainly explored 12 existing log compressing approaches and several empirical studies. As shown in Table~\ref{table:log_compression}, to clearly demonstrate the advantages and disadvantages of these approaches, we list several characteristics that concern the log compressors based on the following descriptions in the surveyed papers.
\textit{(1) Use of general compressors (GC).} Some log compressors utilize the general compressor as the backend technique after reformatting the logs. Specifically, these reformattings can transform raw logs to better formats that achieve a higher compression ratio. 
\textit{(2) General applicability} denotes whether the log compressor can be generally applicable to various log data formats. A general log compressor shows better utility in different scenarios. For example, as shown in the ``target data'' of Table~\ref{table:log_compression}, the log compressor proposed in~\cite{deorowicz2008sub} was specialized for Apache web log while other log compressors~\cite{christensen2013adaptive} can be applied to different log data types. 
\textit{(3) Scalability.} Since logs are often of a great volume in practice, the efficiency of log compression is crucial to the practical employment. A highly scalable compressor could save both the compression and decompression time, which thereby saves the log query time and supports real-time analysis.
\textit{(4) Heuristics.} The deployment would be more efficient and smoother if the log compressor requires little prior knowledge, \textit{e.g.}, heuristic rules on log formats and preprocessing procedures. The less heuristics a log compressor needs, the more generally applicable it would be.

\begin{table}[t]
\begin{center}
\caption{Summary of log compression approaches.}
\label{table:log_compression}
 \begin{tabular}{c|c|c|c|c|c}
 \toprule
\multicolumn{2}{c|}{\bf Methods} & GC	& Target Data	& Scalability	& Heuristics	\\ \midrule
  \multirow{5}{*}{{{\bf Bucket}}} 
& Balakrishnan \textit{et al.}~\cite{balakrishnan2006lossless} & Yes		& Blue Gene/L logs	& High	 & Yes \\
& Skibinski \textit{et al.}~\cite{skibinski2007fast}  & Yes		& All logs	& Low	& No	\\
& Hassan \textit{et al.}~\cite{hassan2008industrial} & No		& Telecom logs	& High	& Yes	 \\
& Christensen \textit{et al.}~\cite{christensen2013adaptive}  & Yes		& All logs	& High	& Yes \\
& Feng \textit{et al.}~\cite{feng2016mlc} & Yes		& All logs	& Middle & No \\
\midrule
  \multirow{4}{*}{{{\bf Dictionary}}} 
& Deorowicz \textit{et al.}~\cite{deorowicz2008sub} & No		& Apache web log	& Middle	& Yes \\
& Lin \textit{et al.}~\cite{lin2015cowic} & No	  	& Structured Logs	& High	& Yes	\\
& Mell \textit{et al.}~\cite{mell2014lightweight}  & Yes		& Structured Logs	& High	& No \\
& Racz \textit{et al.}~\cite{racz2004high} & Yes		& Web log 	& High	& Yes	\\
\midrule
  \multirow{3}{*}{{{\bf Statistics}}} 
& Hatonen \textit{et al.}~\cite{hatonen2003comprehensive} & No		& Structured Logs	& High	& Yes\\
& Meinig \textit{et al.}~\cite{meinig2019rough} & No  & All logs & High & Yes \\
& Liu \textit{et al.}~\cite{liu2019logzip} & Yes	& Structured Logs	& Low	& No	\\
\midrule
  \multirow{2}{*}{{{\bf Industry}}} 
& Splunk~\cite{splunk} & Yes	 	& All logs	& Low	& No    \\
& ElasticSearch~\cite{elk} & Yes	& All logs	& Low	& No	\\
\bottomrule
 \end{tabular} 
\end{center}
\vspace{-3pt}
\end{table}

\subsection{Log Compression Approaches}
\label{sec:log_compress_methods}
According to the techniques employed above, there are three categories of log compression methods: (1) \textit{bucket-based compression} that divides the log data into different blocks and compresses each block in parallel; (2) \textit{dictionary-based compression} that builds a dictionary for fields in the log and replaces strings by referring to the dictionary; and (3) \textit{statistics-based compressors} that apply complex statistical models to find correlations between logs before compression. Note that some log compression methods may involve two or more techniques, in which case we took the major compression technique as their category.

\subsubsection{Bucket-based Compression}
Bucket-based compression methods break the log data into different blocks by some log-specific characteristics and then compresses these blocks in parallel. For instance, many logs show the strong temporal locality or high similarity in certain fields, therefore, these logs can be gathered and then compressed. To tackle the problem that log data were often heterogeneous with varying patterns over time, Christensen \textit{et al.}~\cite{christensen2013adaptive} proposed to partition the log data into different buckets by leveraging the temporal locality and then compress these buckets in parallel. Hassan \textit{et al.}~\cite{hassan2008industrial} adopted a similar idea for telecom logs; however the objective was log event sequence instead of log messages. The log data is firstly broken into equal sized periods and compressed separately. Different from other studies, they use the compressed log event sequences to identify noteworthy usage scenarios for operational profile customization.

Besides, the bucket partition often serves as an initial step towards achieving a high compression ratio. For example, LogPack~\cite{skibinski2007fast} was a multi-tiered log compression method, in which each tier addressed one notion of redundancy. The first tier handled the local redundancy between neighboring lines. The second tier handled the global repetitiveness of tokens while the third tier handled all the remaining redundancy by employing a general compressor. Similarly, Balakrishnan \textit{et al.}~\cite{balakrishnan2006lossless} first observed a number of trends in the Blue Gene/L system logs and proposed several solution to compress the logs accordingly. For example, one trend was that most columns in adjacent logs tended to be the same. At last, the generic compression utilities were applied to further compress the log file.  Hence, they utilized a variant of delta encoding method which compared the log with preceding logs and encoded the differences only. In addition, Multi-level Log Compression (MLC)~\cite{feng2016mlc} divided logs with redundancy into different buckets by calculating the Jaccard similarity among logs. Then, the logs were condensed by a variant of delta encoding, followed by a general compressor to further improve the compression ratio.

\subsubsection{Dictionary-based Compression}
Dictionary-based compression removes the redundancy in log files by replacing repetitive strings with references to a dictionary. For example, some frequent strings (\textit{e.g.}, IP addresses) can be mapped to a condensed string in the dictionary. 
Lin \textit{et al.}~\cite{lin2015cowic} presented a column-wise independent compression for structured logs, which was also very similar to the work presented in~\cite{racz2004high, deorowicz2008sub} though their focuses were web logs. The general idea of these methods is as follows: At first, it splits a log entry into several columns by its fields. After observing  common properties in columns (\textit{e.g.}, common prefixes and suffixes), the method builds a dictionary-based model with Huffman code to compress each column separately. Many general compression techniques can then be applied to achieve a higher compression ratio, including move-to-front coding, phrase sequence substitution, etc. Similarly, Mell \textit{et al.}~\cite{mell2014lightweight} proposed a multi-step method for log compression. It first separated and sorted logs according to particular properties (\textit{e.g.}, value of the first column), and then it created a dictionary where logs are stored by hashing the field name, followed by serializing the dictionary and compressing logs with a general compressor.

\subsubsection{Statistics-based Compression}
Statistics-based methods realize the log compression by building a statistical model to identify possible redundancy in the log data. Unlike the above two categories of methods which are supported by  manually defined compression logic, the statistics-based compression automatically mines the compression rules. Hatonen \textit{et al.}~\cite{hatonen2003comprehensive} demonstrated a Comprehensive Log Compression (CLC) method to dynamically characterize and combine log data entries. Particularly, the method first identified frequently occurring patterns from dense log data by frequent pattern mining, and then linked patterns to the data as a data directory. 
Meinig \textit{et al.}~\cite{meinig2019rough} proposed an approach adopted from the rough set in the uncertainty theory. It treated the log file as a decision table with removable attributes identified by a one-time analysis of log data. Specifically, attributes that could roughly express each other are then collapsed to one representative attribute. The proposed method belongs to the set of lossy compression approaches which allow the information loss after compression. Lossy compression usually sacrifices the information completeness in raw logs for the high efficiency of log compression. Liu \textit{et al.}~\cite{liu2019logzip} proposed the Logzip method based on log parsing introduced in Section~\ref{sec:log_parsing}. Logzip 
first automatically extracted log templates from raw log messages by a statistical log parsing model. Then Logzip structuralized log templates and other information into intermediate representations, which were finally fed into a general compressor. 

\subsubsection{Others}
Recently, Yao \textit{et al.}~\cite{yaostudy} provided an empirical study on comparing the performance of general compressors on compressing log data against natural language data. Although the study was targeted on general compressors instead of specialized log compressors, their findings could potentially guide the design of tailored log compressors. Besides, Otten \textit{et al.}~\cite{otten2008using} presented a systematic review on general compression techniques, based on which they investigated the use of compression for log data reduction and the use of semantic knowledge to improve data compression. Different from the classic log compression, Hamou-Lhadj et al.~\cite{hamou2006summarizing} proposed to summarize the content of logs for the purpose of understanding software system behaviors. In detail, their approach removes implementation details such as utility information from the execution traces by ranking system components under a well-defined metric.

\section{Log Parsing}
\label{sec:log_parsing}

\begin{figure}[t!]
\centering
\includegraphics[scale=0.5]{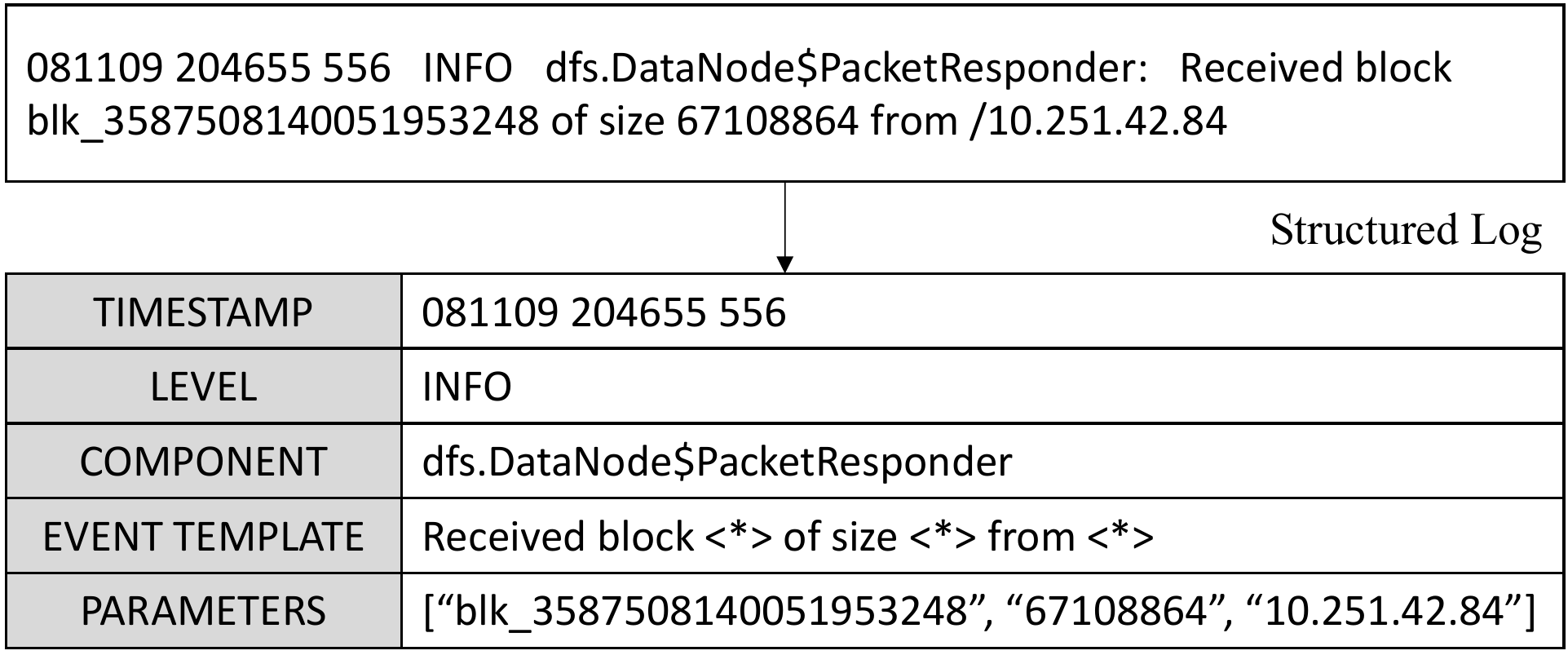}
\caption{A log parsing example.}
\label{fig:parse}
\vspace{-5pt}
\end{figure}

After log collection, log messages will be input into different downstream log mining tasks (\textit{e.g.}, anomaly detection) for further analysis. However, most of the existing log mining tools~\cite{xu2009detecting, he2018identifying} require structured input data (\textit{e.g.}, a list of structured log events or a matrix). Thus, a crucial step of automated log analysis is to parse the semi-structured log messages into structured log events. 


Fig.~\ref{fig:parse} presents a log parsing example, where the input is a log message collected from Hadoop Distributed File System (HDFS)~\cite{xu2009detecting}. A log message is composed of message header and message content. The message header is determined by the logging framework and thus it is relatively easy to extract, such as verbosity levels (\textit{e.g.}, "INFO"). In contrast, it is difficult to extract key information from the message content because it is mainly written by developers in free-form natural language. Typically, the message content contains \textit{constants} and \textit{variables}. Constants are the fixed text written by the developers (\textit{e.g.}, "Received") and describe a system event, while variables are the values of the program variables which carry dynamic runtime information. The goal of log parsing is to distinguish between \textit{constants} and \textit{variables}. All the constants form the \textit{event template}. The output of a log parser is a structured log message, containing an event template and the key parameters.

\subsection{Challenges for Log Parsing}\label{sec:parse_challenge}
As a key component, log parsing has become an appealing selling-point~\cite{rapid_autoparsing,logz_autoparsing,loggly_autoparsing} of many industrial log management solutions~\cite{DBLP:conf/icse/ZhuHLHXZL19} (\textit{e.g.}, Splunk~\cite{splunk}). However, these industrial solutions only support common log types such as Apache logs~\cite{rapid_autoparsing}. When parsing general log messages, they rely on regexs and ad-hoc scripts provided by developers. These scripts separate log messages into different groups, where log messages in the same group have the same event template. In modern software engineering, even with these existing industrial solutions, log parsing is still a challenging task due to three main reasons: (1) the large volume of logs and thus the great effort on manual regex construction; (2) the complexity of software and thus the diversity of event templates; and (3) the frequency of software updates and thus the frequent update of logging statements.





\subsection{Log Parser Characteristics}
We explore 15 automated log parsing approaches. To provide a clear glimpse of these parsers, we focus on three main characteristics: mode, coverage, and preprocessing. We follow the definition introduced by Zhu \textit{et al.}~\cite{DBLP:conf/icse/ZhuHLHXZL19}. Note that different from their work, which was an evaluation study providing benchmarks and open-source implementation of existing log parsers, this survey focuses on the methodological comparison of these parsers. In addition, while parsing rules can be constructed by analyzing source code~\cite{xu2009detecting}, we focus on parsers that only utilize logs as input.

\subsubsection*{Mode} Mode is the most important log parsing characteristic to consider. Depending on the usage scenarios of parsing, log parsers can be categorized into two modes: offline and online. Offline log parsers require all the log messages beforehand and parse log messages in a batch manner. To cope with the frequent software update, developers need to periodically re-run the offline parser to obtain the newest event templates. In contrast, online parsers parse the log messages in a streaming manner, which work seamlessly with the downstream log mining tasks.

\subsubsection*{Coverage} We denote coverage as the capability of a log parser to match all input log messages with event templates. Note that this is orthogonal to whether the matched event templates are correct or not. In Table~\ref{tab:parsesum}, ``Partial" indicates the parser can only parse part of the log messages (\textit{i.e.}, some logs will be matched with no event templates). For example, SLCT~\cite{vaarandi2003data} matches a log message with an event template only if the log message contains a specific ``frequent pattern" and thus leaving the remaining log messages unparsed. Therefore, SLCT achieves “Partial” coverage. ``Partial" coverage might lead to the neglection of crucial system anomalies.

\subsubsection*{Preprocessing} Preprocessing removes some \textit{variables} or replaces them by \textit{constants} based on domain knowledge. For example, IP addresses (\textit{e.g.}, 10.251.42.84) are typical variables in cloud systems' log messages. This step requires some manual efforts (\textit{i.e.}, constructing regexs for these variables).


\begin{table}[t]
\begin{center}
\caption{Summary of log parsing approaches.}
\label{tab:parsesum}
 \begin{tabular}{c|c|c|c|c}
 \toprule
\multicolumn{2}{c|}{\bf Methods} & Coverage & Preprocessing	& Technique	\\ \midrule
  \multirow{10}{*}{\rotatebox[origin=c]{90}{{\bf Offline}}} 
& SLCT~\cite{vaarandi2003data} & Partial	&  No	& Frequent pattern mining	 \\
& AEL~\cite{AEL_1}  & All	&  Yes	& Heuristics		\\
& LKE~\cite{qfu09} & All	&  Yes	& Clustering		 \\
& LFA~\cite{LFA_10}  & All	& No	& Frequent pattern mining	 \\
& LogSig~\cite{tang2011logsig} & All	&  No	& Clustering	 \\
& IPLoM~\cite{IPLoM09, IPLoM12} & All	&  No	& Iterative partitioning	 \\
& LogCluster~\cite{logcluster15} & Partial	&  No	& Frequent pattern mining	 \\
& LogMine~\cite{logmine16} & All	&  Yes	& Clustering	 \\
& POP~\cite{He2017TDSC} & All	&  Yes	& Iterative partitioning	 \\
& MoLFI~\cite{molfi18} & All	&  Yes	& Evolutionary algorithms	 \\

\midrule
  \multirow{5}{*}{\rotatebox[origin=c]{90}{{\bf Online}}} 
& SHISO~\cite{SHISO_13} & All	& No	& Clustering	 \\
& LenMa~\cite{lenma16} & All	  &  No	& Clustering		\\
& Spell~\cite{DuTKDE18}  & All	&  No	& Longest common subsequence	 \\
& Drain~\cite{He17ICWS, He2018Drain} & All	&  Yes	& Heuristics		\\
& Logram~\cite{Dai20Logram} & All	&  Yes	& Frequent pattern mining	 \\
\bottomrule

 \end{tabular} 
\end{center}
\vspace{-5pt}
\end{table} 

\subsection{Offline Log Parsing Approaches}\label{sec:parse_offline}

\subsubsection*{Frequent Pattern Mining} 
\textit{SLCT} (Simple Logfile Clustering Tool) \cite{vaarandi2003data} is the first research paper on automated log parsing. SLCT conducted two passes in total to obtain associated words. In the first pass, SLCT counted the occurrence of all the tokens and marked down the frequent words. These frequent words were utilized in the second pass to find out associated frequent words. Finally, for a log message, if it contained a pattern of associated frequent words, these words would be regarded as constants and employed to generate event templates. Otherwise, the log message would be placed into an outlier cluster without matched event templates.
\textit{LFA}~\cite{LFA_10} adopted a similar strategy as SLCT. Differently, LFA could cover all the log messages.

\subsubsection*{Clustering} \textit{LogCluster}~\cite{logcluster15} is similar to SLCT~\cite{vaarandi2003data}. Differently, LogCluster allowed variable length of parameters in between via a clustering algorithm. Thus, compared with SLCT, LogCluster is better at handling log messages of which the parameter length is flexible. For example, ``Download Facebook and install" and "Download Whats App and install" have the same event template ``Download <*> and install" while the length of the parameter (\textit{i.e.}, an app name) is flexible. \textit{LKE} (Log Key Extraction)~\cite{qfu09} was developed by Microsoft. LKE adopted a hierarchical clustering algorithms with a customized weighted edit distance metric. Additionally, the clusters were further partitioned by heuristic rules. \textit{LogSig} \cite{tang2011logsig} was a more recent clustering-based parser than LKE. Instead of directly clustering log messages, LogSig transformed each log message into a set of word pairs and clustered logs based on the corresponding pairs. \textit{LogMine}~\cite{logmine16} adopted an agglomerative clustering algorithm. It was implemented in map-reduce framework for better efficiency.

\subsubsection*{Heuristics} \textit{AEL}~\cite{AEL_1} employed a list of specialized heuristic rules. For example, for all the pairs like ``word=value," AEL regarded the ``value" as a variable and replaced it with a ``\$v" symbol. 


\subsubsection*{Evolutionary Algorithms} \textit{MoLFI}~\cite{molfi18} formulate log parsing as a multi-objective optimization problem and propose an evolutionary algorithm-based approach. Specifically, MoLFI employs the Non-dominated Sorting Genetic Algorithm II~\cite{deb2002fast} to search for a Pareto optimal set of event templates. Compared with other log parsers, the strength of MoLFI is that it requires little parameter tuning effort because the four parameters required by MoLFI has effective default values. However, because of the adoption of evolutionary algorithm, the MoLFI is slower than most of the parsers.

\subsubsection*{Iterative Partitioning} \textit{IPLoM} \cite{IPLoM09, IPLoM12} contained three steps and partitioned log messages into groups in a hierarchical manner. (1) Partition by log message length. (2) Partition by \textit{token position}. The position containing the least number of unique words is ``token position". Partitioning was conducted according to the words in the token position. (3) Partition by mapping. Mapping relationships were searched between the set of unique tokens in two token positions, which were selected using a heuristic criterion. \textit{POP}~\cite{He2017TDSC} is a parallel log parser that utilizes distributed computing to accelerate the parsing of large-scale software logs. POP can parse 200 million HDFS log messages in 7 mins, while most of the parsers (\textit{e.g.}, LogSig) failed to terminate in reasonable time.


\subsection{Online Log Parsing Approaches}\label{sec:parse_online}

\subsubsection*{Clustering} \textit{SHISO}~\cite{SHISO_13} is the first online log parsing approach. SHISO used a tree-form structure to guide the parsing process, where each node was correlated with a log group and an event template. The numbers of children nodes in all the layers were the same and were manually configured beforehand. During the parsing process, SHISO traversed the tree to find the most suitable log group by comparing the log message and the event templates in the corresponding log groups. SHISO is sensitive to path explosion and thus its efficiency is often unsatisfactory. \textit{LenMa}~\cite{lenma16} is similar to SHISO. LenMa encodes each log message into a \textit{length vector}, where each dimension records the number of characters of a token. For example, ``Receive a file." would be vectorized as [7, 1, 5]. During parsing, LenMa would compare the length vectors of the log messages.

\subsubsection*{Longest Common Subsequence} Similar to SHISO and LenMa, \textit{Spell}~\cite{DuTKDE18} maintained a list of log groups. To accelerate the parsing process, Spell utilized specialized data structures: prefix tree and inverted index. In addition, Spell provided a parallel implementation.

\subsubsection*{Heuristics} \textit{Drain}~\cite{He17ICWS} maintained log groups via the leaf nodes in the tree. The internal nodes of the tree embedded different heuristic rules. The extended version of Drain~\cite{He2018Drain} was based on a directed acyclic graph that allowed log group online merging. In addition, it provided the first automated parameter tuning mechanism for log parsing.


\subsubsection*{Frequent Pattern Mining} \textit{Logram}~\cite{Dai20Logram} is the current state-of-the-art parser. Different from the existing approaches that count frequent tokens, Logram considered frequent n-gram. The core insight of Logram is: frequent n-gram are more likely to be constants. Note that Logram assumed developers had some log messages on hand to construct the dictionary.


\section{Log Mining}
\label{sec:log_mining}

Log mining employs statistics, data mining, and machine learning techniques for automatically exploring and analyzing large volume of log data to glean meaningful patterns and informative trends.  The extracted patterns and knowledge could guide and facilitate monitoring, administering, and troubleshooting of software systems.
In this section, we first elaborate on challenges encountered in log mining (Section~\ref{sec:log_mining_challenges}). Then, we describe the general workflow of log mining (Section~\ref{sec:log_mining_workflow}). Finally, we introduce three major log mining tasks for reliability engineering, including \textit{anomaly detection} (Section~\ref{sec:log_anomaly_detection}), \textit{failure prediction} (Section~\ref{sec:log_failure_prediction}), and \textit{failure diagnosis} (Section~\ref{sec:log_failure_diagnosis}). Other relevant studies with relatively lesser popularity are laid out in Section~\ref{sec:log_mining_others}.

\subsection{Challenges of Log Mining}
\label{sec:log_mining_challenges}

Traditionally, engineers perform simple keyword search (such as "error", "exception", and "failed") to mine suspicious logs that might be associated with software problems, \textit{e.g.}, component failures. Some rule-based tools~\cite{rouillard2004real,prewett2003analyzing,hansen1993automated} have been developed to detect software problems by comparing logs against a set of manually defined rules which describe normal software behaviors. However, due to the ever-increasing volume, variety, and velocity of logs produced by modern software, such approaches fall short for being labor-intensive and error-prone. Moreover, suspicious logs are often overwhelmed by logs generated during software normal executions. Manually sifting through a massive amount of logs to identify failure-relevant ones is like finding a needle in a haystack.

Additionally, inspecting logs for software troubleshooting often requires engineers to possess descent knowledge about the software. However, modern software systems usually consist of many components developed by different engineers, leading to the generation of heterogeneous logs, which makes troubleshooting beyond the ability of a single engineer. Moreover, due to the high complexity of modern software systems, failures could stem from various sources of software and hardware issues. Examples include software bugs, hardware damage, OS crash, service exception, etc. In addition, promptly pinpointing to the root cause by inspecting logs highly relies on engineers' expertise and experience. However, such knowledge is often not well accumulated, organized, and documented. Therefore, sophisticated ways to conduct automatic log mining are in high demand.



\subsection{A General Workflow of Log Mining}
\label{sec:log_mining_workflow}

\begin{figure}[t!b]
\centerline{\includegraphics[width=1.0\columnwidth]{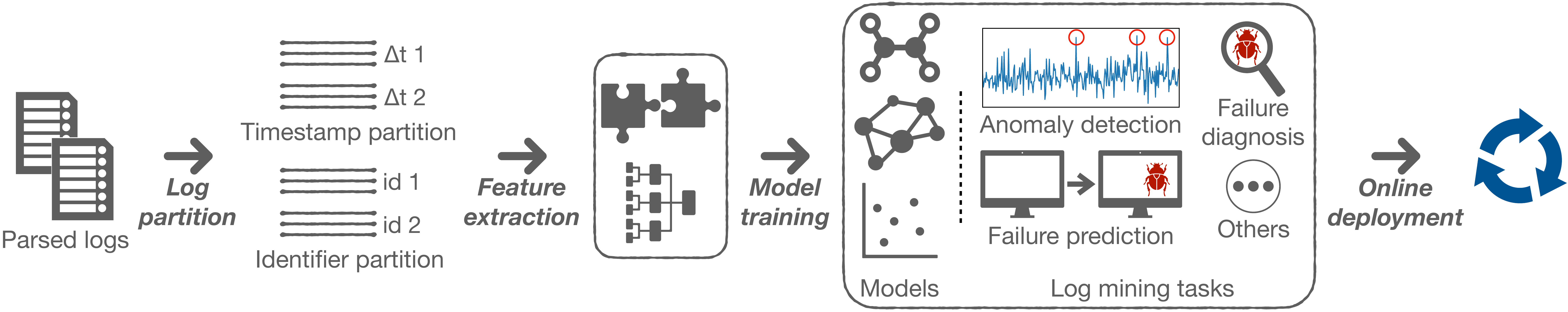}}
\caption{A general workflow of log mining.}
\label{fig:lm_workflow}
\vspace{-5pt}
\end{figure}

The general workflow of log mining is illustrated in Fig.~\ref{fig:lm_workflow}, which mainly consists of four steps, \textit{i.e.}, log partition, feature extraction, model training, and online deployment.

\subsubsection{Log Partition}

Modern software systems often adopt a microservice architecture and comprise a large number of modules operating in a multi-threaded environment. Different microservices or modules often aggregate their execution logs into a single log file, which hinders automated log mining. To tackle the problem, interleaved logs should be partitioned into different groups, each of which represents the execution of individual system tasks. The partitioned log group is the basic unit in extracting features before constructing log mining models. Moreover, studies~\cite{he2016experience} show that the way to partition logs impacts the log mining performance. In the literature, we observe that two elements often serve as the log partitioner, namely, \textit{timestamp} and \textit{log identifier}.

\textbf{Timestamp}. It records the occurrence time of each log, which is a fundamental feature supported by many logging libraries, such as Log4j. Through regexs, the timestamp with different formats can be easily extracted from raw logs in log parsing phase (Section~\ref{sec:log_parsing}).
In general, two strategies are often adopted to conduct timestamp-based log partition, \textit{i.e.}, fixed window and sliding window. Fixed window has a predefined window size, which means the time span or time interval (\textit{e.g.}, 30 minutes) used to split chronologically sorted logs. Extended from the fixed window, sliding window allows the overlapping between two consecutive fixed windows. The sliding window has two attributes, \textit{i.e.}, window size and step size. The step size indicates the forwarding distance of the window alone the time axis to generate log partitions, which is often smaller than the window size.

    

\textbf{Log Identifier}. It is a token identifying a series of related operations or message exchanges of the system. For instance, HDFS logs adopt \textit{block\_id} to record the operations (\textit{e.g.}, allocation, replication, and deletion) on a specific block. Common log identifiers include \textit{user ID}, \textit{task/session/job ID}, \textit{variable/component name}, etc., which can be extracted by log parsing. Compared to timestamp, log identifier is a clearer and more definite signal for partitioning logs. Therefore, many research studies~\cite{qfu09} employed the log identifier for its ability in distinguishing logs of different task executions.
However, the log identifiers might be non-unique in representing distinct system entities in different logs, such as virtual machines, physical machines, and networks~\cite{yu2016cloudseer}. This is because, for example, identifiers may not be propagated and synchronized across different services, one thread or process may serve more than one request through multiplexing, etc. Therefore, dedicated algorithms have been developed to address this problem and we elaborate on them later.

\subsubsection{Feature Extraction}
To analyze logs automatically, textual logs in a log partition should be transformed into appropriate formats that could fit machine learning algorithms. Through reviewing the literature, we identified two categories of log-based features, namely, numerical feature and graphical feature. Particularly, numerical feature is the mainstreaming feature which is widely used in the log analysis community.

\textbf{Numerical Feature}. It represents log's statistical properties including numerical and categorical fields that can be directly extracted from logs. It conveys the information of a log partition into a numerical vector representation. Particularly, numerical features employed by most of the existing work are similar and share some typical forms. In the following, we compactly summarize four types of such features.




\begin{itemize}
	\item \textit{Log event sequence}: A sequence of log events recording system's activities. Particularly, each element can simply be the log event ID or log embedding vector, \textit{e.g.}, learned by word2vec algorithms~\cite{mikolov2013distributed}.
	
	\item \textit{Log event count vector}: A feature vector recording the log events occurrence in a log partition, where each feature denotes a log event type and the value counts the number of occurrence.
	
	\item \textit{Parameter value vector}: A vector recording the value of parameters (\textit{i.e.}, variables extracted by log parsing) that appear in logs.
	
	\item \textit{Ad-hoc features}: A set of relevant and representative features extracted from logs, which are defined using domain knowledge on the object software system and problem context. For example, Zhou \textit{et al.}~\cite{zhou2019latent} manually identified various types of features from system trace logs, including configuration (\textit{e.g.}, memory limit, CPU limit), resource (\textit{e.g.}, memory consumption, CPU consumption), etc. These features profile the typical health states of a system.
\end{itemize}

\textbf{Graphical Feature}. To discover the hierarchical and sequential relations (\textit{e.g.}, dependency and co-occurrence) between system components and events with logs, the graphical feature usually produces a directed graph model characterizing system behaviors, \textit{e.g.}, the execution path of a process.
The graphical features serve as the foundation for a variety of downstream log mining tasks (such as monitoring~\cite{zhao2016non,yu2016cloudseer} and diagnosis~\cite{nandi2016anomaly,amar2018using,bao2019statistical}). For example, log-based behavioral differencing~\cite{goldstein2017experience} can identify system executions that are derived from system normal behaviors, which has various applications in system evolution, testing, and security~\cite{bao2019statistical}.
In this section, we briefly introduce the algorithms used by existing work for graphical feature extraction, which can be roughly categorized into two lines of work.

The first line of work leverages the objects identified in logs, \textit{e.g.}, process ID and system component name, to construct graphs for system state monitoring. As objects often demonstrate complicated hierarchical relations, it usually needs a more sophisticated algorithm for log partition. For instance, to represent execution structure and object hierarchy in logs, Zhao \textit{et al.}~\cite{zhao2016non} constructed a stack structure graph by considering the 1:1, 1:n, and n:m mappings among different objects. 
To tackle the challenge that unique identifiers are often unavailable, Yu \textit{et al.}~\cite{yu2016cloudseer} proposed to group interleaved logs based on a common set of identifiers. Each log contains an identifier set that acts as a state node of the graph, and transitions are added by examining the subset and superset relations of different identifier sets.

The other line of work makes use of the underlying statistical distribution of log events, \textit{e.g.}, the order of log events, the temporal and spatial locality of dependent log events, to build various graphs for system behavior analysis. Particularly, some work aims at recovering the exact behavioral model from log traces. Log partition therefore requires that identifiers can tie together the set of events associated with a program execution. For example, Amar \textit{et al.}~\cite{amar2018using} presented two variants of k-Tails~\cite{biermann1972synthesis}, namely 2KDiff and nKDiff, to build Finite State Machine (FSM) for log differencing. 2KDiff computes and highlights the differences between two log files containing a set of partitioned logs; while nKDiff is able to conduct the comparison for multiple log files at once. Later Bao \textit{et al.}~\cite{bao2019statistical} extended their work by proposing s2KDiff and snKDiff, which take into consideration the frequencies of different behaviors. Busany \textit{et al.}~\cite{busany2016behavioral} studied the scalability problem of existing behavioral log analysis algorithms by extracting Finite State Automaton (FSA) models or temporal properties from logs. We refer readers to this paper for more related studies.

However, as previously discussed, a unique identifier for each execution trace is not always available, especially in distributed applications and systems. Therefore, some work attempts to mine behavioral models using interleaved logs. For example, Lou \textit{et al.}~\cite{lou2010mining_b} employed statistical inference to learn the temporal dependencies among log events, which is also studied in similar work including~\cite{beschastnikh2014inferring,beschastnikh2011leveraging}. Nandi \textit{et al.}~\cite{nandi2016anomaly} computed the nearest neighbor groups to capture the temporal co-occurrence of log events more accurately. The resulted model is a program Control Flow Graph (CFG) spanning distributed components. Particularly, the correlation between two components is calculated using either the Jaccard similarity or the Bayesian conditional probability approach. Du \textit{et al.}~\cite{du2017deeplog} described two methods to capture service executions by FSA models. The first method leverages the trained log anomaly detection model, whose predictions encode the underlying path of task execution. The second method builds a matrix where each entry represents the co-occurrence probability of two log keys appearing together within a predefined distance.

\subsubsection{Model Training}

In this process, appropriate algorithms are selected based on the problem at hand and selected models are trained based on the extracted features. A variety of machine learning algorithms have been proposed, and we leave more details in the following sections. This step is often conducted in an offline manner.
Moreover, a fundamental assumption of various log mining tasks is that the majority of logs should exhibit patterns that conform to system's normal behaviors. For example, in anomaly detection, different models are trained to capture various patterns from different perspectives and used to detect anomalies that lack the desireable properties. 

\subsubsection{Online Deployment}

Once the model is trained offline, it could be deployed to real-world software systems for various log mining tasks. For example, an anomaly detection model can be integrated into software products to detect malicious system behaviors and raise alarms in real time. To tackle the challenge of pattern change caused by system upgrades, several studies~\cite{du2017deeplog,lin2016log} supported online update of a previously trained model to adapt to unprecedented log patterns.

\begin{figure}[t!b]
\centerline{\includegraphics[width=0.6\textwidth]{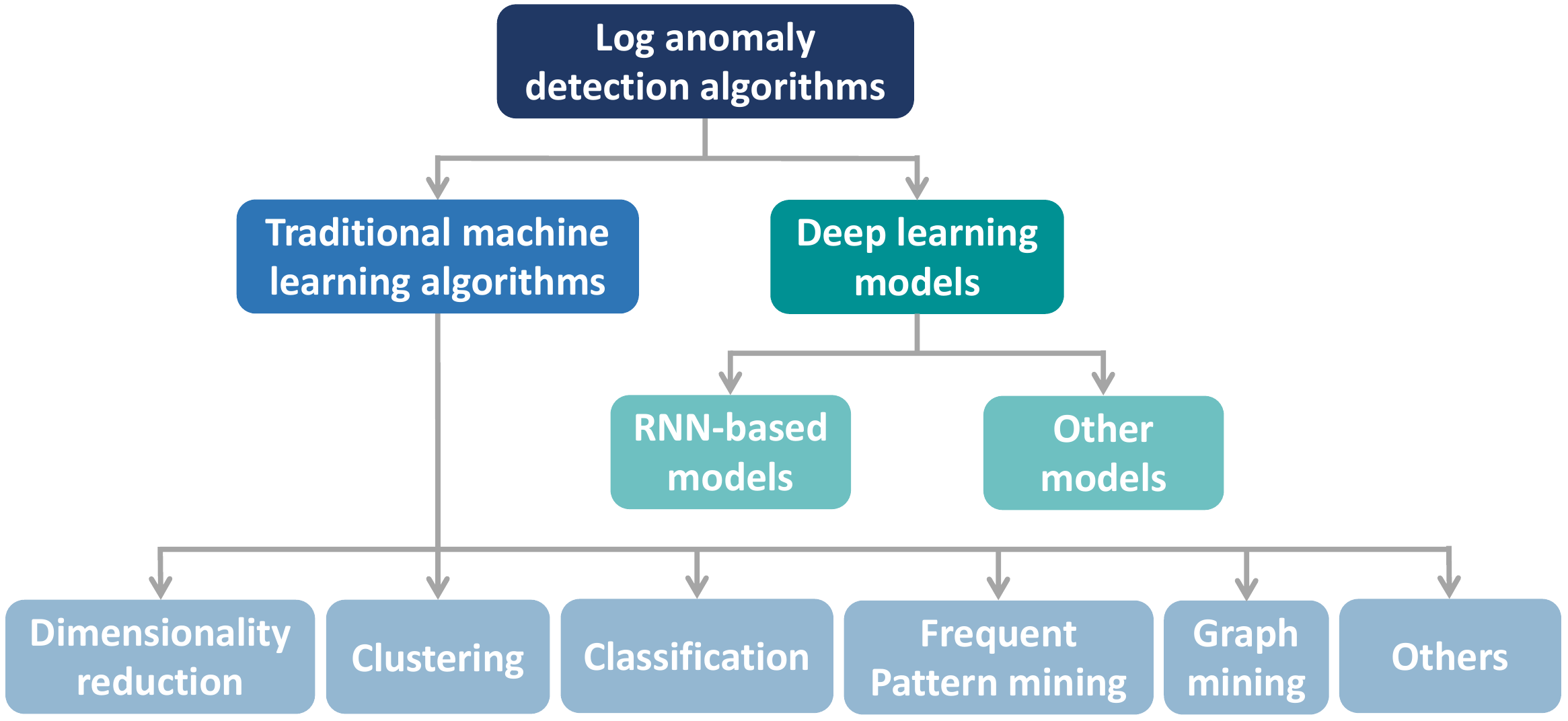}}
\caption{A taxonomy of log anomaly detection algorithms.}
\label{fig:ad_algorithms}
\vspace{-5pt}
\end{figure}

\subsection{Anomaly Detection}
\label{sec:log_anomaly_detection}
\subsubsection{Problem Formulation} Anomaly detection is the task of identifying system anomalous patterns that do not conform to expected behaviors on log data. Typical anomalies often indicate possible error, fault or failure in software systems. In this section, we elaborate on existing research work in this field based on approaches adopted therein. As shown in Fig.~\ref{fig:ad_algorithms}, we categorize the approaches into two broad classes: traditional machine learning algorithms and deep learning models. The surveyed approaches are listed in Table~\ref{tab:anomaly_detection_summary} together with several interesting properties. Specifically, we summarize the algorithm/model and feature used by an approach and whether or not each approach is unsupervised and online. Particularly, unsupervised approaches do not require labels for model training; and an online approach is one that can process its input piece-by-piece in a streaming fashion. In this paper, we mainly focus on general types of anomalies that are natural imperfections of software systems and will directly impact system's reliability. External attacks and intrusions (such as cyber attacks and malicious activities) that are more relevant to system's security go out of the scope for this paper.


\begin{table}[t]
\begin{center}
\caption{Summary of log anomaly detection approaches.}
\label{tab:anomaly_detection_summary}
 \begin{tabular}{c|c|c|c|c|c}
  \toprule
\multicolumn{2}{c|}{\bf Methods} & Algorithm/Model & Feature & Unsupervised & Online \\
\midrule
  \multirow{13}{*}{\rotatebox[origin=c]{90}{{\textbf{Traditional machine learning}}}} 
& Xu \textit{et al.}~\cite{xu2009detecting} & PCA & $\star$~$\dagger$ & Yes & No \\
& Lin \textit{et al.}~\cite{lin2016log} & Clustering & $\ast$ & Yes & No \\
& He \textit{et al.}~\cite{he2018identifying} & Clustering & $\ast$~$\star$ & Yes & No \\
& Liang \textit{et al.}~\cite{liang2007failure} & SVM & $\ddagger$ & No & No \\
& Kimura \textit{et al.}~\cite{kimura2018proactive} & SVM & $\ddagger$ & No & No \\
& Xu \textit{et al.}~\cite{xu2009online} & Frequent pattern mining & $\ast$~$\star$ & Yes & Yes \\
& Shang \textit{et al.}~\cite{shang2013assisting} & Frequent pattern mining & $\ast$ & Yes & No \\
& Lou \textit{et al.}~\cite{lou2010mining_a} & Frequent pattern mining & $\star$ & Yes & No \\
& Farshchi \textit{et al.}~\cite{farshchi2015experience} & Frequent pattern mining & $\star$ & Yes & No \\
& Nandi \textit{et al.}~\cite{nandi2016anomaly} & Graph mining & $\P$ & Yes & No \\
& Lou \textit{et al.}~\cite{lou2010mining_b} & Graph mining & $\P$ & Yes & No \\
& Yamanishi \textit{et al.}~\cite{yamanishi2005dynamic} & Statistical model & $\ast$ & Yes & No \\
& He \textit{et al.}~\cite{he2016experience} & Logistic regression & $\star$ & No & No \\

\midrule
  \multirow{6}{*}{\rotatebox[origin=c]{90}{{\textbf{Deep learning}}}} 
& Du \textit{et al.}~\cite{du2017deeplog} & LSTM model & $\ast$~$\dagger$ & Yes & Yes \\
& Zhang \textit{et al.}~\cite{zhang2019robust} & LSTM classification model & $\ast$ & No & No \\
& Meng \textit{et al.}~\cite{meng2019loganomaly} & LSTM model & $\ast$~$\star$ & Yes & Yes \\
& Xia \textit{et al.}~\cite{xia2020loggan} & LSTM-based GAN model & $\ast$ & Yes & Yes \\
& Lu \textit{et al.}~\cite{lu2018detecting} & CNN model & $\ast$ & No & No \\
& Liu \textit{et al.}~\cite{liu2019log2vec} & Graph embedding model & $\P$ & Yes & No \\
\midrule

\multicolumn{6}{l}{$\ast$~Log event sequence, $\star$~Log event count vector, $\dagger$~Parameter value vector} \\
\multicolumn{6}{l}{$\ddagger$~Ad hoc features, $\P$~Graphical feature}\\
\bottomrule
 \end{tabular} 
\end{center}
\vspace{-4pt}
\end{table}

\subsubsection{Traditional Machine Learning Algorithms}

Traditional machine learning algorithms usually perform on top of features explicitly provided by practitioners, \textit{e.g.}, log event count vector. Particularly, the anomaly detection task can be formalized into different types and solved with different algorithms such as clustering, classification, regression, etc.

\textbf{Dimensionality Reduction}. It transforms data of high dimension into a low-dimension representation such that some meaningful properties of the original data can be retained in the low-dimension space. Principal Components Analysis (PCA) is one of the most popular algorithms of this kind. By projecting data points to the first \textit{k} principal components, anomalies can be identified if the projected distance is larger than a threshold. 
PCA was first applied by Xu \textit{et al.}~\cite{xu2009detecting,xu2009online} to mine system problems from console logs. Particularly, both log event count vector and parameter value vector are constructed and fed into a PCA model for anomaly detection.

\textbf{Clustering}. As an unsupervised method, clustering-based anomaly detection groups log-based feature vectors into different clusters such that vectors in the same cluster are more similar to each other (and as dissimilar as possible to vectors from other clusters). Clusters that contain very few data instances tend to be anomalous. For example, Lin \textit{et al.}~\cite{lin2016log} proposed LogCluster, which clusters log sequences and recommends a representative sequence to assist developers quickly identify potential problems. The representative sequence is selected by calculating the cluster centroid.
Furthermore, He \textit{et al.}~\cite{he2018identifying} proposed the Log3C framework to incorporate system KPIs into the identification of impactful problems in service systems. In particular, they proposed a cascading clustering algorithm to group numerous log sequences promptly. They finally use a multivariate linear regression model to identify impactful problems that lead to KPI degradation.




\textbf{Classification}. Anomaly detection with classification categorizes the log partition into either normal or anomalous type, where anomalous examples are derived from normal ones in terms of some statistical properties. Support Vector Machine (SVM) is a supervised classification method that is commonly used for log anomaly detection. In~\cite{liang2007failure}, Liang et al. vectorized log partitions by identifying six types of features, including events number in a time window, accumulated events number, etc. Based on these features, they applied four classification models for anomaly detection, \textit{i.e.} including SVM and nearest neighbor predictor. In addition, Kimura et al.~\cite{kimura2018proactive} proposed a log analysis method for proactive failure detection based on logs' characteristics, including frequency, periodicity, burstiness, and correlation with maintenance and failures. A SVM model with the Gaussian kernel is employed for detecting failures.

\textbf{Frequent Pattern Mining}. It aims to discover the most frequent item sets and sub-sequences in a log dataset that characterize system's normal behaviors. Data instances that do not conform to the frequent patterns will be reported as anomalies. The presence of specific log events and the order of log events can both constitute patterns. For example, Xu \textit{et al.}~\cite{xu2009online} mined sets of log messages that often co-occur to detect abnormal execution traces in an online setting. Compared to offline approaches, online pattern matching can quickly identify benign system executions, thus striking a balance between accuracy and efficiency.
Similarly, Lim \textit{et al.}~\cite{lim2008log} searched for item sets that were commonly associated with different failure types. Such item sets can assist developers in predicting and characterizing failures.
Other approaches mine the sequential patterns by log events to discover anomalies or bugs~\cite{shang2013assisting,lu2018cloudraid}.


Additionally, Lou et al. ~\cite{lou2010mining_a} were the first to consider mining invariants among log messages for system anomaly detection.
Two kinds of invariants that model the relations between the number of different log messages are derived: (1) invariants in textual logs that describe the equivalence relation; and (2) invariants as a linear equation, which is the linear independence relation. Farshchi \textit{et al.}~\cite{farshchi2015experience} proposed a similar approach which mines the correlation and causation relations between log events and cloud system metric changes. Specifically, they employed a regression-based approach to learn a set of assertions modeling the linear relations. Anomaly detection is then conducted by monitoring log event streams and checking the compliance of metrics against the assertions.

\textbf{Graph Mining}. The set of approaches mainly employs graphical features, \textit{i.e.}, various graph models (Section~\ref{sec:log_mining_workflow}), to identify behavioral changes of complex systems, which can be used for early detection of anomalies and allowing proactive actions for correction. For example, Nandi \textit{et al.}~\cite{nandi2016anomaly} introduced a CFG mining approach to detect anomalous runtime behaviors of distributed applications from execution logs, including both sequence anomaly and distribution anomaly. A sequence anomaly is raised when an expected child is missing for a parent node within the given time interval; a distribution anomaly is raised when an edge probability is violated. In~\cite{qfu09}, Fu et al. modeled the execution behaviors of each system module as a state transition graph by learning an FSA from log sequences. 
Each transition in the learned FSAs corresponds to a log key. For each state transition, two types of information (\textit{i.e.}, the time consumed and circulation number) are recorded and applied to detect two performance issues: low transition time and low transition loop. By employing the Gaussian distribution to model the state transition in a distributed system,  low-performance transitions can be automatically identified by setting a proper threshold.

\textbf{Other Statistical Models}. There are some algorithms that do not belong to the above categories. For example, Yamanishi \textit{et al.}~\cite{yamanishi2005dynamic} employed a mixture of Hidden Markov Models to monitor syslog behaviors. Particularly, the model is learned with an online discounting learning algorithm by dynamically selecting an optimal number of mixture components. Anomaly scores are assigned using universal test statistics whose threshold can be dynamically optimized.
He \textit{et al.}~\cite{he2016experience} employed a logistic regression model to detect anomalies. They adopted event count vectors as the feature and trained the model with a set of labeled data. A testing instance is declared as anomalous when the probability estimated by the logistic function is greater than 0.5.
Nagaraj \textit{et al.}~\cite{nagaraj2012structured} diagnosed performance issues for large-scale distributed systems by finding a set of inter-related log event occurrences and variable values that exhibit the largest divergence across logs sets. Specifically, they first separated logs into two sets according to some performance metrics (\textit{e.g.}, runtime). Then, they recommended log events or state variables that contribute the most to the performance difference via t-tests.

\subsubsection{Deep learning models} Deep learning uses a  multiple-layer architecture (\textit{i.e.}, neural networks) to progressively extract features from inputs with different layers addressing different levels of feature abstraction. Due to the exceptional ability in modeling complex relationships, neural networks are widely applied in log-based anomaly detection. In the literature, we observed a significant portion of papers adopting the Recurrent Neural Network (RNN) model and its variants. Hence, we classify the deep learning  models into RNN-based models and other models.


\textbf{RNN-based Models}. Models belonging to the RNN family (such as LSTM model and GRU model) are commonly used to automatically learn the sequential patterns in log data. Anomalies are raised when log patterns deviate from the model's predictions. For example, Du \textit{et al.}~\cite{du2017deeplog} proposed DeepLog, which utilizes an LSTM model to learn system's normal execution patterns by predicting the next log event given a sequence of preceding log events. However, some anomalies may manifest themselves as an irregular parameter value instead of a deviation from a normal execution path. Hence, DeepLog also applies an LSTM model for checking the validity of parameter value vectors.
Many existing studies assume that the log data are stable over time and the set of distinct log events is fixed and known. However, Zhang \textit{et al.}~\cite{zhang2019robust} found that log data often contain previously unseen log events or log sequences, demonstrating log instability.
To tackle this problem, they proposed LogRobust to extract the semantic information of log events by leveraging off-the-shelf word vectors and then applied a bidirectional LSTM model to detect anomalies.

In terms of capturing log's semantics, Meng \textit{et al.}~\cite{meng2019loganomaly} found existing word2vec models did not distinguish well between synonyms and antonyms. Therefore, they trained a word embedding model to explicitly consider the information of synonyms and antonyms. Meng \textit{et al.}~\cite{mengsemantic} further extended their work by proposing a semantic-aware representation framework for online log analysis. The issues of log-specific word embedding and out-of-vocabulary (OOV) are both addressed.
Recently, Zuo \textit{et al.}~\cite{zuo2020intelligent} combined the transaction-level topic modeling for learning the embedding of logs, where a transaction is a group of logs sequentially occurring in a time window. To address the problem of insufficient labels, Chen \textit{et al.}~\cite{chen2020cross} applied transfer learning to share anomalous knowledge between two software systems. Specifically, they first trained an LSTM model on the data with sufficient anomaly labels to extract sequential log features, which were then fed into fully connected layers for anomaly classification. Next, the LSTM model was fine-tuned with logs from another system with limited labels, while the fully connected layers were fixed.

\textbf{Other Deep Learning Models}. Besides the RNN-based models, other model architectures also play a role in detecting anomalies with logs. For example, 
Xia \textit{et al.}~\cite{xia2020loggan} proposed LogGAN, an LSTM-based Generative Adversarial Network (GAN). Like all GAN-style models, LogGAN comprises a generator and a discriminator. The generator attempts to capture the distribution of real training data and synthesizes plausible examples;
while the discriminator tries to distinguish fake instances from real and synthetic data. 
Effort is also devoted to explore the feasibility of Convolutional Neural Networks (CNN) for anomaly detection. Specifically, Lu \textit{et al.}~\cite{lu2018detecting} first utilized a word embedding technique to encode logs into two-dimension feature matrices, upon which CNN models with different filters were then applied for anomaly detection. Liu \textit{et al.}~\cite{liu2019log2vec} proposed log2vec, a graph embedding based method for cyber threat detection. Specifically, they first converted logs into a heterogeneous graph using heuristic rules and then learned the embedding of each log entry by a graph representation learning approach. Based on the embedding vectors, logs were grouped into clusters, whose size smaller than a threshold would be reported as malicious.

\subsection{Failure Prediction}
\label{sec:log_failure_prediction}
\subsubsection{Problem Formulation}
Anomaly detection aims to detect anomalous status or unexpected behavior patterns which may or may not cause failures. Differently, failure prediction attempts to proactively generate early warnings to prevent server failures, which often lead to unrecoverable states.
The objective and role of anomaly detection and failure prediction are different yet both are important. 
Therefore, the exploration of failure prediction techniques is significant to reliability engineering.
Generally, when software systems deviate from fulfilling a required system  function~\cite{laprie1995dependable}, a failure occurs and it often shows human-perceivable symptoms. The failure could lead to unintended results and user dissatisfaction, especially for large-scale software systems. Traditional ways of failure management (such as anomaly detection) are mostly passive, which deal with failures after they have happened. In contrast, failure prediction aims to proactively predict the failure before it happens. A common practice is to leverage the valuable logs to proactively predict failures. For example, in~\cite{zhang2018prefix}, switch failures in data center networks are predicted from current status and curated historical hardware failure cases. Usually, input data of the predictive model are system logs, which record the system status, changes in configuration, and operational maintenance, etc.






According to the source of failures, failure prediction could mainly be categorized into two scenarios: 
a) Prediction of independent failures in homogeneous systems, \textit{e.g.}, high-performance computing (HPC) systems.
Most existing approaches focus on how to leverage the sequential information to predict the failure of each single component, \textit{i.e.}, sequence-based methods.
b) Prediction of outages caused by a collection of heterogeneous devices or components, which is widely observed in large-scale cloud systems.
The mainstreaming approach tends to explore useful hints in the relationship between multiple heterogeneous components and makes prediction from relation-mining algorithms.
\subsubsection{Homogeneous Systems} In homogeneous systems, the mainstreaming approaches are based on modeling the sequential information that represents the system status.
A typical homogeneous system is large-scale supercomputers which may encounter faults (e.g., component failures) every day.
Exascale systems are expected to combat with higher fault rates due to enormous number of system components and higher workload~\cite{das2018desh}.
However, triggering resilience-mitigating mechanisms is difficult since there are no obvious, well-defined failure indicators, which rely on a deeper understanding of the faults caused by hardware and software components.
Therefore, logs become the most reliable information source to monitor the system health.
Furthermore, to catch early warning signs of these failures, there are additional applications that scan through the monitoring data from the system logs and proactively generate alerts.
Some of these alerts are generated from the manually defined thresholds on the collected system performance data (\textit{e.g.}, CPU utilization rate) or error logs (\textit{e.g.}, certain types of errors appearing too frequently within some time range).
The objective of such alerts is to provide early warnings so that protective actions (\textit{e.g.}, virtual machine migration or software rejuvenation) can be performed to mitigate or minimize the impact of such failures.
In this context, efficient failure prediction via system log mining can enable proactive recovery mechanisms to increase reliability.

In homogeneous systems, component failures or node soft lock-ups typically lead to crashes of user jobs scheduled on the affected nodes, and they may cause undesired downtime.
One approach to mitigate such problems is to predict node failures with a sufficient lead time in order to take proactive measures.
Sahoo \textit{et al.}~\cite{sahoo2003critical} first addressed this problem by collecting system logs representing the components' health status. Then, several time series models are employed to predict the system health of each node through indication metrics, such as the percentage of system utilization, usage of network IO, and system idle time.
Russo \textit{et al.}~\cite{russo2015mining} considered the log sequence as multi-dimensional vector and employed three kinds of support vector machines to predict defective log sequences.
Das \textit{et al.}~\cite{das2018desh} proposed Desh to identify failure indicators in each node with enhanced training and classification for failure event log chain. Desh is a deep learning-based method that contains three phases:
(1) It first recognizes chains of log events leading to a failure. 
(2) Then it re-trains chain recognition of events augmented with expected lead times to failure; 
(3) Finally Desh predicts lead times during testing/inference deployment to predict which specific node fails in how many minutes.
To speed up the recognition of the failure chain from log events, Das \textit{et al.}~\cite{das2020aarohi} proposed a new node failure predicting method called Aarohi to extend their previous work to online learning setting.
Aarohi first trains an offline deep learning model with  log parsing, then utilizes grammar-based rules to provide online testing. Another approach from feature engineering tries to extract specific patterns about a certain time frame before a critical event.
This anomaly pattern can serve as a proactive alerts.
Klinkenberg \textit{et al.}~\cite{klinkenberg2017data} adopted descriptive statistics and supervised machine learning techniques to create such proactive alert from the system logs. They trained a binary classification model to detect the potential node failure based on a given time sequence of monitoring data collected from each node. 
Berrocal \textit{et al.}~\cite{berrocal2014exploring} used environmental logs to extract numerical indicators and conducted a Void Search (VS) algorithm on these numerical values for the failure prediction task. 

\subsubsection{Heterogeneous Systems}  In heterogeneous systems, modeling the relationship among multiple components is the key, which is also the core part for the dominating relation mining algorithms.
Different from homogeneous systems which focus on predicting failures based on raw signals sealed in the event logs or trace logs from single component, the prediction of critical failures (also called outage) in heterogeneous systems relies on the failure signals from log information collected from diversified system components.
In current real-time heterogeneous systems, such as the cloud system, outage prediction is an important and challenging task to perform. On one hand, outages are critical system failures that could lead to severe consequences.
On the other hand, outages occur without a significant alerting signals pattern, which makes it hard to predict. Moreover, the scope of impacting signals is a complex process to define.
Therefore, compared to homogeneous system settings, it is more valuable to detect node outages and distinguish regular internal failures from those caused by external factors, such as maintenance and human errors.

From the perspective of methodology, the advancement of failure prediction in heterogeneous systems relies on effective detection of the relationship between early alerting signals from system logs and node/component outage.
Chen \textit{et al.}~\cite{chen2019outage} proposed AirAlert framework based on Bayesian network to find the conditional dependence between the alerting signal extracted from system logs and the outage.
Then, given several alert signals, AirAlert employed gradient boosting tree method to conduct outage prediction.
Lin \textit{et al.}~\cite{lin2018predicting} designed MING framework to find the relationship between complex failure-indicating alerts and the outage from temporal and spatial features.
They carefully designed a ranking model that combines the intermediate results from LSTM model capturing temporal signals and a Random Forest model incorporating spatial data.
Zhou \textit{et al.}~\cite{zhou2019latent} proposed MEPFL to narrow down the scope of failure prediction into microservice level.
Based on a set of manual selected features defined on the system trace logs, MEPFL trains prediction models at both the trace level and the microservice level to predict three common types of microservice application faults: multi-instance faults, configuration faults, and asynchronous interaction faults.

\subsection{Failure Diagnosis}
\label{sec:log_failure_diagnosis}
\subsubsection{Problem Formulation} Unlike anomaly detection and failure prediction that are usually formalized as a classification task, failure diagnosis targets to identify the underlying causes leading to a failure that has affected end users. It is often closely related to the root cause analysis.
Specifically, although anomaly detection and failure prediction can pinpoint whether a problem occurs or will occur, there is a huge gap between the detection and removal of a problem or a failure. To completely resolve the problems, failure diagnosis is a crucial step, but the diagnosis process is notoriously expensive and inefficient. It is reported that failure diagnosis takes over $100$ billion dollars, and developers spend more than half of their time on debugging~\cite{zhang2019inflection}. Following the concept of error, fault, and failure, as defined in~\cite{laprie1995dependable}, failure diagnosis aims to identify the fault that has led to the user-perceived impairment in a software system. In the broad domain of failure diagnosis, log-based failure diagnosis is now a standard practice for software developers.
However, failure diagnosis is very challenging. The complexity of modern software systems grows rapidly, where different services, software, and hardware are tightly coupled. It is too complex to correctly and efficiently disentangle the relations among the fault, failure, and human-observed symptoms. 
Moreover, as software systems become more mature, failures are becoming more and more hard to detect and diagnose, from perceptible software functionality issues to imperceptible problems~\cite{kavulya2012failure}, \textit{e.g.}, performance issues~\cite{aguilera2003performance, mi2013toward, yu2014comprehending}.

To tackle these challenges, in recent decades, techniques to automate the diagnosis process have been widely developed, for example, using the informative logs~\cite{chuah2010diagnosing, jiang2017causes, zhou2019latent}. Jiang \textit{et al.}~\cite{jiang2009understanding} provided one of the first characteristic studies of log-based problem troubleshooting on real-world cases. They concluded that problem troubleshooting is time-consuming and challenging, which can be significantly facilitated by logs. In addition, they appealed to engineers to employ automated methods to speedup the problem resolution time, which is also the focus of this survey. Likewise, Zhou \textit{et al.}~\cite{zhou2018fault} empirically investigated the faults and debugging practices in microservice systems. The results show that proper tracing and visualization techniques can improve the diagnosis, which also suggest the strong needs for more intelligent log analysis. Other empirical studies include understanding failures with logs in high-performance computing (HPC) systems~\cite{el2013reading}, cloud management systems~\cite{cotroneo2019bad}, web servers~\cite{huynh2009another} and industrial air traffic control system~\cite{cinque2014logs}.

In this section, we review recent work on automated log-based failure diagnosis. Particularly, these research studies mainly focus on the large-scale software systems where failure diagnosis is burdensome, such as general distributed systems, storage systems, big data systems, microservice systems. As the characteristic and functionality may vary, these systems demonstrate different failure behaviors. However, the overall methodology for failure diagnosis still shares similar techniques, which can be categorized into four types, \textit{i.e.}, execution replay, model-based, statistics-based, 
and retrieval-based methods. There are also studies on diagnosing the hardware fault by traces (\textit{e.g.}, \cite{li2008trace}), which are beyond the scope of this survey.

\subsubsection{Execution Replay Methods}
Traditional rule-based methods heavily rely on a set of predefined rules (\textit{e.g.}, in the format of "if-then") from the expert knowledge to diagnose failures. However, the methods cannot be well generalized to unseen failures that are not included in the rules. On the contrary, execution replay methods aim to automatically infer the execution flow from logs and trace back to the software system failure, as previously introduced in Section~\ref{sec:log_mining_workflow}. The set of methods is human-interpretable and automated. Yuan \textit{et al.}~\cite{yuan2010sherlog} proposed the SherLog, which analyzes source code by run-time logs to represent the detailed execution process of a failure. Specifically, SherLog infers both control and data value information of a failed execution to help developers understand the failure. Similarly, LogMap~\cite{chen2019empirical} first retrieves log messages from bug reports and then applies static analysis technique to identify corresponding logging lines in source code. At last, it traverses through logging lines to derive the potential code paths, which help reconstruct the execution path and assist the debugging process. 

\subsubsection{Model-based Methods.} Model-based methods utilize logs to build the reference model (\textit{e.g.}, execution path) for a software system and then check which log events violate the reference model. In short, it sets up standards for normal software system executions and diagnoses failures by detecting the possible inconsistency. The majority of existing model-based studies leverage the log information to reconstruct the system execution flow as a graph representation~\cite{babenko2009ava, jia2017approach, jia2017logsed}. To achieve so, Jia et al. proposed to mine a time-weighted control flow graphs (TCFG)~\cite{jia2017logsed} and the service topology~\cite{jia2017approach} from interleaved logs during the offline phase. In the online phase, a failure can be easily diagnosed by observing the deviation between execution log sequences and the mined graph models. After observing that log sequences generated by a normal execution are consistent across multiple runs, Tak \textit{et al.}~\cite{tak2016logan} proposed the LOGAN to capture normal log patterns by log grouping, log parsing, and log alignment. When a failure occurs, LOGAN highlights the divergence of current log sequence from a reference model and suggests the possible root cause. Different from these research tasks, to detect concurrency bugs in distributed systems, Lu \textit{et al.}~\cite{lu2018cloudraid} proposed to mine logs from historical executions to uncover feasible but undertested log message orders. The log message orders represent possible execution flow that are likely to expose errors.

\subsubsection{Statistics-based Methods.} Since software systems generate logs to record normal and abnormal executions, it is intuitive to employ some statistical techniques (\textit{e.g.}, statistical distribution, correlation analysis) to capture the relationships between logs and the consequent failures.

Chuah \textit{et al.}~\cite{chuah2010diagnosing} developed a diagnostics tool, FDiag, to parse the log messages and employs statistical correlation analysis to attribute the observed failure to a possible root cause. However, the diagnostic capability of the proposed method is limited to known failures. To overcome the difficulty, Chuah \textit{et al.}~\cite{chuah2015insights} further extended their approach to an advanced FDiagV3 method, in which a PCA and ICA-based correlation approach is additionally employed. Differently, FDiagV3 can identify those unknown failures by automatically extracting the outlier issues. In~\cite{fu2014digging}, a three-step approach is designed to establish the causal dependency graph through log events, which helps identify the process that a failure occurs. The general idea resembles execution replay methods, but the causal dependency graph is statistically mined by grouping similar log events and utilizing the temporal order information. Similarly, Yu \textit{et al.}~\cite{yu2014comprehending} proposed to comprehend performance problem of device drivers in Windows. The method narrows down the diagnosis scope and pattern mining by measuring the impact of suspicious components on performance and representing the behavior pattern with the signature set tuple. Then, thresholds are set to identify highly suspicious and high-impact components that are likely to cause performance problems.

Lu \textit{et al.}~\cite{lu2017log} proposed to detect and diagnose anomalies in the Spark system. In complementary to previous studies, the proposed method builds statistical models from the data distribution of several task-related features. It then diagnoses the anomaly by setting threshold and analyzes the root causes with weighted factors. Similar to Xu et al.~\cite{xu2009detecting}, SCMiner~\cite{zaman2019scminer} was proposed to utilize a PCA method to detect abnormal system call sequences. Then, it maps the abnormal sequence to application functions by frequent pattern mining on system call traces.  At last, SCMiner identifies and ranks buggy functions by matching with the function call traces. Likewise, CloudDiag~\cite{mi2013toward} employs a statistical technique (\textit{i.e.}, based on the data distribution) to identify the category of a fine-grained performance problem. Then, a fast matrix recovery algorithm, RPCA, is adopted to identify the root cause (\textit{i.e.}, method invocations) for the performance failure.

\subsubsection{Retrieval-based Methods.} In practice, failures that previously occurred are valuable since they can aid developers in better diagnosing newly-occurred failures. Retrieval-based methods, as indicated by the name, retrieve similar failures in a knowledge base composed of failures in history or populated by injected faults in the test environment.
Shang \textit{et al.}~\cite{shang2013assisting} focused on diagnosing big data analytics applications in Hadoop system by injecting failures manually and analyzing the logs. Particularly, they proposed to uncover differences between pseudo and cloud deployments by log parsing, execution sequence recovery, and sequence comparison. 
Pham \textit{et al.}~\cite{pham2016failure} adopted a similar idea but they targeted on the general distributed system. Under the assumption that similar faults generate similar failures, we can locate the root cause of a reported failure by inspecting matched failures in the knowledge base. The method first reconstructs execution flows between system components, computes the similarity of the reconstructed flows, and performs precise fault injection. Nagaraj \textit{et al.}~\cite{nagaraj2012structured} proposed to diagnose the performance failures by comparing the logs of system behaviors in good and bad performance. Based on the two collected sets of logs, they applied a number of machine learning techniques to automatically compare and infer the strongest associations between performance and system components, where failures were likely to happen.

Similarly, CAM~\cite{jiang2017causes} applies the same idea to the cause analysis for test alarms in system and integration testing. In detail, the failure matching is achieved by using the K nearest neighbors (KNN) algorithm to find similar attribute vectors, which are built on test log terms extracted by term frequency–inverse document frequency (TF-IDF). 
Furthermore, Amar et al~\cite{amar2019mining} extended the CAM by keeping failing logs and removing logs that passed the test. Then, the most relevant logs in historical failures were extracted by a modified TF-IDF and then vectorized. The vectors were utilized to train an exclusive version of KNN to identify possible log lines that led to the failure. Yuan \textit{et al.}~\cite{yuan2019approach} facilitated the failure diagnosis by leveraging historical failures. They built four classifiers by vectorizing historical failures logs using natural language processing techniques. When a new failure occurs, the corresponding classifier was employed to identify the root cause. Moreover, some failures are indistinguishable because they generate very similar logs. To tackle the problem, Ikeuchi et al. ~\cite{ikeuchi2018root} proposed to utilize user actions. At first, a log-based failure database is constructed by performing various user actions. In the online deployment phase, when a user action is performed, operators match the failures with the database to identify the root cause.



\subsection{Others}
\label{sec:log_mining_others}

There are other valuable topics to broaden the log mining literature. In this section, we mainly introduce two reliability engineering tasks: specification mining and log-based software testing.

\subsubsection*{Specification Mining.}
Specification mining is the task of extracting specifications (e.g., program invariants) from program execution traces. Typical specifications impose constraints on the sequencing of program executions (ordering constraint) and program values (value constraint)~\cite{lo2007mining}. Such specifications play an important role in system comprehension, verification, and evolution. 
Ernst \textit{et al.}~\cite{ernst2007daikon,ernst2001dynamically} leveraged dynamic techniques to discover program invariants from execution traces. Specifically, they instrumented programs to write variables at critical program points (\textit{e.g.}, procedure entries and exits) and checked them against a series of potential invariant rules. They also released an open-source toolkit named Daikon~\cite{Daikon} for public reuse, which supports dynamic detection of likely invariants from program computes. Lo~\textit{et al.}~\cite{lo2007mining} mined modal scenario-based specifications in the framework of Damm and Harel’s Live Sequence Charts (LSC)~\cite{damm2001lscs} from execution traces, which describe the relations among caller/callee object identifier, and method signature. Lo \textit{et al.}~\cite{lo2008mining} further extended this model to explore the trigger-and-effect relationship of different charts in LSCs. Moreover, they found the combination of scenario-based and value-based specification mining is able to yield more rich specifications~\cite{lo2012scenario}.
Using the ITU-T standard Use Case Maps language, Hassine \textit{et al.}~\cite{hassine2018framework} visualized execution traces to recover availability scenarios. Such availability scenarios describe the degree to which a system or a component is operational and accessible.
In modern user-intensive applications, understanding user behaviors is crucial for their design. To this end, Ghezzi et al.~\cite{ghezzi2014mining} inferred a set of Markov models from user interaction logs to capture their behaviors probabilistically. Lemieux \textit{et al.}~\cite{lemieux2015general} introduced a tool named Texada to discover temporal behaviors of programs that hold true on log traces.


\subsubsection*{Log-based Software Testing.}
Log-based software testing aims to stimulate the overall software workflow under testing by the generated log data.
Many existing work solves this problem by proposing a variety of state-machine-based formalism, similar to the techniques introduced in Section~\ref{sec:log_mining_workflow}. 
A pioneer work~\cite{andrews1998testing} discussed the availability of using formal log analyzer to enhance software reliability.
The majority of log-based testing approaches utilize cues from log files to improve testing coverage or completeness, 
Chen \textit{et al.}~\cite{chen2018automated} designed a framework called LogCoCo to estimate code coverage measures using the readily available execution logs.
Chen \textit{et al.}~\cite{chen2019experience} derived the state machine framework to conduct workload testing by extracting representative user behaviors from large-scale systems.
Cohen et al. \cite{cohen2015have} proposed a concept called log confidence to evaluate the completeness of any log-based dynamic specification mining task.

\section{Open-source Toolkits and Datasets}
\label{sec:log_tools}

\begin{table}[t]
\begin{center}
\caption{A list of open-source log analysis tools and log datasets.}
\label{table:log_tools}
 \begin{tabular}{c|c|c|c}
 \toprule
 \multicolumn{2}{c|}{\bf Category}  & Description & Link \\
  \midrule
\multirow{8}{*}{{\rotatebox[origin=c]{90}{\bf Tools}}} 
  & GrayLog & A centralized log collection and real-time analysis tool. & \cite{GrayLog} \\
  & GoAccess &  A real-time terminal-based log analyzer for web logs. &  \cite{GoAccess} \\
  & Fluentd & A data collector that unifies multiple logging infrastructures.  &\cite{Fluentd}  \\
 & Logstash & A server-side processor that ingests and transforms log data.  &\cite{Logstash}  \\
 & Logalyze & A centralized log management and network monitoring software. & \cite{Logalyze} \\
 & Prometheus & A monitoring solution for metrics collection and alerting. &\cite{Prometheus} \\
 & Syslog-ng & A scalable log collector and processor across infrastructures.&\cite{Syslog-ng}  \\
 & LogPAI & A comprehensive automated log analysis toolkit &\cite{LogPAI}  \\
 \midrule
\multirow{5}{*}{{\rotatebox[origin=c]{90}{\bf Datasets}}}
  & Loghub & A large collection of log datasets from various systems. &\cite{Loghub}  \\
  & Failure dataset& Error logs produced by OpenStack cloud system. & \cite{failure_data} \\
  & CFDR & Computer failure data repository of different systems. &\cite{CFDR}  \\
  & SecRepo & A list of security log data such as malware and system logs. &\cite{SecRepo}\\
  & EDGAR & Apache log files that record and store user access statistic. &\cite{EDGAR}  \\
  \bottomrule
 \end{tabular} 
\end{center}
\vspace{-4pt}
\end{table}


In recent years, there are many research studies focusing on automated log analysis. Meanwhile, a number of log management tools have been developed to facilitate the real-world deployment in industry. In this section, we review existing open-source tools and datasets for automated log analysis. To select the desired tools, we define the following criteria: First, we did not consider commercial log analysis tools, such as Splunk~\cite{splunk} and Loggly~\cite{loggly}; Second, the selected tools should cover multiple log analysis phases presented in the paper. After carefully searching, we curated a list of open-source log analysis toolkits, as shown in Table~\ref{table:log_tools}. These log analysis tools are GrayLog, GoAccess, Fluentd, Logstash, Logalyze, Prometheus, Syslog-ng, and LogPAI. In common, these tools support crucial features such as log collection, searching, routing, parsing, visualization, alerting and automated analysis. Among these tools, LogPAI is an open-source project built upon pure research outputs while other tools are packed as open-source enterprise products. Additionally, most of existing tools focus on phases before the log mining (\textit{e.g.}, log parsing), and the resulting outcomes (\textit{e.g.}, visualization) are presented to developers for manual analysis. Particularly, LogPAI provides a set of automated log analysis tools such as log parsing and anomaly detection.
%


Besides the open-source toolkits, a number of log datasets collected from various software systems are available for research studies in academia. Table~\ref{table:log_tools} shows the list of public log datasets and their descriptions. The  Loghub~\cite{he2020loghub} maintains a large log collection that is freely accessible for research purposes. Some log datasets are production data released from previous studies, while others are collected in the lab environment. Another research study~\cite{cotroneo2019bad} presented error logs produced by injecting failures in the OpenStack cloud management system. The computer failure data repository (CFDR) focuses mainly on component failure logs in a variety of large production systems. Besides these log datasets, there are also logs collected for other purposes, such as security and web search. SecRepo is a curated list of security data such as malware and system logs.  The EDGAR dataset contains Apache logs that record user access statistic through internet search.

\section{Conclusion}

Recent years have witnessed the blossom of log analysis. Given the importance of software logs in reliability engineering, extensive efforts have been contributed to efficient and effective log analysis. This survey mainly explores the four main steps in automated log analysis framework: logging, log compression, log parsing, and log mining. Additionally, we introduce the available open-source toolkits and datasets. Our article enables the outsiders to step in this promising and practical field, and allows the experts to fill in the gaps of their knowledge background. Based on the investigation of these recent advances, we proposed new insights and discussed several future directions, including how to make automated log analysis more feasible under the modern agile and distributed development style, and the concept of a next-generation log analysis framework.

\begin{acks}
The authors appreciate informative and insightful comments provided by anonymous reviewers,
which significantly improve the quality of this survey.
\end{acks}





\bibliographystyle{ACM-Reference-Format}
\bibliography{bibliography}


\begin{thebibliography}{205}


\ifx \showCODEN    \undefined \def \showCODEN     #1{\unskip}     \fi
\ifx \showDOI      \undefined \def \showDOI       #1{#1}\fi
\ifx \showISBNx    \undefined \def \showISBNx     #1{\unskip}     \fi
\ifx \showISBNxiii \undefined \def \showISBNxiii  #1{\unskip}     \fi
\ifx \showISSN     \undefined \def \showISSN      #1{\unskip}     \fi
\ifx \showLCCN     \undefined \def \showLCCN      #1{\unskip}     \fi
\ifx \shownote     \undefined \def \shownote      #1{#1}          \fi
\ifx \showarticletitle \undefined \def \showarticletitle #1{#1}   \fi
\ifx \showURL      \undefined \def \showURL       {\relax}        \fi
\providecommand\bibfield[2]{#2}
\providecommand\bibinfo[2]{#2}
\providecommand\natexlab[1]{#1}
\providecommand\showeprint[2][]{arXiv:#2}

\bibitem[\protect\citeauthoryear{Aguilera, Mogul, Wiener, Reynolds, and
  Muthitacharoen}{Aguilera et~al\mbox{.}}{2003}]%
        {aguilera2003performance}
\bibfield{author}{\bibinfo{person}{Marcos~K Aguilera},
  \bibinfo{person}{Jeffrey~C Mogul}, \bibinfo{person}{Janet~L Wiener},
  \bibinfo{person}{Patrick Reynolds}, {and} \bibinfo{person}{Athicha
  Muthitacharoen}.} \bibinfo{year}{2003}\natexlab{}.
\newblock \showarticletitle{Performance debugging for distributed systems of
  black boxes}.
\newblock \bibinfo{journal}{\emph{ACM SIGOPS Operating Systems Review}}
  \bibinfo{volume}{37}, \bibinfo{number}{5} (\bibinfo{year}{2003}),
  \bibinfo{pages}{74--89}.
\newblock


\bibitem[\protect\citeauthoryear{Amar and Rigby}{Amar and Rigby}{2019}]%
        {amar2019mining}
\bibfield{author}{\bibinfo{person}{Anunay Amar} {and} \bibinfo{person}{Peter~C
  Rigby}.} \bibinfo{year}{2019}\natexlab{}.
\newblock \showarticletitle{Mining historical test logs to predict bugs and
  localize faults in the test logs}. In \bibinfo{booktitle}{\emph{Proc. of the
  IEEE/ACM 41st International Conference on Software Engineering (ICSE)}}.
  \bibinfo{pages}{140--151}.
\newblock


\bibitem[\protect\citeauthoryear{Amar, Bao, Busany, Lo, and Maoz}{Amar
  et~al\mbox{.}}{2018}]%
        {amar2018using}
\bibfield{author}{\bibinfo{person}{Hen Amar}, \bibinfo{person}{Lingfeng Bao},
  \bibinfo{person}{Nimrod Busany}, \bibinfo{person}{David Lo}, {and}
  \bibinfo{person}{Shahar Maoz}.} \bibinfo{year}{2018}\natexlab{}.
\newblock \showarticletitle{Using finite-state models for log differencing}. In
  \bibinfo{booktitle}{\emph{Proc. of the 26th ACM Joint European Software
  Engineering Conference and Symposium on the Foundations of Software
  Engineering (ESEC/FSE)}}. \bibinfo{pages}{49--59}.
\newblock


\bibitem[\protect\citeauthoryear{Andrews}{Andrews}{1998}]%
        {andrews1998testing}
\bibfield{author}{\bibinfo{person}{James~H. Andrews}.}
  \bibinfo{year}{1998}\natexlab{}.
\newblock \showarticletitle{Testing using log file analysis: tools, methods,
  and issues}. In \bibinfo{booktitle}{\emph{Proc. of the thirteenth {IEEE}
  conference on automated software engineering ASE}}. \bibinfo{pages}{157}.
\newblock


\bibitem[\protect\citeauthoryear{AspectJ}{AspectJ}{2020}]%
        {AspectJ}
\bibfield{author}{\bibinfo{person}{AspectJ}.} \bibinfo{year}{2020}\natexlab{}.
\newblock \bibinfo{title}{Eclipse AspectJ}.
\newblock
\newblock
\urldef\tempurl%
\url{https://www.eclipse.org/aspectj/}
\showURL{%
Retrieved September 1, 2020 from \tempurl}


\bibitem[\protect\citeauthoryear{Babenko, Mariani, and Pastore}{Babenko
  et~al\mbox{.}}{2009}]%
        {babenko2009ava}
\bibfield{author}{\bibinfo{person}{Anton Babenko}, \bibinfo{person}{Leonardo
  Mariani}, {and} \bibinfo{person}{Fabrizio Pastore}.}
  \bibinfo{year}{2009}\natexlab{}.
\newblock \showarticletitle{AVA: automated interpretation of dynamically
  detected anomalies}. In \bibinfo{booktitle}{\emph{Proc. of the eighteenth
  international symposium on Software testing and analysis (ISSTA)}}.
  \bibinfo{pages}{237--248}.
\newblock


\bibitem[\protect\citeauthoryear{Balakrishnan and Sahoo}{Balakrishnan and
  Sahoo}{2006}]%
        {balakrishnan2006lossless}
\bibfield{author}{\bibinfo{person}{Raju Balakrishnan} {and}
  \bibinfo{person}{Ramendra~K Sahoo}.} \bibinfo{year}{2006}\natexlab{}.
\newblock \showarticletitle{Lossless compression for large scale cluster logs}.
  In \bibinfo{booktitle}{\emph{Proc. 20th {IEEE} International Parallel \&
  Distributed Processing Symposium (IPDPS)}}. \bibinfo{pages}{7}.
\newblock


\bibitem[\protect\citeauthoryear{Bao, Busany, Lo, and Maoz}{Bao
  et~al\mbox{.}}{2019}]%
        {bao2019statistical}
\bibfield{author}{\bibinfo{person}{Lingfeng Bao}, \bibinfo{person}{Nimrod
  Busany}, \bibinfo{person}{David Lo}, {and} \bibinfo{person}{Shahar Maoz}.}
  \bibinfo{year}{2019}\natexlab{}.
\newblock \showarticletitle{Statistical log differencing}. In
  \bibinfo{booktitle}{\emph{Proc. of the 34th IEEE/ACM International Conference
  on Automated Software Engineering (ASE)}}. IEEE, \bibinfo{pages}{851--862}.
\newblock


\bibitem[\protect\citeauthoryear{Barik, DeLine, Drucker, and Fisher}{Barik
  et~al\mbox{.}}{2016}]%
        {DBLP:conf/icse/BarikDDF16}
\bibfield{author}{\bibinfo{person}{Titus Barik}, \bibinfo{person}{Robert
  DeLine}, \bibinfo{person}{Steven Drucker}, {and} \bibinfo{person}{Danyel
  Fisher}.} \bibinfo{year}{2016}\natexlab{}.
\newblock \showarticletitle{The bones of the system: A case study of logging
  and telemetry at microsoft}. In \bibinfo{booktitle}{\emph{2016 IEEE/ACM 38th
  International Conference on Software Engineering Companion (ICSE-C)}}. IEEE,
  \bibinfo{pages}{92--101}.
\newblock


\bibitem[\protect\citeauthoryear{Berrocal, Yu, Wallace, Papka, and
  Lan}{Berrocal et~al\mbox{.}}{2014}]%
        {berrocal2014exploring}
\bibfield{author}{\bibinfo{person}{Eduardo Berrocal}, \bibinfo{person}{Li Yu},
  \bibinfo{person}{Sean Wallace}, \bibinfo{person}{Michael~E. Papka}, {and}
  \bibinfo{person}{Zhiling Lan}.} \bibinfo{year}{2014}\natexlab{}.
\newblock \showarticletitle{Exploring void search for fault detection on
  extreme scale systems}. In \bibinfo{booktitle}{\emph{Proc. of the {IEEE}
  international conference on cluster computing (CLUSTER)}}.
  \bibinfo{pages}{1--9}.
\newblock


\bibitem[\protect\citeauthoryear{Beschastnikh, Brun, Ernst, and
  Krishnamurthy}{Beschastnikh et~al\mbox{.}}{2014}]%
        {beschastnikh2014inferring}
\bibfield{author}{\bibinfo{person}{Ivan Beschastnikh}, \bibinfo{person}{Yuriy
  Brun}, \bibinfo{person}{Michael~D Ernst}, {and} \bibinfo{person}{Arvind
  Krishnamurthy}.} \bibinfo{year}{2014}\natexlab{}.
\newblock \showarticletitle{Inferring models of concurrent systems from logs of
  their behavior with CSight}. In \bibinfo{booktitle}{\emph{Proc. of the
  IEEE/ACM 36th International Conference on Software Engineering (ICSE)}}.
  \bibinfo{pages}{468--479}.
\newblock


\bibitem[\protect\citeauthoryear{Beschastnikh, Brun, Schneider, Sloan, and
  Ernst}{Beschastnikh et~al\mbox{.}}{2011}]%
        {beschastnikh2011leveraging}
\bibfield{author}{\bibinfo{person}{Ivan Beschastnikh}, \bibinfo{person}{Yuriy
  Brun}, \bibinfo{person}{Sigurd Schneider}, \bibinfo{person}{Michael Sloan},
  {and} \bibinfo{person}{Michael~D Ernst}.} \bibinfo{year}{2011}\natexlab{}.
\newblock \showarticletitle{Leveraging existing instrumentation to
  automatically infer invariant-constrained models}. In
  \bibinfo{booktitle}{\emph{Proc. of the 19th ACM Joint European Software
  Engineering Conference and Symposium on the Foundations of Software
  Engineering (ESEC/FSE)}}. \bibinfo{pages}{267--277}.
\newblock


\bibitem[\protect\citeauthoryear{Biermann and Feldman}{Biermann and
  Feldman}{1972}]%
        {biermann1972synthesis}
\bibfield{author}{\bibinfo{person}{Alan~W Biermann} {and}
  \bibinfo{person}{Jerome~A Feldman}.} \bibinfo{year}{1972}\natexlab{}.
\newblock \showarticletitle{On the synthesis of finite-state machines from
  samples of their behavior}.
\newblock \bibinfo{journal}{\emph{IEEE transactions on Computers}}
  (\bibinfo{year}{1972}), \bibinfo{pages}{592--597}.
\newblock


\bibitem[\protect\citeauthoryear{Busany and Maoz}{Busany and Maoz}{2016}]%
        {busany2016behavioral}
\bibfield{author}{\bibinfo{person}{Nimrod Busany} {and} \bibinfo{person}{Shahar
  Maoz}.} \bibinfo{year}{2016}\natexlab{}.
\newblock \showarticletitle{Behavioral log analysis with statistical
  guarantees}. In \bibinfo{booktitle}{\emph{Proc. of the IEEE/ACM 38th
  International Conference on Software Engineering (ICSE)}}.
  \bibinfo{pages}{877--887}.
\newblock


\bibitem[\protect\citeauthoryear{CFDR}{CFDR}{2020}]%
        {CFDR}
\bibfield{author}{\bibinfo{person}{CFDR}.} \bibinfo{year}{2020}\natexlab{}.
\newblock \bibinfo{title}{Computer failure data repository}.
\newblock
\newblock
\urldef\tempurl%
\url{https://www.usenix.org/cfdr}
\showURL{%
Retrieved September 1, 2020 from \tempurl}


\bibitem[\protect\citeauthoryear{Chen}{Chen}{2019}]%
        {chen2019empirical}
\bibfield{author}{\bibinfo{person}{An~Ran Chen}.}
  \bibinfo{year}{2019}\natexlab{}.
\newblock \showarticletitle{An empirical study on leveraging logs for debugging
  production failures}. In \bibinfo{booktitle}{\emph{Proc. of the IEEE/ACM 41st
  International Conference on Software Engineering: Companion
  (ICSE-Companion)}}. \bibinfo{pages}{126--128}.
\newblock


\bibitem[\protect\citeauthoryear{Chen and Jiang}{Chen and Jiang}{2017a}]%
        {DBLP:conf/icse/ChenJ17}
\bibfield{author}{\bibinfo{person}{Boyuan Chen} {and} \bibinfo{person}{Zhen
  Ming~(Jack) Jiang}.} \bibinfo{year}{2017}\natexlab{a}.
\newblock \showarticletitle{Characterizing and detecting anti-patterns in the
  logging code}. In \bibinfo{booktitle}{\emph{Proceedings of the 39th
  International Conference on Software Engineering (ICSE)}}.
  \bibinfo{pages}{71--81}.
\newblock


\bibitem[\protect\citeauthoryear{Chen and Jiang}{Chen and Jiang}{2017b}]%
        {DBLP:journals/ese/ChenJ17}
\bibfield{author}{\bibinfo{person}{Boyuan Chen} {and} \bibinfo{person}{Zhen
  Ming~(Jack) Jiang}.} \bibinfo{year}{2017}\natexlab{b}.
\newblock \showarticletitle{Characterizing logging practices in Java-based open
  source software projects - a replication study in Apache Software
  Foundation}.
\newblock \bibinfo{journal}{\emph{Empirical Software Engineering}}
  (\bibinfo{year}{2017}), \bibinfo{pages}{330--374}.
\newblock


\bibitem[\protect\citeauthoryear{Chen and Jiang}{Chen and Jiang}{2019}]%
        {chen2019extracting}
\bibfield{author}{\bibinfo{person}{Boyuan Chen} {and} \bibinfo{person}{Zhen
  Ming~Jack Jiang}.} \bibinfo{year}{2019}\natexlab{}.
\newblock \showarticletitle{Extracting and studying the
  Logging-Code-Issue-Introducing changes in Java-based large-scale open source
  software systems}.
\newblock \bibinfo{journal}{\emph{Empirical Software Engineering}}
  (\bibinfo{year}{2019}), \bibinfo{pages}{2285--2322}.
\newblock


\bibitem[\protect\citeauthoryear{Chen and Jiang}{Chen and Jiang}{2020}]%
        {chen2020studying}
\bibfield{author}{\bibinfo{person}{Boyuan Chen} {and} \bibinfo{person}{Zhen
  Ming~(Jack) Jiang}.} \bibinfo{year}{2020}\natexlab{}.
\newblock \showarticletitle{Studying the use of java logging utilities in the
  wild}. In \bibinfo{booktitle}{\emph{Proc. of the ACM/IEEE 42nd International
  Conference on Software Engineering (ICSE)}}. \bibinfo{pages}{397–408}.
\newblock


\bibitem[\protect\citeauthoryear{Chen, Song, Xu, Hu, and Jiang}{Chen
  et~al\mbox{.}}{2018}]%
        {chen2018automated}
\bibfield{author}{\bibinfo{person}{Boyuan Chen}, \bibinfo{person}{Jian Song},
  \bibinfo{person}{Peng Xu}, \bibinfo{person}{Xing Hu}, {and}
  \bibinfo{person}{Zhen~Ming Jiang}.} \bibinfo{year}{2018}\natexlab{}.
\newblock \showarticletitle{An automated approach to estimating code coverage
  measures via execution logs}. In \bibinfo{booktitle}{\emph{Proc. of the 33rd
  ACM/IEEE International Conference on Automated Software Engineering (ASE)}}.
  \bibinfo{pages}{305--316}.
\newblock


\bibitem[\protect\citeauthoryear{Chen, Shang, Hassan, Wang, and Lin}{Chen
  et~al\mbox{.}}{2019a}]%
        {chen2019experience}
\bibfield{author}{\bibinfo{person}{Jinfu Chen}, \bibinfo{person}{Weiyi Shang},
  \bibinfo{person}{Ahmed~E. Hassan}, \bibinfo{person}{Yong Wang}, {and}
  \bibinfo{person}{Jiangbin Lin}.} \bibinfo{year}{2019}\natexlab{a}.
\newblock \showarticletitle{An experience report of generating load tests using
  log-recovered workloads at varying granularities of user behaviour}. In
  \bibinfo{booktitle}{\emph{Proc. of the 34th {IEEE}/{ACM} international
  conference on automated software engineering (ASE)}}.
  \bibinfo{pages}{669--681}.
\newblock


\bibitem[\protect\citeauthoryear{Chen, Schach, Yu, Offutt, and Heller}{Chen
  et~al\mbox{.}}{2004}]%
        {chen2004open}
\bibfield{author}{\bibinfo{person}{Kai Chen}, \bibinfo{person}{Stephen~R
  Schach}, \bibinfo{person}{Liguo Yu}, \bibinfo{person}{Jeff Offutt}, {and}
  \bibinfo{person}{Gillian~Z Heller}.} \bibinfo{year}{2004}\natexlab{}.
\newblock \showarticletitle{Open-source change logs}.
\newblock \bibinfo{journal}{\emph{Empirical Software Engineering}}
  (\bibinfo{year}{2004}), \bibinfo{pages}{197--210}.
\newblock


\bibitem[\protect\citeauthoryear{Chen, Zhang, Li, Zhang, Guo, Meng, Pei, Zhang,
  Chen, and Liu}{Chen et~al\mbox{.}}{2020}]%
        {chen2020cross}
\bibfield{author}{\bibinfo{person}{Rui Chen}, \bibinfo{person}{Shenglin Zhang},
  \bibinfo{person}{Dongwen Li}, \bibinfo{person}{Yuzhe Zhang},
  \bibinfo{person}{Fangrui Guo}, \bibinfo{person}{Weibin Meng},
  \bibinfo{person}{Dan Pei}, \bibinfo{person}{Yuzhi Zhang}, \bibinfo{person}{Xu
  Chen}, {and} \bibinfo{person}{Yuqing Liu}.} \bibinfo{year}{2020}\natexlab{}.
\newblock \showarticletitle{Cross-system log anomaly detection for software
  systems}. In \bibinfo{booktitle}{\emph{Proc. of the 31th International
  Symposium on Software Reliability Engineering (ISSRE)}}.
  \bibinfo{pages}{37--47}.
\newblock


\bibitem[\protect\citeauthoryear{Chen, Xu, Li, Kang, Yang, Lin, Zhang, Gao, Xu,
  Dang, Zhang, and Dong}{Chen et~al\mbox{.}}{2019b}]%
        {chen2019outage}
\bibfield{author}{\bibinfo{person}{Yujun Chen}, \bibinfo{person}{Yong Xu},
  \bibinfo{person}{Hao Li}, \bibinfo{person}{Yu Kang}, \bibinfo{person}{Xian
  Yang}, \bibinfo{person}{Qingwei Lin}, \bibinfo{person}{Hongyu Zhang},
  \bibinfo{person}{Feng Gao}, \bibinfo{person}{Zhangwei Xu},
  \bibinfo{person}{Yingnong Dang}, \bibinfo{person}{Dongmei Zhang}, {and}
  \bibinfo{person}{Hang Dong}.} \bibinfo{year}{2019}\natexlab{b}.
\newblock \showarticletitle{Outage prediction and diagnosis for cloud service
  systems}. In \bibinfo{booktitle}{\emph{Proc. of the World Wide Web Conference
  (WWW)}}. \bibinfo{pages}{2659--2665}.
\newblock


\bibitem[\protect\citeauthoryear{Chowdhury, Nardo, Hindle, and Jiang}{Chowdhury
  et~al\mbox{.}}{2018}]%
        {DBLP:journals/ese/ChowdhuryNHJ18}
\bibfield{author}{\bibinfo{person}{Shaiful~Alam Chowdhury},
  \bibinfo{person}{Silvia~Di Nardo}, \bibinfo{person}{Abram Hindle}, {and}
  \bibinfo{person}{Zhen Ming~(Jack) Jiang}.} \bibinfo{year}{2018}\natexlab{}.
\newblock \showarticletitle{An exploratory study on assessing the energy impact
  of logging on Android applications}.
\newblock \bibinfo{journal}{\emph{Empirical Software Engineering}}
  (\bibinfo{year}{2018}), \bibinfo{pages}{1422--1456}.
\newblock


\bibitem[\protect\citeauthoryear{Christensen and Li}{Christensen and
  Li}{2013}]%
        {christensen2013adaptive}
\bibfield{author}{\bibinfo{person}{Robert Christensen} {and}
  \bibinfo{person}{Feifei Li}.} \bibinfo{year}{2013}\natexlab{}.
\newblock \showarticletitle{Adaptive log compression for massive log data}. In
  \bibinfo{booktitle}{\emph{Proc. of the SIGMOD Conference}}.
  \bibinfo{pages}{1283--1284}.
\newblock


\bibitem[\protect\citeauthoryear{Chuah, Jhumka, Browne, Barth, and
  Narasimhamurthy}{Chuah et~al\mbox{.}}{2015}]%
        {chuah2015insights}
\bibfield{author}{\bibinfo{person}{Edward Chuah}, \bibinfo{person}{Arshad
  Jhumka}, \bibinfo{person}{James~C Browne}, \bibinfo{person}{Bill Barth},
  {and} \bibinfo{person}{Sai Narasimhamurthy}.}
  \bibinfo{year}{2015}\natexlab{}.
\newblock \showarticletitle{Insights into the diagnosis of system failures from
  cluster message logs}. In \bibinfo{booktitle}{\emph{Proc. of European
  Dependable Computing Conference (EDCC)}}. \bibinfo{pages}{225--232}.
\newblock


\bibitem[\protect\citeauthoryear{Chuah, Kuo, Hiew, Tjhi, Lee, Hammond,
  Michalewicz, Hung, and Browne}{Chuah et~al\mbox{.}}{2010}]%
        {chuah2010diagnosing}
\bibfield{author}{\bibinfo{person}{Edward Chuah}, \bibinfo{person}{Shyh-hao
  Kuo}, \bibinfo{person}{Paul Hiew}, \bibinfo{person}{William-Chandra Tjhi},
  \bibinfo{person}{Gary Lee}, \bibinfo{person}{John Hammond},
  \bibinfo{person}{Marek~T Michalewicz}, \bibinfo{person}{Terence Hung}, {and}
  \bibinfo{person}{James~C Browne}.} \bibinfo{year}{2010}\natexlab{}.
\newblock \showarticletitle{Diagnosing the root-causes of failures from cluster
  log files}. In \bibinfo{booktitle}{\emph{Proc. of the International
  Conference on High Performance Computing (HPC)}}. \bibinfo{pages}{1--10}.
\newblock


\bibitem[\protect\citeauthoryear{Cinque, Cotroneo, Della~Corte, and
  Pecchia}{Cinque et~al\mbox{.}}{2014}]%
        {cinque2014logs}
\bibfield{author}{\bibinfo{person}{Marcello Cinque}, \bibinfo{person}{Domenico
  Cotroneo}, \bibinfo{person}{Raffaele Della~Corte}, {and}
  \bibinfo{person}{Antonio Pecchia}.} \bibinfo{year}{2014}\natexlab{}.
\newblock \showarticletitle{What logs should you look at when an application
  fails? insights from an industrial case study}. In
  \bibinfo{booktitle}{\emph{Proc. of the 44th Annual IEEE/IFIP International
  Conference on Dependable Systems and Networks (DSN)}}.
  \bibinfo{pages}{690--695}.
\newblock


\bibitem[\protect\citeauthoryear{Cinque, Cotroneo, Natella, and Pecchia}{Cinque
  et~al\mbox{.}}{2010}]%
        {DBLP:conf/dsn/CinqueCNP10}
\bibfield{author}{\bibinfo{person}{Marcello Cinque}, \bibinfo{person}{Domenico
  Cotroneo}, \bibinfo{person}{Roberto Natella}, {and} \bibinfo{person}{Antonio
  Pecchia}.} \bibinfo{year}{2010}\natexlab{}.
\newblock \showarticletitle{Assessing and improving the effectiveness of logs
  for the analysis of software faults}. In \bibinfo{booktitle}{\emph{Proc. of
  the 2010 {IEEE/IFIP} International Conference on Dependable Systems and
  Networks (DSN)}}. \bibinfo{pages}{457--466}.
\newblock


\bibitem[\protect\citeauthoryear{Cinque, Cotroneo, and Pecchia}{Cinque
  et~al\mbox{.}}{2013}]%
        {DBLP:journals/tse/CinqueCP13}
\bibfield{author}{\bibinfo{person}{Marcello Cinque}, \bibinfo{person}{Domenico
  Cotroneo}, {and} \bibinfo{person}{Antonio Pecchia}.}
  \bibinfo{year}{2013}\natexlab{}.
\newblock \showarticletitle{Event logs for the analysis of software failures: a
  rule-based approach}.
\newblock \bibinfo{journal}{\emph{{IEEE} Transactions on Software Engineering}}
  (\bibinfo{year}{2013}), \bibinfo{pages}{806--821}.
\newblock


\bibitem[\protect\citeauthoryear{Cohen and Maoz}{Cohen and Maoz}{2015}]%
        {cohen2015have}
\bibfield{author}{\bibinfo{person}{Hila Cohen} {and} \bibinfo{person}{Shahar
  Maoz}.} \bibinfo{year}{2015}\natexlab{}.
\newblock \showarticletitle{Have we seen enough traces?}. In
  \bibinfo{booktitle}{\emph{Proc. of the 30th {IEEE}/{ACM} international
  conference on automated software engineering (ASE)}}.
  \bibinfo{pages}{93--103}.
\newblock


\bibitem[\protect\citeauthoryear{Cotroneo, De~Simone, Liguori, Natella, and
  Bidokhti}{Cotroneo et~al\mbox{.}}{2019}]%
        {cotroneo2019bad}
\bibfield{author}{\bibinfo{person}{Domenico Cotroneo}, \bibinfo{person}{Luigi
  De~Simone}, \bibinfo{person}{Pietro Liguori}, \bibinfo{person}{Roberto
  Natella}, {and} \bibinfo{person}{Nematollah Bidokhti}.}
  \bibinfo{year}{2019}\natexlab{}.
\newblock \showarticletitle{How bad can a bug get? an empirical analysis of
  software failures in the OpenStack cloud computing platform}. In
  \bibinfo{booktitle}{\emph{Proc. of the 27th ACM Joint Meeting on European
  Software Engineering Conference and Symposium on the Foundations of Software
  Engineering (ESEC/FSE)}}. \bibinfo{pages}{200--211}.
\newblock


\bibitem[\protect\citeauthoryear{Dai, Li, Shang, Chen, and Chen}{Dai
  et~al\mbox{.}}{2020}]%
        {Dai20Logram}
\bibfield{author}{\bibinfo{person}{Hetong Dai}, \bibinfo{person}{Heng Li},
  \bibinfo{person}{Weiyi Shang}, \bibinfo{person}{Tse-Hsun Chen}, {and}
  \bibinfo{person}{Che-Shao Chen}.} \bibinfo{year}{2020}\natexlab{}.
\newblock \showarticletitle{Logram: efficient log parsing using n-gram
  dictionaries}.
\newblock \bibinfo{journal}{\emph{IEEE Transactions on Software Engineering}}
  (\bibinfo{year}{2020}).
\newblock


\bibitem[\protect\citeauthoryear{Daikon}{Daikon}{2020}]%
        {Daikon}
\bibfield{author}{\bibinfo{person}{Daikon}.} \bibinfo{year}{2020}\natexlab{}.
\newblock \bibinfo{title}{A dynamic invariant detector}.
\newblock
\newblock
\urldef\tempurl%
\url{http://plse.cs.washington.edu/daikon/}
\showURL{%
Retrieved September 1, 2020 from \tempurl}


\bibitem[\protect\citeauthoryear{Damm and Harel}{Damm and Harel}{2001}]%
        {damm2001lscs}
\bibfield{author}{\bibinfo{person}{Werner Damm} {and} \bibinfo{person}{David
  Harel}.} \bibinfo{year}{2001}\natexlab{}.
\newblock \showarticletitle{LSCs: Breathing life into message sequence charts}.
\newblock \bibinfo{journal}{\emph{Formal methods in system design}}
  (\bibinfo{year}{2001}), \bibinfo{pages}{45--80}.
\newblock


\bibitem[\protect\citeauthoryear{Das, Mueller, and Rountree}{Das
  et~al\mbox{.}}{2020}]%
        {das2020aarohi}
\bibfield{author}{\bibinfo{person}{Anwesha Das}, \bibinfo{person}{Frank
  Mueller}, {and} \bibinfo{person}{Barry Rountree}.}
  \bibinfo{year}{2020}\natexlab{}.
\newblock \showarticletitle{Aarohi: making real-time node failure prediction
  feasible}. In \bibinfo{booktitle}{\emph{Proc. pf the {IEEE} international
  parallel and distributed processing symposium ({IPDPS})}}.
  \bibinfo{pages}{1092--1101}.
\newblock


\bibitem[\protect\citeauthoryear{Das, Mueller, Siegel, and Vishnu}{Das
  et~al\mbox{.}}{2018}]%
        {das2018desh}
\bibfield{author}{\bibinfo{person}{Anwesha Das}, \bibinfo{person}{Frank
  Mueller}, \bibinfo{person}{Charles Siegel}, {and} \bibinfo{person}{Abhinav
  Vishnu}.} \bibinfo{year}{2018}\natexlab{}.
\newblock \showarticletitle{Desh: deep learning for system health prediction of
  lead times to failure in {HPC}}. In \bibinfo{booktitle}{\emph{Proc. of the
  27th International Symposium on High-Performance Parallel and Distributed
  Computing (HPDC)}}. \bibinfo{pages}{40--51}.
\newblock


\bibitem[\protect\citeauthoryear{dataset}{dataset}{2020}]%
        {failure_data}
\bibfield{author}{\bibinfo{person}{Failure dataset}.}
  \bibinfo{year}{2020}\natexlab{}.
\newblock \bibinfo{title}{Error logs produced by OpenStack}.
\newblock
\newblock
\urldef\tempurl%
\url{https://figshare.com/articles/Failure_dataset/7732268/2}
\showURL{%
Retrieved September 1, 2020 from \tempurl}


\bibitem[\protect\citeauthoryear{Deb, Pratap, Agarwal, and Meyarivan}{Deb
  et~al\mbox{.}}{2002}]%
        {deb2002fast}
\bibfield{author}{\bibinfo{person}{Kalyanmoy Deb}, \bibinfo{person}{Amrit
  Pratap}, \bibinfo{person}{Sameer Agarwal}, {and} \bibinfo{person}{TAMT
  Meyarivan}.} \bibinfo{year}{2002}\natexlab{}.
\newblock \showarticletitle{A fast and elitist multiobjective genetic
  algorithm: NSGA-II}.
\newblock \bibinfo{journal}{\emph{IEEE transactions on evolutionary
  computation}} \bibinfo{volume}{6}, \bibinfo{number}{2}
  (\bibinfo{year}{2002}), \bibinfo{pages}{182--197}.
\newblock


\bibitem[\protect\citeauthoryear{Deorowicz and Grabowski}{Deorowicz and
  Grabowski}{2008}]%
        {deorowicz2008sub}
\bibfield{author}{\bibinfo{person}{Sebastian Deorowicz} {and}
  \bibinfo{person}{Szymon Grabowski}.} \bibinfo{year}{2008}\natexlab{}.
\newblock \showarticletitle{Sub-atomic field processing for improved web log
  compression}. In \bibinfo{booktitle}{\emph{Proc. of the {IEEE} International
  Conference on "Modern Problems of Radio Engineering, Telecommunications and
  Computer Science" (TCSET)}}. \bibinfo{pages}{551--556}.
\newblock


\bibitem[\protect\citeauthoryear{Ding, Fu, Lou, Lin, Zhang, and Xie}{Ding
  et~al\mbox{.}}{2014}]%
        {ding2014mining}
\bibfield{author}{\bibinfo{person}{Rui Ding}, \bibinfo{person}{Qiang Fu},
  \bibinfo{person}{Jian-Guang Lou}, \bibinfo{person}{Qingwei Lin},
  \bibinfo{person}{Dongmei Zhang}, {and} \bibinfo{person}{Tao Xie}.}
  \bibinfo{year}{2014}\natexlab{}.
\newblock \showarticletitle{Mining historical issue repositories to heal
  large-scale online service systems}. In \bibinfo{booktitle}{\emph{Proc. of
  the 44th Annual {IEEE/IFIP} International Conference on Dependable Systems
  and Networks (DSN)}}. \bibinfo{pages}{311--322}.
\newblock


\bibitem[\protect\citeauthoryear{Ding, Zhou, Lou, Zhang, Lin, Fu, Zhang, and
  Xie}{Ding et~al\mbox{.}}{2015}]%
        {DBLP:conf/usenix/DingZLZLFZX15}
\bibfield{author}{\bibinfo{person}{Rui Ding}, \bibinfo{person}{Hucheng Zhou},
  \bibinfo{person}{Jian{-}Guang Lou}, \bibinfo{person}{Hongyu Zhang},
  \bibinfo{person}{Qingwei Lin}, \bibinfo{person}{Qiang Fu},
  \bibinfo{person}{Dongmei Zhang}, {and} \bibinfo{person}{Tao Xie}.}
  \bibinfo{year}{2015}\natexlab{}.
\newblock \showarticletitle{Log2: {A} cost-aware logging mechanism for
  performance diagnosis}. In \bibinfo{booktitle}{\emph{Proc. of the 2015
  {USENIX} Annual Technical Conference (ATC)}}. \bibinfo{pages}{139--150}.
\newblock


\bibitem[\protect\citeauthoryear{Du and Li}{Du and Li}{2018}]%
        {DuTKDE18}
\bibfield{author}{\bibinfo{person}{Min Du} {and} \bibinfo{person}{Feifei Li}.}
  \bibinfo{year}{2018}\natexlab{}.
\newblock \showarticletitle{Spell: online streaming parsing of large
  unstructured system logs}.
\newblock \bibinfo{journal}{\emph{IEEE Transactions on Knowledge and Data
  Engineering (TKDE)}} (\bibinfo{year}{2018}), \bibinfo{pages}{2213--2227}.
\newblock


\bibitem[\protect\citeauthoryear{Du, Li, Zheng, and Srikumar}{Du
  et~al\mbox{.}}{2017}]%
        {du2017deeplog}
\bibfield{author}{\bibinfo{person}{Min Du}, \bibinfo{person}{Feifei Li},
  \bibinfo{person}{Guineng Zheng}, {and} \bibinfo{person}{Vivek Srikumar}.}
  \bibinfo{year}{2017}\natexlab{}.
\newblock \showarticletitle{Deeplog: anomaly detection and diagnosis from
  system logs through deep learning}. In \bibinfo{booktitle}{\emph{Proc. of the
  {ACM} SIGSAC Conference on Computer and Communications Security (CCS)}}.
  \bibinfo{pages}{1285--1298}.
\newblock


\bibitem[\protect\citeauthoryear{EDGAR}{EDGAR}{2020}]%
        {EDGAR}
\bibfield{author}{\bibinfo{person}{EDGAR}.} \bibinfo{year}{2020}\natexlab{}.
\newblock \bibinfo{title}{Apache log files}.
\newblock
\newblock
\urldef\tempurl%
\url{https://www.sec.gov/dera/data/edgar-log-file-data-set.html}
\showURL{%
Retrieved September 1, 2020 from \tempurl}


\bibitem[\protect\citeauthoryear{El-Sayed and Schroeder}{El-Sayed and
  Schroeder}{2013}]%
        {el2013reading}
\bibfield{author}{\bibinfo{person}{Nosayba El-Sayed} {and}
  \bibinfo{person}{Bianca Schroeder}.} \bibinfo{year}{2013}\natexlab{}.
\newblock \showarticletitle{Reading between the lines of failure logs:
  Understanding how HPC systems fail}. In \bibinfo{booktitle}{\emph{Proc. of
  the 43rd annual IEEE/IFIP international conference on dependable systems and
  networks (DSN)}}. \bibinfo{pages}{1--12}.
\newblock


\bibitem[\protect\citeauthoryear{ELK}{ELK}{2012}]%
        {elk}
\bibfield{author}{\bibinfo{person}{ELK}.} \bibinfo{year}{2012}\natexlab{}.
\newblock \bibinfo{title}{ElasticSearch}.
\newblock
\newblock
\urldef\tempurl%
\url{https://www.elastic.co/elk-stack}
\showURL{%
Retrieved September 1, 2020 from \tempurl}


\bibitem[\protect\citeauthoryear{Ernst, Cockrell, Griswold, and Notkin}{Ernst
  et~al\mbox{.}}{2001}]%
        {ernst2001dynamically}
\bibfield{author}{\bibinfo{person}{Michael~D Ernst}, \bibinfo{person}{Jake
  Cockrell}, \bibinfo{person}{William~G Griswold}, {and} \bibinfo{person}{David
  Notkin}.} \bibinfo{year}{2001}\natexlab{}.
\newblock \showarticletitle{Dynamically discovering likely program invariants
  to support program evolution}.
\newblock \bibinfo{journal}{\emph{IEEE Transactions on Software Engineering}}
  (\bibinfo{year}{2001}), \bibinfo{pages}{99--123}.
\newblock


\bibitem[\protect\citeauthoryear{Ernst, Perkins, Guo, McCamant, Pacheco,
  Tschantz, and Xiao}{Ernst et~al\mbox{.}}{[n.d.]}]%
        {ernst2007daikon}
\bibfield{author}{\bibinfo{person}{Michael~D Ernst}, \bibinfo{person}{Jeff~H
  Perkins}, \bibinfo{person}{Philip~J Guo}, \bibinfo{person}{Stephen McCamant},
  \bibinfo{person}{Carlos Pacheco}, \bibinfo{person}{Matthew~S Tschantz}, {and}
  \bibinfo{person}{Chen Xiao}.} \bibinfo{year}{[n.d.]}\natexlab{}.
\newblock \showarticletitle{The Daikon system for dynamic detection of likely
  invariants}.
\newblock \bibinfo{journal}{\emph{Science of computer programming}}
  (\bibinfo{year}{[n.\,d.]}), \bibinfo{pages}{35--45}.
\newblock


\bibitem[\protect\citeauthoryear{Facebook}{Facebook}{2019a}]%
        {downtimeGeneral}
\bibfield{author}{\bibinfo{person}{Facebook}.}
  \bibinfo{year}{2019}\natexlab{a}.
\newblock \bibinfo{title}{Downtime, outages and failures - understanding their
  true costs}.
\newblock
\newblock
\urldef\tempurl%
\url{http://www.evolven.com/blog/downtime-outages-and-failures-understanding-their-true-costs.html}
\showURL{%
\tempurl}


\bibitem[\protect\citeauthoryear{Facebook}{Facebook}{2019b}]%
        {downtimeFacebook}
\bibfield{author}{\bibinfo{person}{Facebook}.}
  \bibinfo{year}{2019}\natexlab{b}.
\newblock \bibinfo{title}{Facebook loses \$24,420 a minute during outages}.
\newblock
\newblock
\urldef\tempurl%
\url{https://www.theatlantic.com/technology/archive/2014/10/facebook-is-losing-24420-per-minute/382054/}
\showURL{%
\tempurl}


\bibitem[\protect\citeauthoryear{Farshchi, Schneider, Weber, and
  Grundy}{Farshchi et~al\mbox{.}}{2015}]%
        {farshchi2015experience}
\bibfield{author}{\bibinfo{person}{Mostafa Farshchi}, \bibinfo{person}{Jean-Guy
  Schneider}, \bibinfo{person}{Ingo Weber}, {and} \bibinfo{person}{John
  Grundy}.} \bibinfo{year}{2015}\natexlab{}.
\newblock \showarticletitle{Experience report: Anomaly detection of cloud
  application operations using log and cloud metric correlation analysis}. In
  \bibinfo{booktitle}{\emph{Proc. of the 26th International Symposium on
  Software Reliability Engineering (ISSRE)}}. \bibinfo{pages}{24--34}.
\newblock


\bibitem[\protect\citeauthoryear{Feng, Wu, and Li}{Feng et~al\mbox{.}}{2016}]%
        {feng2016mlc}
\bibfield{author}{\bibinfo{person}{Bo Feng}, \bibinfo{person}{Chentao Wu},
  {and} \bibinfo{person}{Jie Li}.} \bibinfo{year}{2016}\natexlab{}.
\newblock \showarticletitle{MLC: an efficient multi-level log compression
  method for cloud backup systems}. In \bibinfo{booktitle}{\emph{Proc. of the
  2016 {IEEE} Trustcom/BigDataSE/ISPA}}. \bibinfo{pages}{1358--1365}.
\newblock


\bibitem[\protect\citeauthoryear{Fluentd}{Fluentd}{2020}]%
        {Fluentd}
\bibfield{author}{\bibinfo{person}{Fluentd}.} \bibinfo{year}{2020}\natexlab{}.
\newblock \bibinfo{title}{An open source data collector for unified logging
  layer}.
\newblock
\newblock
\urldef\tempurl%
\url{https://www.fluentd.org}
\showURL{%
Retrieved September 1, 2020 from \tempurl}


\bibitem[\protect\citeauthoryear{Fu, Lou, Wang, and Li}{Fu
  et~al\mbox{.}}{2009}]%
        {qfu09}
\bibfield{author}{\bibinfo{person}{Qiang Fu}, \bibinfo{person}{Jian-Guang Lou},
  \bibinfo{person}{Yi Wang}, {and} \bibinfo{person}{Jiang Li}.}
  \bibinfo{year}{2009}\natexlab{}.
\newblock \showarticletitle{Execution Anomaly Detection in Distributed Systems
  through Unstructured Log Analysis}. In \bibinfo{booktitle}{\emph{Proc. of
  International Conference on Data Mining (ICDM)}}. \bibinfo{pages}{149--158}.
\newblock


\bibitem[\protect\citeauthoryear{Fu, Zhu, Hu, Lou, Ding, Lin, Zhang, and
  Xie}{Fu et~al\mbox{.}}{2014b}]%
        {DBLP:conf/icse/FuZHLDLZX14}
\bibfield{author}{\bibinfo{person}{Qiang Fu}, \bibinfo{person}{Jieming Zhu},
  \bibinfo{person}{Wenlu Hu}, \bibinfo{person}{Jian{-}Guang Lou},
  \bibinfo{person}{Rui Ding}, \bibinfo{person}{Qingwei Lin},
  \bibinfo{person}{Dongmei Zhang}, {and} \bibinfo{person}{Tao Xie}.}
  \bibinfo{year}{2014}\natexlab{b}.
\newblock \showarticletitle{Where do developers log? an empirical study on
  logging practices in industry}. In \bibinfo{booktitle}{\emph{Proc. of the
  36th International Conference on Software Engineering (ICSE)}}.
  \bibinfo{pages}{24--33}.
\newblock


\bibitem[\protect\citeauthoryear{Fu, Ren, McKee, Zhan, and Sun}{Fu
  et~al\mbox{.}}{2014a}]%
        {fu2014digging}
\bibfield{author}{\bibinfo{person}{Xiaoyu Fu}, \bibinfo{person}{Rui Ren},
  \bibinfo{person}{Sally~A McKee}, \bibinfo{person}{Jianfeng Zhan}, {and}
  \bibinfo{person}{Ninghui Sun}.} \bibinfo{year}{2014}\natexlab{a}.
\newblock \showarticletitle{Digging deeper into cluster system logs for failure
  prediction and root cause diagnosis}. In \bibinfo{booktitle}{\emph{Proc. of
  International Conference on Cluster Computing (CLUSTER)}}.
  \bibinfo{pages}{103--112}.
\newblock


\bibitem[\protect\citeauthoryear{Ghezzi, Pezz{\`e}, Sama, and
  Tamburrelli}{Ghezzi et~al\mbox{.}}{2014}]%
        {ghezzi2014mining}
\bibfield{author}{\bibinfo{person}{Carlo Ghezzi}, \bibinfo{person}{Mauro
  Pezz{\`e}}, \bibinfo{person}{Michele Sama}, {and} \bibinfo{person}{Giordano
  Tamburrelli}.} \bibinfo{year}{2014}\natexlab{}.
\newblock \showarticletitle{Mining behavior models from user-intensive web
  applications}. In \bibinfo{booktitle}{\emph{Proceedings of the 36th
  International Conference on Software Engineering}}.
  \bibinfo{pages}{277--287}.
\newblock


\bibitem[\protect\citeauthoryear{GoAccess}{GoAccess}{2020}]%
        {GoAccess}
\bibfield{author}{\bibinfo{person}{GoAccess}.} \bibinfo{year}{2020}\natexlab{}.
\newblock \bibinfo{title}{A fast, terminal-based log analyzer}.
\newblock
\newblock
\urldef\tempurl%
\url{https://goaccess.io}
\showURL{%
Retrieved September 1, 2020 from \tempurl}


\bibitem[\protect\citeauthoryear{Goldstein, Raz, and Segall}{Goldstein
  et~al\mbox{.}}{2017}]%
        {goldstein2017experience}
\bibfield{author}{\bibinfo{person}{Maayan Goldstein}, \bibinfo{person}{Danny
  Raz}, {and} \bibinfo{person}{Itai Segall}.} \bibinfo{year}{2017}\natexlab{}.
\newblock \showarticletitle{Experience report: Log-based behavioral
  differencing}. In \bibinfo{booktitle}{\emph{Proc. of the 28th IEEE
  International Symposium on Software Reliability Engineering (ISSRE)}}.
  \bibinfo{pages}{282--293}.
\newblock


\bibitem[\protect\citeauthoryear{GrayLog}{GrayLog}{2020}]%
        {GrayLog}
\bibfield{author}{\bibinfo{person}{GrayLog}.} \bibinfo{year}{2020}\natexlab{}.
\newblock \bibinfo{title}{A leading centralized log management solution}.
\newblock
\newblock
\urldef\tempurl%
\url{https://www.graylog.org}
\showURL{%
Retrieved September 1, 2020 from \tempurl}


\bibitem[\protect\citeauthoryear{Hamooni, Debnath, Xu, Zhang, Jiang, and
  Mueen}{Hamooni et~al\mbox{.}}{2016}]%
        {logmine16}
\bibfield{author}{\bibinfo{person}{Hossein Hamooni}, \bibinfo{person}{Biplob
  Debnath}, \bibinfo{person}{Jianwu Xu}, \bibinfo{person}{Hui Zhang},
  \bibinfo{person}{Guofei Jiang}, {and} \bibinfo{person}{Abdullah Mueen}.}
  \bibinfo{year}{2016}\natexlab{}.
\newblock \showarticletitle{{LogMine}: fast pattern recognition for log
  analytics}. In \bibinfo{booktitle}{\emph{Proc. of the 25th ACM international
  conference on Information and knowledge management (CIKM)}}.
  \bibinfo{pages}{1573--1582}.
\newblock


\bibitem[\protect\citeauthoryear{Hamou-Lhadj and Lethbridge}{Hamou-Lhadj and
  Lethbridge}{2006}]%
        {hamou2006summarizing}
\bibfield{author}{\bibinfo{person}{Abdelwahab Hamou-Lhadj} {and}
  \bibinfo{person}{Timothy Lethbridge}.} \bibinfo{year}{2006}\natexlab{}.
\newblock \showarticletitle{Summarizing the content of large traces to
  facilitate the understanding of the behaviour of a software system}. In
  \bibinfo{booktitle}{\emph{Proc. of the 14th IEEE International Conference on
  Program Comprehension (ICPC)}}. \bibinfo{pages}{181--190}.
\newblock


\bibitem[\protect\citeauthoryear{Hansen and Atkins}{Hansen and Atkins}{1993}]%
        {hansen1993automated}
\bibfield{author}{\bibinfo{person}{Stephen~E Hansen} {and}
  \bibinfo{person}{E~Todd Atkins}.} \bibinfo{year}{1993}\natexlab{}.
\newblock \showarticletitle{Automated system monitoring and notification with
  swatch.}. In \bibinfo{booktitle}{\emph{Proc. of the 7th {USENIX} Large
  Installation System Administration Conference (LISA)}},
  Vol.~\bibinfo{volume}{93}. \bibinfo{pages}{145--152}.
\newblock


\bibitem[\protect\citeauthoryear{Hassan, Martin, Flora, Mansfield, and
  Dietz}{Hassan et~al\mbox{.}}{2008}]%
        {hassan2008industrial}
\bibfield{author}{\bibinfo{person}{Ahmed~E Hassan}, \bibinfo{person}{Daryl~J
  Martin}, \bibinfo{person}{Parminder Flora}, \bibinfo{person}{Paul Mansfield},
  {and} \bibinfo{person}{Dave Dietz}.} \bibinfo{year}{2008}\natexlab{}.
\newblock \showarticletitle{An industrial case study of customizing operational
  profiles using log compression}. In \bibinfo{booktitle}{\emph{Proc. of the
  30th international conference on software engineering (ICSE)}}.
  \bibinfo{pages}{713--723}.
\newblock


\bibitem[\protect\citeauthoryear{Hassani, Shang, Shihab, and Tsantalis}{Hassani
  et~al\mbox{.}}{2018}]%
        {DBLP:journals/ese/HassaniSST18}
\bibfield{author}{\bibinfo{person}{Mehran Hassani}, \bibinfo{person}{Weiyi
  Shang}, \bibinfo{person}{Emad Shihab}, {and} \bibinfo{person}{Nikolaos
  Tsantalis}.} \bibinfo{year}{2018}\natexlab{}.
\newblock \showarticletitle{Studying and detecting log-related issues}.
\newblock \bibinfo{journal}{\emph{Empirical Software Engineering}}
  (\bibinfo{year}{2018}), \bibinfo{pages}{3248--3280}.
\newblock


\bibitem[\protect\citeauthoryear{Hassine, Hamou-Lhadj, and Alawneh}{Hassine
  et~al\mbox{.}}{2018}]%
        {hassine2018framework}
\bibfield{author}{\bibinfo{person}{Jameleddine Hassine},
  \bibinfo{person}{Abdelwahab Hamou-Lhadj}, {and} \bibinfo{person}{Luay
  Alawneh}.} \bibinfo{year}{2018}\natexlab{}.
\newblock \showarticletitle{A framework for the recovery and visualization of
  system availability scenarios from execution traces}.
\newblock \bibinfo{journal}{\emph{Information and Software Technology}}
  (\bibinfo{year}{2018}), \bibinfo{pages}{78--93}.
\newblock


\bibitem[\protect\citeauthoryear{H{\"a}t{\"o}nen, Boulicaut, Klemettinen,
  Miettinen, and Masson}{H{\"a}t{\"o}nen et~al\mbox{.}}{2003}]%
        {hatonen2003comprehensive}
\bibfield{author}{\bibinfo{person}{Kimmo H{\"a}t{\"o}nen},
  \bibinfo{person}{Jean~Fran{\c{c}}ois Boulicaut}, \bibinfo{person}{Mika
  Klemettinen}, \bibinfo{person}{Markus Miettinen}, {and}
  \bibinfo{person}{Cyrille Masson}.} \bibinfo{year}{2003}\natexlab{}.
\newblock \showarticletitle{Comprehensive log compression with frequent
  patterns}. In \bibinfo{booktitle}{\emph{Proc. of the International Conference
  on Data Warehousing and Knowledge Discovery (DaWaK)}}. Springer,
  \bibinfo{pages}{360--370}.
\newblock


\bibitem[\protect\citeauthoryear{He, Chen, He, and Lyu}{He
  et~al\mbox{.}}{2018a}]%
        {DBLP:conf/kbse/HeCHL18}
\bibfield{author}{\bibinfo{person}{Pinjia He}, \bibinfo{person}{Zhuangbin
  Chen}, \bibinfo{person}{Shilin He}, {and} \bibinfo{person}{Michael~R. Lyu}.}
  \bibinfo{year}{2018}\natexlab{a}.
\newblock \showarticletitle{Characterizing the natural language descriptions in
  software logging statements}. In \bibinfo{booktitle}{\emph{Proc. of the 33rd
  {ACM/IEEE} International Conference on Automated Software Engineering
  (ASE)}}. \bibinfo{pages}{178--189}.
\newblock


\bibitem[\protect\citeauthoryear{He, Zhu, He, Li, and Lyu}{He
  et~al\mbox{.}}{2017a}]%
        {He2017TDSC}
\bibfield{author}{\bibinfo{person}{Pinjia He}, \bibinfo{person}{Jieming Zhu},
  \bibinfo{person}{Shilin He}, \bibinfo{person}{Jian Li}, {and}
  \bibinfo{person}{Michael~R. Lyu}.} \bibinfo{year}{2017}\natexlab{a}.
\newblock \showarticletitle{Towards automated log parsing for large-scale log
  data analysis}.
\newblock \bibinfo{journal}{\emph{IEEE Transactions on Dependable and Secure
  Computing}} (\bibinfo{year}{2017}), \bibinfo{pages}{931--944}.
\newblock


\bibitem[\protect\citeauthoryear{He, Zhu, Xu, Zheng, and Lyu}{He
  et~al\mbox{.}}{2018c}]%
        {He2018Drain}
\bibfield{author}{\bibinfo{person}{Pinjia He}, \bibinfo{person}{Jieming Zhu},
  \bibinfo{person}{Pengcheng Xu}, \bibinfo{person}{Zibin Zheng}, {and}
  \bibinfo{person}{Michael~R. Lyu}.} \bibinfo{year}{2018}\natexlab{c}.
\newblock \showarticletitle{A directed acyclic graph approach to online log
  parsing}.
\newblock \bibinfo{journal}{\emph{arXiv preprint arXiv:1806.04356}}
  (\bibinfo{year}{2018}).
\newblock


\bibitem[\protect\citeauthoryear{He, Zhu, Zheng, and Lyu}{He
  et~al\mbox{.}}{2017b}]%
        {He17ICWS}
\bibfield{author}{\bibinfo{person}{Pinjia. He}, \bibinfo{person}{Jieming. Zhu},
  \bibinfo{person}{Zibin. Zheng}, {and} \bibinfo{person}{Michael~R. Lyu}.}
  \bibinfo{year}{2017}\natexlab{b}.
\newblock \showarticletitle{Drain: an online log parsing approach with fixed
  depth tree}. In \bibinfo{booktitle}{\emph{Proc. of the 24th International
  Conference on Web Services (ICWS)}}. \bibinfo{pages}{33--40}.
\newblock


\bibitem[\protect\citeauthoryear{He, Lin, Lou, Zhang, Lyu, and Zhang}{He
  et~al\mbox{.}}{2018b}]%
        {he2018identifying}
\bibfield{author}{\bibinfo{person}{Shilin He}, \bibinfo{person}{Qingwei Lin},
  \bibinfo{person}{Jian-Guang Lou}, \bibinfo{person}{Hongyu Zhang},
  \bibinfo{person}{Michael~R. Lyu}, {and} \bibinfo{person}{Dongmei Zhang}.}
  \bibinfo{year}{2018}\natexlab{b}.
\newblock \showarticletitle{Identifying impactful service system problems via
  log analysis}. In \bibinfo{booktitle}{\emph{Proc. of the 26th ACM Joint
  European Software Engineering Conference and Symposium on the Foundations of
  Software Engineering (ESEC/FSE)}}. \bibinfo{pages}{60--70}.
\newblock


\bibitem[\protect\citeauthoryear{He, Zhu, He, and Lyu}{He
  et~al\mbox{.}}{2016}]%
        {he2016experience}
\bibfield{author}{\bibinfo{person}{Shilin He}, \bibinfo{person}{Jieming Zhu},
  \bibinfo{person}{Pinjia He}, {and} \bibinfo{person}{Michael~R. Lyu}.}
  \bibinfo{year}{2016}\natexlab{}.
\newblock \showarticletitle{Experience report: System log analysis for anomaly
  detection}. In \bibinfo{booktitle}{\emph{Proc. of the 27th {IEEE}
  International Symposium on Software Reliability Engineering (ISSRE)}}.
  \bibinfo{pages}{207--218}.
\newblock


\bibitem[\protect\citeauthoryear{He, Zhu, He, and Lyu}{He
  et~al\mbox{.}}{2020}]%
        {he2020loghub}
\bibfield{author}{\bibinfo{person}{Shilin He}, \bibinfo{person}{Jieming Zhu},
  \bibinfo{person}{Pinjia He}, {and} \bibinfo{person}{Michael~R. Lyu}.}
  \bibinfo{year}{2020}\natexlab{}.
\newblock \showarticletitle{Loghub: A Large Collection of System Log Datasets
  towards Automated Log Analytics}.
\newblock \bibinfo{journal}{\emph{arXiv preprint arXiv:2008.06448}}
  (\bibinfo{year}{2020}).
\newblock


\bibitem[\protect\citeauthoryear{Huynh and Miller}{Huynh and Miller}{2009}]%
        {huynh2009another}
\bibfield{author}{\bibinfo{person}{Toan Huynh} {and} \bibinfo{person}{James
  Miller}.} \bibinfo{year}{2009}\natexlab{}.
\newblock \showarticletitle{Another viewpoint on “evaluating web software
  reliability based on workload and failure data extracted from server
  logs”}.
\newblock \bibinfo{journal}{\emph{Empirical Software Engineering}}
  (\bibinfo{year}{2009}), \bibinfo{pages}{371--396}.
\newblock


\bibitem[\protect\citeauthoryear{Ikeuchi, Watanabe, Kawata, and
  Kawahara}{Ikeuchi et~al\mbox{.}}{2018}]%
        {ikeuchi2018root}
\bibfield{author}{\bibinfo{person}{Hiroki Ikeuchi}, \bibinfo{person}{Akio
  Watanabe}, \bibinfo{person}{Takehiro Kawata}, {and} \bibinfo{person}{Ryoichi
  Kawahara}.} \bibinfo{year}{2018}\natexlab{}.
\newblock \showarticletitle{Root-cause diagnosis using logs generated by user
  actions}. In \bibinfo{booktitle}{\emph{Proc. of the IEEE Global
  Communications Conference (GLOBECOM)}}. \bibinfo{pages}{1--7}.
\newblock


\bibitem[\protect\citeauthoryear{Jia, Chen, Yang, Li, Meng, and Xu}{Jia
  et~al\mbox{.}}{2017a}]%
        {jia2017approach}
\bibfield{author}{\bibinfo{person}{Tong Jia}, \bibinfo{person}{Pengfei Chen},
  \bibinfo{person}{Lin Yang}, \bibinfo{person}{Ying Li},
  \bibinfo{person}{Fanjing Meng}, {and} \bibinfo{person}{Jingmin Xu}.}
  \bibinfo{year}{2017}\natexlab{a}.
\newblock \showarticletitle{An approach for anomaly diagnosis based on hybrid
  graph model with logs for distributed services}. In
  \bibinfo{booktitle}{\emph{Proc. of the IEEE International Conference on Web
  Services (ICWS)}}. \bibinfo{pages}{25--32}.
\newblock


\bibitem[\protect\citeauthoryear{Jia, Yang, Chen, Li, Meng, and Xu}{Jia
  et~al\mbox{.}}{2017b}]%
        {jia2017logsed}
\bibfield{author}{\bibinfo{person}{Tong Jia}, \bibinfo{person}{Lin Yang},
  \bibinfo{person}{Pengfei Chen}, \bibinfo{person}{Ying Li},
  \bibinfo{person}{Fanjing Meng}, {and} \bibinfo{person}{Jingmin Xu}.}
  \bibinfo{year}{2017}\natexlab{b}.
\newblock \showarticletitle{Logsed: anomaly diagnosis through mining
  time-weighted control flow graph in logs}. In \bibinfo{booktitle}{\emph{Proc.
  of the IEEE 10th International Conference on Cloud Computing (CLOUD)}}.
  \bibinfo{pages}{447--455}.
\newblock


\bibitem[\protect\citeauthoryear{Jiang, Li, Yang, and Xuan}{Jiang
  et~al\mbox{.}}{2017}]%
        {jiang2017causes}
\bibfield{author}{\bibinfo{person}{He Jiang}, \bibinfo{person}{Xiaochen Li},
  \bibinfo{person}{Zijiang Yang}, {and} \bibinfo{person}{Jifeng Xuan}.}
  \bibinfo{year}{2017}\natexlab{}.
\newblock \showarticletitle{What causes my test alarm? Automatic cause analysis
  for test alarms in system and integration testing}. In
  \bibinfo{booktitle}{\emph{Proc. of the IEEE/ACM 39th International Conference
  on Software Engineering (ICSE)}}. \bibinfo{pages}{712--723}.
\newblock


\bibitem[\protect\citeauthoryear{Jiang, Hu, Pasupathy, Kanevsky, Li, and
  Zhou}{Jiang et~al\mbox{.}}{2009}]%
        {jiang2009understanding}
\bibfield{author}{\bibinfo{person}{Weihang Jiang}, \bibinfo{person}{Chongfeng
  Hu}, \bibinfo{person}{Shankar Pasupathy}, \bibinfo{person}{Arkady Kanevsky},
  \bibinfo{person}{Zhenmin Li}, {and} \bibinfo{person}{Yuanyuan Zhou}.}
  \bibinfo{year}{2009}\natexlab{}.
\newblock \showarticletitle{Understanding customer problem troubleshooting from
  storage system logs}. In \bibinfo{booktitle}{\emph{Proc. of the 7th
  conference on File and storage technologies (FAST)}}.
  \bibinfo{pages}{43--56}.
\newblock


\bibitem[\protect\citeauthoryear{Jiang, Hassan, Flora, and Hamann}{Jiang
  et~al\mbox{.}}{2008a}]%
        {AEL_1}
\bibfield{author}{\bibinfo{person}{Zhen~Ming Jiang}, \bibinfo{person}{Ahmed~E
  Hassan}, \bibinfo{person}{Parminder Flora}, {and} \bibinfo{person}{Gilbert
  Hamann}.} \bibinfo{year}{2008}\natexlab{a}.
\newblock \showarticletitle{Abstracting execution logs to execution events for
  enterprise applications}. In \bibinfo{booktitle}{\emph{Proc. of the Eighth
  International Conference on Quality Software (QSIC)}}.
\newblock


\bibitem[\protect\citeauthoryear{Jiang, Hassan, Hamann, and Flora}{Jiang
  et~al\mbox{.}}{2008b}]%
        {DBLP:conf/icsm/JiangHHF08}
\bibfield{author}{\bibinfo{person}{Zhen~Ming Jiang}, \bibinfo{person}{Ahmed~E.
  Hassan}, \bibinfo{person}{Gilbert Hamann}, {and} \bibinfo{person}{Parminder
  Flora}.} \bibinfo{year}{2008}\natexlab{b}.
\newblock \showarticletitle{Automatic identification of load testing problems}.
  In \bibinfo{booktitle}{\emph{Proc. of the 24th {IEEE} International
  Conference on Software Maintenance (ICSM)}}. \bibinfo{pages}{307--316}.
\newblock


\bibitem[\protect\citeauthoryear{Kabinna, Bezemer, Shang, and Hassan}{Kabinna
  et~al\mbox{.}}{2016}]%
        {DBLP:conf/msr/KabinnaBSH16}
\bibfield{author}{\bibinfo{person}{Suhas Kabinna}, \bibinfo{person}{Cor{-}Paul
  Bezemer}, \bibinfo{person}{Weiyi Shang}, {and} \bibinfo{person}{Ahmed~E.
  Hassan}.} \bibinfo{year}{2016}\natexlab{}.
\newblock \showarticletitle{Logging library migrations: a case study for the
  apache software foundation projects}. In \bibinfo{booktitle}{\emph{Proc. of
  the 13th International Conference on Mining Software Repositories (MSR)}}.
  \bibinfo{pages}{154--164}.
\newblock


\bibitem[\protect\citeauthoryear{Kabinna, Bezemer, Shang, Syer, and
  Hassan}{Kabinna et~al\mbox{.}}{2018}]%
        {DBLP:journals/ese/KabinnaBSSH18}
\bibfield{author}{\bibinfo{person}{Suhas Kabinna}, \bibinfo{person}{Cor{-}Paul
  Bezemer}, \bibinfo{person}{Weiyi Shang}, \bibinfo{person}{Mark~D. Syer},
  {and} \bibinfo{person}{Ahmed~E. Hassan}.} \bibinfo{year}{2018}\natexlab{}.
\newblock \showarticletitle{Examining the stability of logging statements}.
\newblock \bibinfo{journal}{\emph{Empirical Software Engineering}}
  (\bibinfo{year}{2018}), \bibinfo{pages}{290--333}.
\newblock


\bibitem[\protect\citeauthoryear{Kavulya, Joshi, Di~Giandomenico, and
  Narasimhan}{Kavulya et~al\mbox{.}}{2012}]%
        {kavulya2012failure}
\bibfield{author}{\bibinfo{person}{Soila~P Kavulya}, \bibinfo{person}{Kaustubh
  Joshi}, \bibinfo{person}{Felicita Di~Giandomenico}, {and}
  \bibinfo{person}{Priya Narasimhan}.} \bibinfo{year}{2012}\natexlab{}.
\newblock \showarticletitle{Failure diagnosis of complex systems}.
\newblock In \bibinfo{booktitle}{\emph{Resilience assessment and evaluation of
  computing systems}}. \bibinfo{publisher}{Springer},
  \bibinfo{pages}{239--261}.
\newblock


\bibitem[\protect\citeauthoryear{Khan, Gani, Wahab, Bagiwa, Shiraz, Khan,
  Buyya, and Zomaya}{Khan et~al\mbox{.}}{2016}]%
        {khan2016cloud}
\bibfield{author}{\bibinfo{person}{Suleman Khan}, \bibinfo{person}{Abdullah
  Gani}, \bibinfo{person}{Ainuddin Wahid~Abdul Wahab},
  \bibinfo{person}{Mustapha~Aminu Bagiwa}, \bibinfo{person}{Muhammad Shiraz},
  \bibinfo{person}{Samee~U Khan}, \bibinfo{person}{Rajkumar Buyya}, {and}
  \bibinfo{person}{Albert~Y Zomaya}.} \bibinfo{year}{2016}\natexlab{}.
\newblock \showarticletitle{Cloud log forensics: Foundations, state of the art,
  and future directions}.
\newblock \bibinfo{journal}{\emph{ACM Computing Surveys (CSUR)}}
  (\bibinfo{year}{2016}), \bibinfo{pages}{178--184}.
\newblock


\bibitem[\protect\citeauthoryear{Kiczales, Lamping, Mendhekar, Maeda, Lopes,
  Loingtier, and Irwin}{Kiczales et~al\mbox{.}}{1997}]%
        {kiczales1997aspect}
\bibfield{author}{\bibinfo{person}{Gregor Kiczales}, \bibinfo{person}{John
  Lamping}, \bibinfo{person}{Anurag Mendhekar}, \bibinfo{person}{Chris Maeda},
  \bibinfo{person}{Cristina Lopes}, \bibinfo{person}{Jean-Marc Loingtier},
  {and} \bibinfo{person}{John Irwin}.} \bibinfo{year}{1997}\natexlab{}.
\newblock \showarticletitle{Aspect-oriented programming}. In
  \bibinfo{booktitle}{\emph{European conference on object-oriented
  programming}}. Springer.
\newblock


\bibitem[\protect\citeauthoryear{Kimura, Watanabe, Toyono, and
  Ishibashi}{Kimura et~al\mbox{.}}{2018}]%
        {kimura2018proactive}
\bibfield{author}{\bibinfo{person}{Tatsuaki Kimura}, \bibinfo{person}{Akio
  Watanabe}, \bibinfo{person}{Tsuyoshi Toyono}, {and} \bibinfo{person}{Keisuke
  Ishibashi}.} \bibinfo{year}{2018}\natexlab{}.
\newblock \showarticletitle{Proactive failure detection learning generation
  patterns of large-scale network logs}.
\newblock \bibinfo{journal}{\emph{IEICE Transactions on Communications}}
  (\bibinfo{year}{2018}).
\newblock


\bibitem[\protect\citeauthoryear{King, Stallings, Riaz, and Williams}{King
  et~al\mbox{.}}{2017}]%
        {king2017log}
\bibfield{author}{\bibinfo{person}{Jason King}, \bibinfo{person}{Jon
  Stallings}, \bibinfo{person}{Maria Riaz}, {and} \bibinfo{person}{Laurie
  Williams}.} \bibinfo{year}{2017}\natexlab{}.
\newblock \showarticletitle{To log, or not to log: using heuristics to identify
  mandatory log events--a controlled experiment}.
\newblock \bibinfo{journal}{\emph{Empirical Software Engineering}}
  (\bibinfo{year}{2017}), \bibinfo{pages}{2684--2717}.
\newblock


\bibitem[\protect\citeauthoryear{Klinkenberg, Terboven, Lankes, and
  Müller}{Klinkenberg et~al\mbox{.}}{2017}]%
        {klinkenberg2017data}
\bibfield{author}{\bibinfo{person}{Jannis Klinkenberg},
  \bibinfo{person}{Christian Terboven}, \bibinfo{person}{Stefan Lankes}, {and}
  \bibinfo{person}{Matthias~S. Müller}.} \bibinfo{year}{2017}\natexlab{}.
\newblock \showarticletitle{Data mining-based analysis of {hpc} center
  operations}. In \bibinfo{booktitle}{\emph{Proc. of the {IEEE} International
  Conference on Cluster Computing (CLUSTER)}}. \bibinfo{pages}{766--773}.
\newblock


\bibitem[\protect\citeauthoryear{Landauer, Skopik, Wurzenberger, and
  Rauber}{Landauer et~al\mbox{.}}{2020}]%
        {landauer2020system}
\bibfield{author}{\bibinfo{person}{Max Landauer}, \bibinfo{person}{Florian
  Skopik}, \bibinfo{person}{Markus Wurzenberger}, {and}
  \bibinfo{person}{Andreas Rauber}.} \bibinfo{year}{2020}\natexlab{}.
\newblock \showarticletitle{System log clustering approaches for cyber security
  applications: A survey}.
\newblock \bibinfo{journal}{\emph{Computers \& Security}}  \bibinfo{volume}{92}
  (\bibinfo{year}{2020}), \bibinfo{pages}{101739}.
\newblock


\bibitem[\protect\citeauthoryear{Laprie}{Laprie}{1995}]%
        {laprie1995dependable}
\bibfield{author}{\bibinfo{person}{Jean-Claude Laprie}.}
  \bibinfo{year}{1995}\natexlab{}.
\newblock \showarticletitle{Dependable computing: Concepts, limits,
  challenges}. In \bibinfo{booktitle}{\emph{Special issue of the 25th
  international symposium on fault-tolerant computing (FTCS)}}.
  \bibinfo{pages}{42--54}.
\newblock


\bibitem[\protect\citeauthoryear{Lemieux, Park, and Beschastnikh}{Lemieux
  et~al\mbox{.}}{2015}]%
        {lemieux2015general}
\bibfield{author}{\bibinfo{person}{Caroline Lemieux}, \bibinfo{person}{Dennis
  Park}, {and} \bibinfo{person}{Ivan Beschastnikh}.}
  \bibinfo{year}{2015}\natexlab{}.
\newblock \showarticletitle{General ltl specification mining (t)}. In
  \bibinfo{booktitle}{\emph{2015 30th IEEE/ACM International Conference on
  Automated Software Engineering (ASE)}}. \bibinfo{pages}{81--92}.
\newblock


\bibitem[\protect\citeauthoryear{Li, Chen, Shang, and Hassan}{Li
  et~al\mbox{.}}{2018}]%
        {DBLP:journals/ese/LiCSH18}
\bibfield{author}{\bibinfo{person}{Heng Li},
  \bibinfo{person}{Tse{-}Hsun~(Peter) Chen}, \bibinfo{person}{Weiyi Shang},
  {and} \bibinfo{person}{Ahmed~E. Hassan}.} \bibinfo{year}{2018}\natexlab{}.
\newblock \showarticletitle{Studying software logging using topic models}.
\newblock \bibinfo{journal}{\emph{Empir. Softw. Eng.}} (\bibinfo{year}{2018}),
  \bibinfo{pages}{2655--2694}.
\newblock


\bibitem[\protect\citeauthoryear{Li, Shang, and Hassan}{Li
  et~al\mbox{.}}{2017a}]%
        {li2017log}
\bibfield{author}{\bibinfo{person}{Heng Li}, \bibinfo{person}{Weiyi Shang},
  {and} \bibinfo{person}{Ahmed~E Hassan}.} \bibinfo{year}{2017}\natexlab{a}.
\newblock \showarticletitle{Which log level should developers choose for a new
  logging statement?}
\newblock \bibinfo{journal}{\emph{Empirical Software Engineering}}
  (\bibinfo{year}{2017}), \bibinfo{pages}{1684--1716}.
\newblock


\bibitem[\protect\citeauthoryear{Li, Shang, Zou, and Hassan}{Li
  et~al\mbox{.}}{2017b}]%
        {li2017towards}
\bibfield{author}{\bibinfo{person}{Heng Li}, \bibinfo{person}{Weiyi Shang},
  \bibinfo{person}{Ying Zou}, {and} \bibinfo{person}{Ahmed~E Hassan}.}
  \bibinfo{year}{2017}\natexlab{b}.
\newblock \showarticletitle{Towards just-in-time suggestions for log changes}.
\newblock \bibinfo{journal}{\emph{Empirical Software Engineering}}
  (\bibinfo{year}{2017}), \bibinfo{pages}{1831--1865}.
\newblock


\bibitem[\protect\citeauthoryear{Li, Ramachandran, Sahoo, Adve, Adve, and
  Zhou}{Li et~al\mbox{.}}{2008}]%
        {li2008trace}
\bibfield{author}{\bibinfo{person}{Man-Lap Li}, \bibinfo{person}{Pradeep
  Ramachandran}, \bibinfo{person}{Swarup~K Sahoo}, \bibinfo{person}{Sarita~V
  Adve}, \bibinfo{person}{Vikram~S Adve}, {and} \bibinfo{person}{Yuanyuan
  Zhou}.} \bibinfo{year}{2008}\natexlab{}.
\newblock \showarticletitle{Trace-based microarchitecture-level diagnosis of
  permanent hardware faults}. In \bibinfo{booktitle}{\emph{Proc. of the IEEE
  International Conference on Dependable Systems and Networks (DSN)}}.
  \bibinfo{pages}{22--31}.
\newblock


\bibitem[\protect\citeauthoryear{Li, Niu, Jia, Liao, Wang, and Li}{Li
  et~al\mbox{.}}{2020b}]%
        {DBLP:journals/ese/LiNJLWL20}
\bibfield{author}{\bibinfo{person}{Shanshan Li}, \bibinfo{person}{Xu Niu},
  \bibinfo{person}{Zhouyang Jia}, \bibinfo{person}{Xiangke Liao},
  \bibinfo{person}{Ji Wang}, {and} \bibinfo{person}{Tao Li}.}
  \bibinfo{year}{2020}\natexlab{b}.
\newblock \showarticletitle{Guiding log revisions by learning from software
  evolution history}.
\newblock \bibinfo{journal}{\emph{Empir. Softw. Eng.}} (\bibinfo{year}{2020}),
  \bibinfo{pages}{2302--2340}.
\newblock


\bibitem[\protect\citeauthoryear{Li, Chen, and Shang}{Li
  et~al\mbox{.}}{2020a}]%
        {li2020wherelog}
\bibfield{author}{\bibinfo{person}{Zhenhao Li}, \bibinfo{person}{Tse-Hsun
  Chen}, {and} \bibinfo{person}{Weiyi Shang}.}
  \bibinfo{year}{2020}\natexlab{a}.
\newblock \showarticletitle{Where shall we log? studying and suggesting logging
  locations in code blocks}. In \bibinfo{booktitle}{\emph{Proc. of the 35rd
  {IEEE/ACM} International Conference on Automated Software Engineering
  (ASE)}}.
\newblock


\bibitem[\protect\citeauthoryear{Li, Chen, Yang, and Shang}{Li
  et~al\mbox{.}}{2019}]%
        {DBLP:conf/icse/LiC0S19}
\bibfield{author}{\bibinfo{person}{Zhenhao Li},
  \bibinfo{person}{Tse{-}Hsun~(Peter) Chen}, \bibinfo{person}{Jinqiu Yang},
  {and} \bibinfo{person}{Weiyi Shang}.} \bibinfo{year}{2019}\natexlab{}.
\newblock \showarticletitle{Dlfinder: characterizing and detecting duplicate
  logging code smells}. In \bibinfo{booktitle}{\emph{Proc. of the 41st
  International Conference on Software Engineering (ICSE)}}.
  \bibinfo{pages}{152--163}.
\newblock


\bibitem[\protect\citeauthoryear{Liang, Zhang, Xiong, and Sahoo}{Liang
  et~al\mbox{.}}{2007}]%
        {liang2007failure}
\bibfield{author}{\bibinfo{person}{Yinglung Liang}, \bibinfo{person}{Yanyong
  Zhang}, \bibinfo{person}{Hui Xiong}, {and} \bibinfo{person}{Ramendra Sahoo}.}
  \bibinfo{year}{2007}\natexlab{}.
\newblock \showarticletitle{Failure prediction in ibm bluegene/l event logs}.
  In \bibinfo{booktitle}{\emph{Proc. of the 7th IEEE International Conference
  on Data Mining (ICDM)}}. \bibinfo{pages}{583--588}.
\newblock


\bibitem[\protect\citeauthoryear{Lim, Singh, and Yajnik}{Lim
  et~al\mbox{.}}{2008}]%
        {lim2008log}
\bibfield{author}{\bibinfo{person}{Chinghway Lim}, \bibinfo{person}{Navjot
  Singh}, {and} \bibinfo{person}{Shalini Yajnik}.}
  \bibinfo{year}{2008}\natexlab{}.
\newblock \showarticletitle{A log mining approach to failure analysis of
  enterprise telephony systems}. In \bibinfo{booktitle}{\emph{Proc. of the 38th
  IEEE/IFIP International Conference on Dependable Systems and Networks
  (DSN)}}.
\newblock


\bibitem[\protect\citeauthoryear{Lin, Zhou, Yao, Guo, and Li}{Lin
  et~al\mbox{.}}{2015}]%
        {lin2015cowic}
\bibfield{author}{\bibinfo{person}{Hao Lin}, \bibinfo{person}{Jingyu Zhou},
  \bibinfo{person}{Bin Yao}, \bibinfo{person}{Minyi Guo}, {and}
  \bibinfo{person}{Jie Li}.} \bibinfo{year}{2015}\natexlab{}.
\newblock \showarticletitle{Cowic: A column-wise independent compression for
  log stream analysis}. In \bibinfo{booktitle}{\emph{Proc. of the 15th
  International Symposium on Cluster, Cloud and Grid Computing (CCGRID)}}.
\newblock


\bibitem[\protect\citeauthoryear{Lin, Yao, Chintalapati, Zhang, Hsieh, Dang,
  Zhang, Sui, Xu, Lou, Li, and Wu}{Lin et~al\mbox{.}}{2018}]%
        {lin2018predicting}
\bibfield{author}{\bibinfo{person}{Qingwei Lin}, \bibinfo{person}{Randolph
  Yao}, \bibinfo{person}{Murali Chintalapati}, \bibinfo{person}{Dongmei Zhang},
  \bibinfo{person}{Ken Hsieh}, \bibinfo{person}{Yingnong Dang},
  \bibinfo{person}{Hongyu Zhang}, \bibinfo{person}{Kaixin Sui},
  \bibinfo{person}{Yong Xu}, \bibinfo{person}{Jian-Guang Lou},
  \bibinfo{person}{Chenggang Li}, {and} \bibinfo{person}{Youjiang Wu}.}
  \bibinfo{year}{2018}\natexlab{}.
\newblock \showarticletitle{Predicting node failure in cloud service systems}.
  In \bibinfo{booktitle}{\emph{Proc. of the 2018 26th {ACM} Joint Meeting on
  European Software Engineering Conference and Symposium on the Foundations of
  Software Engineering ({ESEC}/{FSE})}}. \bibinfo{publisher}{{ACM} Press},
  \bibinfo{pages}{480--490}.
\newblock


\bibitem[\protect\citeauthoryear{Lin, Zhang, Lou, Zhang, and Chen}{Lin
  et~al\mbox{.}}{2016}]%
        {lin2016log}
\bibfield{author}{\bibinfo{person}{Qingwei Lin}, \bibinfo{person}{Hongyu
  Zhang}, \bibinfo{person}{Jian-Guang Lou}, \bibinfo{person}{Yu Zhang}, {and}
  \bibinfo{person}{Xuewei Chen}.} \bibinfo{year}{2016}\natexlab{}.
\newblock \showarticletitle{Log clustering based problem identification for
  online service systems}. In \bibinfo{booktitle}{\emph{Proc. of the 38th
  International Conference on Software Engineering Companion
  (ICSE-Companion)}}. \bibinfo{pages}{102--111}.
\newblock


\bibitem[\protect\citeauthoryear{Liu, Wen, Zhang, Jiang, Xing, and Meng}{Liu
  et~al\mbox{.}}{2019a}]%
        {liu2019log2vec}
\bibfield{author}{\bibinfo{person}{Fucheng Liu}, \bibinfo{person}{Yu Wen},
  \bibinfo{person}{Dongxue Zhang}, \bibinfo{person}{Xihe Jiang},
  \bibinfo{person}{Xinyu Xing}, {and} \bibinfo{person}{Dan Meng}.}
  \bibinfo{year}{2019}\natexlab{a}.
\newblock \showarticletitle{Log2vec: a heterogeneous graph embedding based
  approach for detecting cyber threats within enterprise}. In
  \bibinfo{booktitle}{\emph{Proc. of the 26th ACM SIGSAC Conference on Computer
  and Communications Security (CCS)}}. \bibinfo{pages}{1777--1794}.
\newblock


\bibitem[\protect\citeauthoryear{Liu, Zhu, He, He, Zheng, and Lyu}{Liu
  et~al\mbox{.}}{2019c}]%
        {liu2019logzip}
\bibfield{author}{\bibinfo{person}{Jinyang Liu}, \bibinfo{person}{Jieming Zhu},
  \bibinfo{person}{Shilin He}, \bibinfo{person}{Pinjia He},
  \bibinfo{person}{Zibin Zheng}, {and} \bibinfo{person}{Michael~R. Lyu}.}
  \bibinfo{year}{2019}\natexlab{c}.
\newblock \showarticletitle{Logzip: extracting hidden structures via iterative
  clustering for log compression}. In \bibinfo{booktitle}{\emph{Proc. of the
  34th {IEEE/ACM} International Conference on Automated Software Engineering
  (ASE)}}. \bibinfo{pages}{863--873}.
\newblock


\bibitem[\protect\citeauthoryear{Liu, Xia, Lo, Xing, Hassan, and Li}{Liu
  et~al\mbox{.}}{2019b}]%
        {liu2019variables}
\bibfield{author}{\bibinfo{person}{Zhongxin Liu}, \bibinfo{person}{Xin Xia},
  \bibinfo{person}{David Lo}, \bibinfo{person}{Zhenchang Xing},
  \bibinfo{person}{Ahmed~E Hassan}, {and} \bibinfo{person}{Shanping Li}.}
  \bibinfo{year}{2019}\natexlab{b}.
\newblock \showarticletitle{Which variables should I log?}
\newblock \bibinfo{journal}{\emph{IEEE Transactions on Software Engineering}}
  (\bibinfo{year}{2019}).
\newblock


\bibitem[\protect\citeauthoryear{Lo and Maoz}{Lo and Maoz}{2008}]%
        {lo2008mining}
\bibfield{author}{\bibinfo{person}{David Lo} {and} \bibinfo{person}{Shahar
  Maoz}.} \bibinfo{year}{2008}\natexlab{}.
\newblock \showarticletitle{Mining scenario-based triggers and effects}. In
  \bibinfo{booktitle}{\emph{2008 23rd IEEE/ACM International Conference on
  Automated Software Engineering}}. \bibinfo{pages}{109--118}.
\newblock


\bibitem[\protect\citeauthoryear{Lo and Maoz}{Lo and Maoz}{2012}]%
        {lo2012scenario}
\bibfield{author}{\bibinfo{person}{David Lo} {and} \bibinfo{person}{Shahar
  Maoz}.} \bibinfo{year}{2012}\natexlab{}.
\newblock \showarticletitle{Scenario-based and value-based specification
  mining: better together}.
\newblock \bibinfo{journal}{\emph{Automated Software Engineering}}
  (\bibinfo{year}{2012}), \bibinfo{pages}{423--458}.
\newblock


\bibitem[\protect\citeauthoryear{Lo, Maoz, and Khoo}{Lo et~al\mbox{.}}{2007}]%
        {lo2007mining}
\bibfield{author}{\bibinfo{person}{David Lo}, \bibinfo{person}{Shahar Maoz},
  {and} \bibinfo{person}{Siau-Cheng Khoo}.} \bibinfo{year}{2007}\natexlab{}.
\newblock \showarticletitle{Mining modal scenario-based specifications from
  execution traces of reactive systems}. In
  \bibinfo{booktitle}{\emph{Proceedings of the twenty-second IEEE/ACM
  international conference on Automated software engineering}}.
  \bibinfo{pages}{465--468}.
\newblock


\bibitem[\protect\citeauthoryear{Lockerman, Faleiro, Kim, Sankaran, Abadi,
  Aspnes, Sen, and Balakrishnan}{Lockerman et~al\mbox{.}}{2018}]%
        {DBLP:conf/osdi/LockermanFKSAA018}
\bibfield{author}{\bibinfo{person}{Joshua Lockerman}, \bibinfo{person}{Jose~M.
  Faleiro}, \bibinfo{person}{Juno Kim}, \bibinfo{person}{Soham Sankaran},
  \bibinfo{person}{Daniel~J. Abadi}, \bibinfo{person}{James Aspnes},
  \bibinfo{person}{Siddhartha Sen}, {and} \bibinfo{person}{Mahesh
  Balakrishnan}.} \bibinfo{year}{2018}\natexlab{}.
\newblock \showarticletitle{The fuzzylog: {a} partially ordered shared log}. In
  \bibinfo{booktitle}{\emph{Proc. of the 13th {USENIX} Symposium on Operating
  Systems Design and Implementation (OSDI)}}. \bibinfo{pages}{357--372}.
\newblock


\bibitem[\protect\citeauthoryear{Log4j}{Log4j}{2020}]%
        {Log4j}
\bibfield{author}{\bibinfo{person}{Log4j}.} \bibinfo{year}{2020}\natexlab{}.
\newblock \bibinfo{title}{Apache Log4j}.
\newblock
\newblock
\urldef\tempurl%
\url{http://logging.apache.org/log4j/}
\showURL{%
Retrieved September 1, 2020 from \tempurl}


\bibitem[\protect\citeauthoryear{Logalyze}{Logalyze}{2020}]%
        {Logalyze}
\bibfield{author}{\bibinfo{person}{Logalyze}.} \bibinfo{year}{2020}\natexlab{}.
\newblock \bibinfo{title}{An open source log management and network monitoring
  software.}
\newblock
\newblock
\urldef\tempurl%
\url{http://www.logalyze.com}
\showURL{%
Retrieved September 1, 2020 from \tempurl}


\bibitem[\protect\citeauthoryear{Loggly}{Loggly}{2009a}]%
        {loggly_autoparsing}
\bibfield{author}{\bibinfo{person}{Loggly}.} \bibinfo{year}{2009}\natexlab{a}.
\newblock \bibinfo{title}{Automated parsing log types}.
\newblock
\newblock
\urldef\tempurl%
\url{https://www.loggly.com/docs/automated-parsing}
\showURL{%
Retrieved September 1, 2020 from \tempurl}


\bibitem[\protect\citeauthoryear{Loggly}{Loggly}{2009b}]%
        {loggly}
\bibfield{author}{\bibinfo{person}{Loggly}.} \bibinfo{year}{2009}\natexlab{b}.
\newblock \bibinfo{title}{Loggly - log management by loggly}.
\newblock
\newblock
\urldef\tempurl%
\url{https://www.loggly.com}
\showURL{%
Retrieved September 1, 2020 from \tempurl}


\bibitem[\protect\citeauthoryear{Loghub}{Loghub}{2020}]%
        {Loghub}
\bibfield{author}{\bibinfo{person}{Loghub}.} \bibinfo{year}{2020}\natexlab{}.
\newblock \bibinfo{title}{A large collection of log datasets from various
  systems}.
\newblock
\newblock
\urldef\tempurl%
\url{https://github.com/logpai/loghub/}
\showURL{%
Retrieved September 1, 2020 from \tempurl}


\bibitem[\protect\citeauthoryear{LogPAI}{LogPAI}{2020}]%
        {LogPAI}
\bibfield{author}{\bibinfo{person}{LogPAI}.} \bibinfo{year}{2020}\natexlab{}.
\newblock \bibinfo{title}{A platform for log analytics powered by AI}.
\newblock
\newblock
\urldef\tempurl%
\url{https://www.logpai.com}
\showURL{%
Retrieved September 1, 2020 from \tempurl}


\bibitem[\protect\citeauthoryear{Logstash}{Logstash}{2020}]%
        {Logstash}
\bibfield{author}{\bibinfo{person}{Logstash}.} \bibinfo{year}{2020}\natexlab{}.
\newblock \bibinfo{title}{A server-side processor for log data}.
\newblock
\newblock
\urldef\tempurl%
\url{https://www.elastic.co/logstash}
\showURL{%
Retrieved September 1, 2020 from \tempurl}


\bibitem[\protect\citeauthoryear{logz}{logz}{2014}]%
        {logz_autoparsing}
\bibfield{author}{\bibinfo{person}{logz}.} \bibinfo{year}{2014}\natexlab{}.
\newblock \bibinfo{title}{Log parsing - automated, easy to use, and efficient}.
\newblock
\newblock
\urldef\tempurl%
\url{https://logz.io/product/log-parsing}
\showURL{%
Retrieved September 1, 2020 from \tempurl}


\bibitem[\protect\citeauthoryear{Lou, Fu, Yang, Li, and Wu}{Lou
  et~al\mbox{.}}{2010a}]%
        {lou2010mining_b}
\bibfield{author}{\bibinfo{person}{Jian-Guang Lou}, \bibinfo{person}{Qiang Fu},
  \bibinfo{person}{Shengqi Yang}, \bibinfo{person}{Jiang Li}, {and}
  \bibinfo{person}{Bin Wu}.} \bibinfo{year}{2010}\natexlab{a}.
\newblock \showarticletitle{Mining program workflow from interleaved traces}.
  In \bibinfo{booktitle}{\emph{Proc. of the 16th ACM SIGKDD international
  conference on Knowledge discovery and data mining (SIGKDD)}}.
\newblock


\bibitem[\protect\citeauthoryear{Lou, Fu, Yang, Xu, and Li}{Lou
  et~al\mbox{.}}{2010b}]%
        {lou2010mining_a}
\bibfield{author}{\bibinfo{person}{Jian-Guang Lou}, \bibinfo{person}{Qiang Fu},
  \bibinfo{person}{Shengqi Yang}, \bibinfo{person}{Ye Xu}, {and}
  \bibinfo{person}{Jiang Li}.} \bibinfo{year}{2010}\natexlab{b}.
\newblock \showarticletitle{Mining invariants from console logs for system
  problem detection.}. In \bibinfo{booktitle}{\emph{Proc. of the 2010 {USENIX}
  Annual Technical Conference (ATC)}}. \bibinfo{pages}{1--14}.
\newblock


\bibitem[\protect\citeauthoryear{Lu, Li, Li, and Feng}{Lu
  et~al\mbox{.}}{2018a}]%
        {lu2018cloudraid}
\bibfield{author}{\bibinfo{person}{Jie Lu}, \bibinfo{person}{Feng Li},
  \bibinfo{person}{Lian Li}, {and} \bibinfo{person}{Xiaobing Feng}.}
  \bibinfo{year}{2018}\natexlab{a}.
\newblock \showarticletitle{Cloudraid: hunting concurrency bugs in the cloud
  via log-mining}. In \bibinfo{booktitle}{\emph{Proc. of the 26th ACM Joint
  European Software Engineering Conference and Symposium on the Foundations of
  Software Engineering (ESEC/FSE)}}. \bibinfo{pages}{3--14}.
\newblock


\bibitem[\protect\citeauthoryear{Lu, Rao, Wei, Tak, Wang, and Wang}{Lu
  et~al\mbox{.}}{2017}]%
        {lu2017log}
\bibfield{author}{\bibinfo{person}{Siyang Lu}, \bibinfo{person}{BingBing Rao},
  \bibinfo{person}{Xiang Wei}, \bibinfo{person}{Byungchul Tak},
  \bibinfo{person}{Long Wang}, {and} \bibinfo{person}{Liqiang Wang}.}
  \bibinfo{year}{2017}\natexlab{}.
\newblock \showarticletitle{Log-based abnormal task detection and root cause
  analysis for spark}. In \bibinfo{booktitle}{\emph{Proc. of the IEEE
  International Conference on Web Services (ICWS)}}.
\newblock


\bibitem[\protect\citeauthoryear{Lu, Wei, Li, and Wang}{Lu
  et~al\mbox{.}}{2018b}]%
        {lu2018detecting}
\bibfield{author}{\bibinfo{person}{Siyang Lu}, \bibinfo{person}{Xiang Wei},
  \bibinfo{person}{Yandong Li}, {and} \bibinfo{person}{Liqiang Wang}.}
  \bibinfo{year}{2018}\natexlab{b}.
\newblock \showarticletitle{Detecting anomaly in big data system logs using
  convolutional neural network}. In \bibinfo{booktitle}{\emph{Proc. of the 16th
  {IEEE} International Conference on Dependable, Autonomic and Secure Computing
  (DASC)}}. \bibinfo{pages}{151--158}.
\newblock


\bibitem[\protect\citeauthoryear{Luo, Nath, Sivalingam, Musuvathi, and
  Ceze}{Luo et~al\mbox{.}}{2018}]%
        {DBLP:conf/usenix/LuoNSMC18}
\bibfield{author}{\bibinfo{person}{Liang Luo}, \bibinfo{person}{Suman Nath},
  \bibinfo{person}{Lenin~Ravindranath Sivalingam}, \bibinfo{person}{Madan
  Musuvathi}, {and} \bibinfo{person}{Luis Ceze}.}
  \bibinfo{year}{2018}\natexlab{}.
\newblock \showarticletitle{Troubleshooting transiently-recurring errors in
  production systems with blame-proportional logging}. In
  \bibinfo{booktitle}{\emph{Proc. of the 2018 {USENIX} Annual Technical
  Conference (ATC)}}. \bibinfo{pages}{321--334}.
\newblock


\bibitem[\protect\citeauthoryear{Lyu~(ed.)}{Lyu~(ed.)}{1996}]%
        {lyu1996handbook}
\bibfield{author}{\bibinfo{person}{Michael~R. Lyu~(ed.)}.}
  \bibinfo{year}{1996}\natexlab{}.
\newblock \bibinfo{booktitle}{\emph{Handbook of software reliability
  engineering}}.
\newblock \bibinfo{publisher}{IEEE Computer Society Press}.
\newblock


\bibitem[\protect\citeauthoryear{Makanju, Zincir-Heywood, and Milios}{Makanju
  et~al\mbox{.}}{2012}]%
        {IPLoM12}
\bibfield{author}{\bibinfo{person}{Adetokunbo Makanju}, \bibinfo{person}{A~Nur
  Zincir-Heywood}, {and} \bibinfo{person}{Evangelos~E Milios}.}
  \bibinfo{year}{2012}\natexlab{}.
\newblock \showarticletitle{A lightweight algorithm for message type extraction
  in system application logs}.
\newblock \bibinfo{journal}{\emph{IEEE Transactions on Knowledge and Data
  Engineering (TKDE)}} (\bibinfo{year}{2012}).
\newblock


\bibitem[\protect\citeauthoryear{Makanju, Zincir-Heywood, and Milios}{Makanju
  et~al\mbox{.}}{2009}]%
        {IPLoM09}
\bibfield{author}{\bibinfo{person}{Adetokunbo~AO Makanju},
  \bibinfo{person}{A~Nur Zincir-Heywood}, {and} \bibinfo{person}{Evangelos~E
  Milios}.} \bibinfo{year}{2009}\natexlab{}.
\newblock \showarticletitle{Clustering event logs using iterative
  partitioning}. In \bibinfo{booktitle}{\emph{Proc. of International Conference
  on Knowledge Discovery and Data Mining (KDD)}}. \bibinfo{pages}{1255--1264}.
\newblock


\bibitem[\protect\citeauthoryear{Meinig, Tr{\"o}ger, and Meinel}{Meinig
  et~al\mbox{.}}{2019}]%
        {meinig2019rough}
\bibfield{author}{\bibinfo{person}{Michael Meinig}, \bibinfo{person}{Peter
  Tr{\"o}ger}, {and} \bibinfo{person}{Christoph Meinel}.}
  \bibinfo{year}{2019}\natexlab{}.
\newblock \showarticletitle{Rough logs: a data reduction approach for log
  files}. In \bibinfo{booktitle}{\emph{Proc. of the International Conference on
  Enterprise Information Systems (ICEIS)}}. \bibinfo{pages}{295--302}.
\newblock


\bibitem[\protect\citeauthoryear{Mell and Harang}{Mell and Harang}{2014}]%
        {mell2014lightweight}
\bibfield{author}{\bibinfo{person}{Peter Mell} {and} \bibinfo{person}{Richard~E
  Harang}.} \bibinfo{year}{2014}\natexlab{}.
\newblock \showarticletitle{Lightweight packing of log files for improved
  compression in mobile tactical networks}. In \bibinfo{booktitle}{\emph{Proc.
  of the {IEEE} Military Communications Conference (MILCOM)}}.
  \bibinfo{pages}{192--197}.
\newblock


\bibitem[\protect\citeauthoryear{Meng, Liu, Huang, Zhang, Zaiter, Chen, and
  Pei}{Meng et~al\mbox{.}}{2020}]%
        {mengsemantic}
\bibfield{author}{\bibinfo{person}{Weibin Meng}, \bibinfo{person}{Ying Liu},
  \bibinfo{person}{Yuheng Huang}, \bibinfo{person}{Shenglin Zhang},
  \bibinfo{person}{Federico Zaiter}, \bibinfo{person}{Bingjin Chen}, {and}
  \bibinfo{person}{Dan Pei}.} \bibinfo{year}{2020}\natexlab{}.
\newblock \showarticletitle{A semantic-aware representation framework for
  online log analysis}.
\newblock  (\bibinfo{year}{2020}).
\newblock


\bibitem[\protect\citeauthoryear{Meng, Liu, Zhu, Zhang, Pei, Liu, Chen, Zhang,
  Tao, Sun, et~al\mbox{.}}{Meng et~al\mbox{.}}{2019}]%
        {meng2019loganomaly}
\bibfield{author}{\bibinfo{person}{Weibin Meng}, \bibinfo{person}{Ying Liu},
  \bibinfo{person}{Yichen Zhu}, \bibinfo{person}{Shenglin Zhang},
  \bibinfo{person}{Dan Pei}, \bibinfo{person}{Yuqing Liu},
  \bibinfo{person}{Yihao Chen}, \bibinfo{person}{Ruizhi Zhang},
  \bibinfo{person}{Shimin Tao}, \bibinfo{person}{Pei Sun}, {et~al\mbox{.}}}
  \bibinfo{year}{2019}\natexlab{}.
\newblock \showarticletitle{LogAnomaly: unsupervised detection of sequential
  and quantitative anomalies in unstructured logs.}. In
  \bibinfo{booktitle}{\emph{Proc. of the 2019 International Joint Conferences
  on Artificial Intelligence (IJCAI)}}. \bibinfo{pages}{4739--4745}.
\newblock


\bibitem[\protect\citeauthoryear{Messaoudi, Panichella, Bianculli, Briand, and
  Sasnauskas}{Messaoudi et~al\mbox{.}}{2018}]%
        {molfi18}
\bibfield{author}{\bibinfo{person}{Salma Messaoudi}, \bibinfo{person}{Annibale
  Panichella}, \bibinfo{person}{Domenico Bianculli}, \bibinfo{person}{Lionel
  Briand}, {and} \bibinfo{person}{Raimondas Sasnauskas}.}
  \bibinfo{year}{2018}\natexlab{}.
\newblock \showarticletitle{A search-based approach for accurate identification
  of log message formats}. In \bibinfo{booktitle}{\emph{Proc. of the IEEE/ACM
  26th International Conference on Program Comprehension (ICPC)}}.
  \bibinfo{pages}{167--16710}.
\newblock


\bibitem[\protect\citeauthoryear{Mi, Wang, Zhou, Lyu, and Cai}{Mi
  et~al\mbox{.}}{2013}]%
        {mi2013toward}
\bibfield{author}{\bibinfo{person}{Haibo Mi}, \bibinfo{person}{Huaimin Wang},
  \bibinfo{person}{Yangfan Zhou}, \bibinfo{person}{Michael~R. Lyu}, {and}
  \bibinfo{person}{Hua Cai}.} \bibinfo{year}{2013}\natexlab{}.
\newblock \showarticletitle{Toward fine-grained, unsupervised, scalable
  performance diagnosis for production cloud computing systems}.
\newblock \bibinfo{journal}{\emph{Transactions on Parallel and Distributed
  Systems}} \bibinfo{volume}{24}, \bibinfo{number}{6} (\bibinfo{year}{2013}),
  \bibinfo{pages}{1245--1255}.
\newblock


\bibitem[\protect\citeauthoryear{Microsoft}{Microsoft}{2018}]%
        {ms-event-logging}
\bibfield{author}{\bibinfo{person}{Microsoft}.}
  \bibinfo{year}{2018}\natexlab{}.
\newblock \bibinfo{title}{Event logging}.
\newblock
\newblock
\urldef\tempurl%
\url{https://docs.microsoft.com/en-us/windows/win32/eventlog/event-logging}
\showURL{%
Retrieved September 1, 2020 from \tempurl}


\bibitem[\protect\citeauthoryear{Mikolov, Sutskever, Chen, Corrado, and
  Dean}{Mikolov et~al\mbox{.}}{2013}]%
        {mikolov2013distributed}
\bibfield{author}{\bibinfo{person}{Tomas Mikolov}, \bibinfo{person}{Ilya
  Sutskever}, \bibinfo{person}{Kai Chen}, \bibinfo{person}{Greg~S Corrado},
  {and} \bibinfo{person}{Jeff Dean}.} \bibinfo{year}{2013}\natexlab{}.
\newblock \showarticletitle{Distributed representations of words and phrases
  and their compositionality}. In \bibinfo{booktitle}{\emph{Proc. of the 27th
  Conference on Neural Information Processing Systems (NIPS)}}.
  \bibinfo{pages}{3111--3119}.
\newblock


\bibitem[\protect\citeauthoryear{Mizouchi, Shimari, Ishio, and Inoue}{Mizouchi
  et~al\mbox{.}}{2019}]%
        {DBLP:conf/iwpc/MizouchiSII19}
\bibfield{author}{\bibinfo{person}{Tsuyoshi Mizouchi},
  \bibinfo{person}{Kazumasa Shimari}, \bibinfo{person}{Takashi Ishio}, {and}
  \bibinfo{person}{Katsuro Inoue}.} \bibinfo{year}{2019}\natexlab{}.
\newblock \showarticletitle{{PADLA:} a dynamic log level adapter using online
  phase detection}. In \bibinfo{booktitle}{\emph{Proc. of the 27th
  International Conference on Program Comprehension (ICPC)}}.
  \bibinfo{pages}{135--138}.
\newblock


\bibitem[\protect\citeauthoryear{Mizutani}{Mizutani}{2013}]%
        {SHISO_13}
\bibfield{author}{\bibinfo{person}{Masayoshi Mizutani}.}
  \bibinfo{year}{2013}\natexlab{}.
\newblock \showarticletitle{Incremental mining of system log format}. In
  \bibinfo{booktitle}{\emph{Proc. of the IEEE International Conference on
  Services Computing (SCC)}}. \bibinfo{pages}{595--602}.
\newblock


\bibitem[\protect\citeauthoryear{Nagappan and Vouk}{Nagappan and Vouk}{2010}]%
        {LFA_10}
\bibfield{author}{\bibinfo{person}{Meiyappan Nagappan} {and}
  \bibinfo{person}{Mladen~A Vouk}.} \bibinfo{year}{2010}\natexlab{}.
\newblock \showarticletitle{Abstracting log lines to log event types for mining
  software system logs}. In \bibinfo{booktitle}{\emph{Proc. of the 7th IEEE
  Working Conference on Mining Software Repositories (MSR)}}.
  \bibinfo{pages}{114--117}.
\newblock


\bibitem[\protect\citeauthoryear{Nagaraj, Killian, and Neville}{Nagaraj
  et~al\mbox{.}}{2012}]%
        {nagaraj2012structured}
\bibfield{author}{\bibinfo{person}{Karthik Nagaraj}, \bibinfo{person}{Charles
  Killian}, {and} \bibinfo{person}{Jennifer Neville}.}
  \bibinfo{year}{2012}\natexlab{}.
\newblock \showarticletitle{Structured comparative analysis of systems logs to
  diagnose performance problems}. In \bibinfo{booktitle}{\emph{Proc. of the 9th
  USENIX Symposium on Networked Systems Design and Implementation (NSDI)}}.
  \bibinfo{pages}{353--366}.
\newblock


\bibitem[\protect\citeauthoryear{Nandi, Mandal, Atreja, Dasgupta, and
  Bhattacharya}{Nandi et~al\mbox{.}}{2016}]%
        {nandi2016anomaly}
\bibfield{author}{\bibinfo{person}{Animesh Nandi}, \bibinfo{person}{Atri
  Mandal}, \bibinfo{person}{Shubham Atreja}, \bibinfo{person}{Gargi~B
  Dasgupta}, {and} \bibinfo{person}{Subhrajit Bhattacharya}.}
  \bibinfo{year}{2016}\natexlab{}.
\newblock \showarticletitle{Anomaly detection using program control flow graph
  mining from execution logs}. In \bibinfo{booktitle}{\emph{Proc. of the 22nd
  ACM SIGKDD International Conference on Knowledge Discovery and Data Mining
  (SIGKDD)}}. \bibinfo{pages}{215--224}.
\newblock


\bibitem[\protect\citeauthoryear{Oliner, Ganapathi, and Xu}{Oliner
  et~al\mbox{.}}{2012}]%
        {DBLP:journals/cacm/OlinerGX12}
\bibfield{author}{\bibinfo{person}{Adam~J. Oliner}, \bibinfo{person}{Archana
  Ganapathi}, {and} \bibinfo{person}{Wei Xu}.} \bibinfo{year}{2012}\natexlab{}.
\newblock \showarticletitle{Advances and challenges in log analysis}.
\newblock \bibinfo{journal}{\emph{ACM Communication}} (\bibinfo{year}{2012}),
  \bibinfo{pages}{55--61}.
\newblock


\bibitem[\protect\citeauthoryear{Otten et~al\mbox{.}}{Otten
  et~al\mbox{.}}{2008}]%
        {otten2008using}
\bibfield{author}{\bibinfo{person}{Frederick~John Otten} {et~al\mbox{.}}}
  \bibinfo{year}{2008}\natexlab{}.
\newblock \emph{\bibinfo{title}{Using semantic knowledge to improve compression
  on log files}}.
\newblock \bibinfo{thesistype}{Ph.D. Dissertation}. \bibinfo{school}{Rhodes
  University}.
\newblock


\bibitem[\protect\citeauthoryear{Paccagnella, Datta, Hassan, Bates, Fletcher,
  Miller, and Tian}{Paccagnella et~al\mbox{.}}{2020}]%
        {paccagnella2020custos}
\bibfield{author}{\bibinfo{person}{Riccardo Paccagnella},
  \bibinfo{person}{Pubali Datta}, \bibinfo{person}{Wajih~Ul Hassan},
  \bibinfo{person}{Adam Bates}, \bibinfo{person}{Christopher Fletcher},
  \bibinfo{person}{Andrew Miller}, {and} \bibinfo{person}{Dave Tian}.}
  \bibinfo{year}{2020}\natexlab{}.
\newblock \showarticletitle{Custos: Practical tamper-evident auditing of
  operating systems using trusted execution}. In
  \bibinfo{booktitle}{\emph{Network and Distributed System Security
  Symposium}}.
\newblock


\bibitem[\protect\citeauthoryear{Pecchia, Cinque, Carrozza, and
  Cotroneo}{Pecchia et~al\mbox{.}}{2015}]%
        {DBLP:conf/icse/PecchiaCCC15}
\bibfield{author}{\bibinfo{person}{Antonio Pecchia}, \bibinfo{person}{Marcello
  Cinque}, \bibinfo{person}{Gabriella Carrozza}, {and}
  \bibinfo{person}{Domenico Cotroneo}.} \bibinfo{year}{2015}\natexlab{}.
\newblock \showarticletitle{Industry practices and event logging: assessment of
  a critical software development process}. In \bibinfo{booktitle}{\emph{Proc.
  of the 37th {IEEE/ACM} International Conference on Software Engineering
  (ICSE)}}. \bibinfo{pages}{169--178}.
\newblock


\bibitem[\protect\citeauthoryear{Pham, Wang, Tak, Baset, Tang, Kalbarczyk, and
  Iyer}{Pham et~al\mbox{.}}{2016}]%
        {pham2016failure}
\bibfield{author}{\bibinfo{person}{Cuong Pham}, \bibinfo{person}{Long Wang},
  \bibinfo{person}{Byung~Chul Tak}, \bibinfo{person}{Salman Baset},
  \bibinfo{person}{Chunqiang Tang}, \bibinfo{person}{Zbigniew Kalbarczyk},
  {and} \bibinfo{person}{Ravishankar~K Iyer}.} \bibinfo{year}{2016}\natexlab{}.
\newblock \showarticletitle{Failure diagnosis for distributed systems using
  targeted fault injection}.
\newblock \bibinfo{journal}{\emph{Transactions on Parallel and Distributed
  Systems}} \bibinfo{volume}{28}, \bibinfo{number}{2} (\bibinfo{year}{2016}),
  \bibinfo{pages}{503--516}.
\newblock


\bibitem[\protect\citeauthoryear{Prewett}{Prewett}{2003}]%
        {prewett2003analyzing}
\bibfield{author}{\bibinfo{person}{James~E Prewett}.}
  \bibinfo{year}{2003}\natexlab{}.
\newblock \showarticletitle{Analyzing cluster log files using logsurfer}. In
  \bibinfo{booktitle}{\emph{Proc. of the 4th Annual Conference on Linux
  Clusters}}. Citeseer.
\newblock


\bibitem[\protect\citeauthoryear{Prometheus}{Prometheus}{2020}]%
        {Prometheus}
\bibfield{author}{\bibinfo{person}{Prometheus}.}
  \bibinfo{year}{2020}\natexlab{}.
\newblock \bibinfo{title}{A systems and service monitoring system}.
\newblock
\newblock
\urldef\tempurl%
\url{https://github.com/prometheus}
\showURL{%
Retrieved September 1, 2020 from \tempurl}


\bibitem[\protect\citeauthoryear{R{\'a}cz and Luk{\'a}cs}{R{\'a}cz and
  Luk{\'a}cs}{2004}]%
        {racz2004high}
\bibfield{author}{\bibinfo{person}{Bal{\'a}zs R{\'a}cz} {and}
  \bibinfo{person}{Andr{\'a}s Luk{\'a}cs}.} \bibinfo{year}{2004}\natexlab{}.
\newblock \showarticletitle{High density compression of log files}. In
  \bibinfo{booktitle}{\emph{Proc. of the Data Compression Conference (DCC)}}.
  IEEE, \bibinfo{pages}{557--557}.
\newblock


\bibitem[\protect\citeauthoryear{Rapid7}{Rapid7}{2000}]%
        {rapid_autoparsing}
\bibfield{author}{\bibinfo{person}{Rapid7}.} \bibinfo{year}{2000}\natexlab{}.
\newblock \bibinfo{title}{New automated log parsing}.
\newblock
\newblock
\urldef\tempurl%
\url{https://blog.rapid7.com/2016/03/03/new-automated-log-parsing}
\showURL{%
Retrieved September 1, 2020 from \tempurl}


\bibitem[\protect\citeauthoryear{Rouillard}{Rouillard}{2004}]%
        {rouillard2004real}
\bibfield{author}{\bibinfo{person}{John~P Rouillard}.}
  \bibinfo{year}{2004}\natexlab{}.
\newblock \showarticletitle{Real-time log file analysis using the simple event
  correlator (sec).}. In \bibinfo{booktitle}{\emph{Proc. of the 18th {USENIX}
  Large Installation System Administration Conference (LISA)}},
  Vol.~\bibinfo{volume}{4}. \bibinfo{pages}{133--150}.
\newblock


\bibitem[\protect\citeauthoryear{Russo, Succi, and Pedrycz}{Russo
  et~al\mbox{.}}{2015}]%
        {russo2015mining}
\bibfield{author}{\bibinfo{person}{Barbara Russo}, \bibinfo{person}{Giancarlo
  Succi}, {and} \bibinfo{person}{Witold Pedrycz}.}
  \bibinfo{year}{2015}\natexlab{}.
\newblock \showarticletitle{Mining system logs to learn error predictors: a
  case study of a telemetry system}.
\newblock \bibinfo{journal}{\emph{Empirical Software Engineering}}
  \bibinfo{volume}{20}, \bibinfo{number}{4} (\bibinfo{year}{2015}),
  \bibinfo{pages}{879--927}.
\newblock


\bibitem[\protect\citeauthoryear{Sahoo, Oliner, Rish, Gupta, Moreira, Ma,
  Vilalta, and Sivasubramaniam}{Sahoo et~al\mbox{.}}{2003}]%
        {sahoo2003critical}
\bibfield{author}{\bibinfo{person}{Ramendra~K. Sahoo}, \bibinfo{person}{Adam~J.
  Oliner}, \bibinfo{person}{Irina Rish}, \bibinfo{person}{Manish Gupta},
  \bibinfo{person}{José~E. Moreira}, \bibinfo{person}{Sheng Ma},
  \bibinfo{person}{Ricardo Vilalta}, {and} \bibinfo{person}{Anand
  Sivasubramaniam}.} \bibinfo{year}{2003}\natexlab{}.
\newblock \showarticletitle{Critical event prediction for proactive management
  in large-scale computer clusters}. In \bibinfo{booktitle}{\emph{Proc. of the
  ninth {ACM} {SIGKDD} international conference on knowledge discovery and data
  mining (SIGKDD)}}. \bibinfo{pages}{426--435}.
\newblock


\bibitem[\protect\citeauthoryear{Schroeder and Gibson}{Schroeder and
  Gibson}{2006}]%
        {DBLP:conf/dsn/SchroederG06}
\bibfield{author}{\bibinfo{person}{Bianca Schroeder} {and}
  \bibinfo{person}{Garth~A. Gibson}.} \bibinfo{year}{2006}\natexlab{}.
\newblock \showarticletitle{{A} large-scale study of failures in
  high-performance computing systems}. In \bibinfo{booktitle}{\emph{Proc. of
  the 2006 International Conference on Dependable Systems and Networks (DSN)}}.
  \bibinfo{pages}{249--258}.
\newblock


\bibitem[\protect\citeauthoryear{SecRepo}{SecRepo}{2020}]%
        {SecRepo}
\bibfield{author}{\bibinfo{person}{SecRepo}.} \bibinfo{year}{2020}\natexlab{}.
\newblock \bibinfo{title}{A list of security log data}.
\newblock
\newblock
\urldef\tempurl%
\url{http://www.secrepo.com}
\showURL{%
Retrieved September 1, 2020 from \tempurl}


\bibitem[\protect\citeauthoryear{Shang}{Shang}{2012}]%
        {DBLP:conf/icse/Shang12}
\bibfield{author}{\bibinfo{person}{Weiyi Shang}.}
  \bibinfo{year}{2012}\natexlab{}.
\newblock \showarticletitle{Bridging the divide between software developers and
  operators using logs}. In \bibinfo{booktitle}{\emph{Proc. of the 34th
  International Conference on Software Engineering (ICSE)}}.
  \bibinfo{pages}{1583--1586}.
\newblock


\bibitem[\protect\citeauthoryear{Shang, Jiang, Hemmati, Adams, Hassan, and
  Martin}{Shang et~al\mbox{.}}{2013}]%
        {shang2013assisting}
\bibfield{author}{\bibinfo{person}{Weiyi Shang}, \bibinfo{person}{Zhen~Ming
  Jiang}, \bibinfo{person}{Hadi Hemmati}, \bibinfo{person}{Brain Adams},
  \bibinfo{person}{Ahmed~E Hassan}, {and} \bibinfo{person}{Patrick Martin}.}
  \bibinfo{year}{2013}\natexlab{}.
\newblock \showarticletitle{Assisting developers of big data analytics
  applications when deploying on hadoop clouds}. In
  \bibinfo{booktitle}{\emph{Proc. of the 35th International Conference on
  Software Engineering (ICSE)}}. \bibinfo{pages}{402--411}.
\newblock


\bibitem[\protect\citeauthoryear{Shang, Nagappan, and Hassan}{Shang
  et~al\mbox{.}}{2015}]%
        {DBLP:journals/ese/ShangNH15}
\bibfield{author}{\bibinfo{person}{Weiyi Shang}, \bibinfo{person}{Meiyappan
  Nagappan}, {and} \bibinfo{person}{Ahmed~E. Hassan}.}
  \bibinfo{year}{2015}\natexlab{}.
\newblock \showarticletitle{Studying the relationship between logging
  characteristics and the code quality of platform software}.
\newblock \bibinfo{journal}{\emph{Empirical Software Engineering}}
  (\bibinfo{year}{2015}), \bibinfo{pages}{1--27}.
\newblock


\bibitem[\protect\citeauthoryear{Shang, Nagappan, Hassan, and Jiang}{Shang
  et~al\mbox{.}}{2014}]%
        {DBLP:conf/icsm/ShangNHJ14}
\bibfield{author}{\bibinfo{person}{Weiyi Shang}, \bibinfo{person}{Meiyappan
  Nagappan}, \bibinfo{person}{Ahmed~E. Hassan}, {and}
  \bibinfo{person}{Zhen~Ming Jiang}.} \bibinfo{year}{2014}\natexlab{}.
\newblock \showarticletitle{Understanding log lines using development
  knowledge}. In \bibinfo{booktitle}{\emph{Proc. of the 30th International
  Conference on Software Maintenance and Evolution (ICSME)}}.
\newblock


\bibitem[\protect\citeauthoryear{Shima}{Shima}{2016}]%
        {lenma16}
\bibfield{author}{\bibinfo{person}{Keiichi Shima}.}
  \bibinfo{year}{2016}\natexlab{}.
\newblock \showarticletitle{Length matters: clustering system log messages
  using length of words}.
\newblock \bibinfo{journal}{\emph{arXiv:1611.03213}} (\bibinfo{year}{2016}).
\newblock


\bibitem[\protect\citeauthoryear{Skibi{\'n}ski and Swacha}{Skibi{\'n}ski and
  Swacha}{2007}]%
        {skibinski2007fast}
\bibfield{author}{\bibinfo{person}{Przemys{\l}aw Skibi{\'n}ski} {and}
  \bibinfo{person}{Jakub Swacha}.} \bibinfo{year}{2007}\natexlab{}.
\newblock \showarticletitle{Fast and efficient log file compression}. In
  \bibinfo{booktitle}{\emph{Proc. of 11th east-European conference on advances
  in databases and information systems (ADBIS)}}. \bibinfo{pages}{56--69}.
\newblock


\bibitem[\protect\citeauthoryear{SLF4J}{SLF4J}{2020}]%
        {SLF4J}
\bibfield{author}{\bibinfo{person}{SLF4J}.} \bibinfo{year}{2020}\natexlab{}.
\newblock \bibinfo{title}{Simple Logging Facade for Java (SLF4J)}.
\newblock
\newblock
\urldef\tempurl%
\url{http://www.slf4j.org/}
\showURL{%
Retrieved September 1, 2020 from \tempurl}


\bibitem[\protect\citeauthoryear{spdlog}{spdlog}{2020}]%
        {spdlog}
\bibfield{author}{\bibinfo{person}{spdlog}.} \bibinfo{year}{2020}\natexlab{}.
\newblock \bibinfo{title}{Spdlog}.
\newblock
\newblock
\urldef\tempurl%
\url{https://github.com/gabime/spdlog}
\showURL{%
Retrieved September 1, 2020 from \tempurl}


\bibitem[\protect\citeauthoryear{Splunk}{Splunk}{2005}]%
        {splunk}
\bibfield{author}{\bibinfo{person}{Splunk}.} \bibinfo{year}{2005}\natexlab{}.
\newblock \bibinfo{title}{Splunk platform}.
\newblock
\newblock
\urldef\tempurl%
\url{http://www.splunk.com}
\showURL{%
Retrieved September 1, 2020 from \tempurl}


\bibitem[\protect\citeauthoryear{Syslog-ng}{Syslog-ng}{2020}]%
        {Syslog-ng}
\bibfield{author}{\bibinfo{person}{Syslog-ng}.}
  \bibinfo{year}{2020}\natexlab{}.
\newblock \bibinfo{title}{A log management solution}.
\newblock
\newblock
\urldef\tempurl%
\url{https://www.syslog-ng.com}
\showURL{%
Retrieved September 1, 2020 from \tempurl}


\bibitem[\protect\citeauthoryear{Tak, Tao, Yang, Zhu, and Ruan}{Tak
  et~al\mbox{.}}{2016}]%
        {tak2016logan}
\bibfield{author}{\bibinfo{person}{Byung~Chul Tak}, \bibinfo{person}{Shu Tao},
  \bibinfo{person}{Lin Yang}, \bibinfo{person}{Chao Zhu}, {and}
  \bibinfo{person}{Yaoping Ruan}.} \bibinfo{year}{2016}\natexlab{}.
\newblock \showarticletitle{Logan: problem diagnosis in the cloud using
  log-based reference models}. In \bibinfo{booktitle}{\emph{Proc. of the IEEE
  International Conference on Cloud Engineering (IC2E)}}.
  \bibinfo{pages}{62--67}.
\newblock


\bibitem[\protect\citeauthoryear{Tang, Li, and Perng}{Tang
  et~al\mbox{.}}{2011}]%
        {tang2011logsig}
\bibfield{author}{\bibinfo{person}{Liang Tang}, \bibinfo{person}{Tao Li}, {and}
  \bibinfo{person}{Chang-Shing Perng}.} \bibinfo{year}{2011}\natexlab{}.
\newblock \showarticletitle{{LogSig}: Generating system events from raw textual
  logs}. In \bibinfo{booktitle}{\emph{Proc. of the 20th ACM international
  conference on Information and knowledge management (CIKM)}}.
  \bibinfo{pages}{785--794}.
\newblock


\bibitem[\protect\citeauthoryear{Team}{Team}{2020}]%
        {downtimeAmazon}
\bibfield{author}{\bibinfo{person}{StatusCake Team}.}
  \bibinfo{year}{2020}\natexlab{}.
\newblock \bibinfo{title}{The Most Expensive Website Downtime Periods in
  History}.
\newblock
\newblock
\urldef\tempurl%
\url{https://www.statuscake.com/the-most-expensive-website-downtime-periods-in-history/}
\showURL{%
\tempurl}


\bibitem[\protect\citeauthoryear{UpGuard}{UpGuard}{2019}]%
        {downtimeexample}
\bibfield{author}{\bibinfo{person}{UpGuard}.} \bibinfo{year}{2019}\natexlab{}.
\newblock \bibinfo{title}{The cost of downtime at the world's biggest online
  retailer}.
\newblock
\newblock
\urldef\tempurl%
\url{https://www.upguard.com/blog/the-cost-of-downtime-at-the-worlds-biggest-online-retailer}
\showURL{%
Retrieved September 1, 2020 from \tempurl}


\bibitem[\protect\citeauthoryear{Vaarandi}{Vaarandi}{2003}]%
        {vaarandi2003data}
\bibfield{author}{\bibinfo{person}{Risto Vaarandi}.}
  \bibinfo{year}{2003}\natexlab{}.
\newblock \showarticletitle{A data clustering algorithm for mining patterns
  from event logs}. In \bibinfo{booktitle}{\emph{Proc. of the 3rd IEEE Workshop
  on IP Operations \& Management (IPOM)}}. \bibinfo{pages}{119--126}.
\newblock


\bibitem[\protect\citeauthoryear{Vaarandi and Pihelgas}{Vaarandi and
  Pihelgas}{2015}]%
        {logcluster15}
\bibfield{author}{\bibinfo{person}{Risto Vaarandi} {and} \bibinfo{person}{Mauno
  Pihelgas}.} \bibinfo{year}{2015}\natexlab{}.
\newblock \showarticletitle{LogCluster - A data clustering and pattern mining
  algorithm for event logs}. In \bibinfo{booktitle}{\emph{Proc. of the 11th
  International conference on network and service management (CNSM)}}.
  \bibinfo{pages}{1--7}.
\newblock


\bibitem[\protect\citeauthoryear{Van Der~Aalst}{Van Der~Aalst}{2012}]%
        {van2012process}
\bibfield{author}{\bibinfo{person}{Wil Van Der~Aalst}.}
  \bibinfo{year}{2012}\natexlab{}.
\newblock \showarticletitle{Process mining}.
\newblock \bibinfo{journal}{\emph{Commun. ACM}} (\bibinfo{year}{2012}),
  \bibinfo{pages}{76--83}.
\newblock


\bibitem[\protect\citeauthoryear{Xia, Bai, Yin, Li, and Xu}{Xia
  et~al\mbox{.}}{2020}]%
        {xia2020loggan}
\bibfield{author}{\bibinfo{person}{Bin Xia}, \bibinfo{person}{Yuxuan Bai},
  \bibinfo{person}{Junjie Yin}, \bibinfo{person}{Yun Li}, {and}
  \bibinfo{person}{Jian Xu}.} \bibinfo{year}{2020}\natexlab{}.
\newblock \showarticletitle{LogGAN: a log-level generative adversarial network
  for anomaly detection using permutation event modeling}.
\newblock \bibinfo{journal}{\emph{Information Systems Frontiers}}
  (\bibinfo{year}{2020}), \bibinfo{pages}{1--14}.
\newblock


\bibitem[\protect\citeauthoryear{Xu}{Xu}{2010}]%
        {weixutheis}
\bibfield{author}{\bibinfo{person}{Wei Xu}.} \bibinfo{year}{2010}\natexlab{}.
\newblock \emph{\bibinfo{title}{System problem detection by mining console
  logs}}.
\newblock \bibinfo{thesistype}{Ph.D. Dissertation}. \bibinfo{school}{University
  of California, Berkeley}.
\newblock


\bibitem[\protect\citeauthoryear{Xu, Huang, Fox, Patterson, and Jordan}{Xu
  et~al\mbox{.}}{2009a}]%
        {xu2009online}
\bibfield{author}{\bibinfo{person}{Wei Xu}, \bibinfo{person}{Ling Huang},
  \bibinfo{person}{Armando Fox}, \bibinfo{person}{David Patterson}, {and}
  \bibinfo{person}{Michael Jordan}.} \bibinfo{year}{2009}\natexlab{a}.
\newblock \showarticletitle{Online system problem detection by mining patterns
  of console logs}. In \bibinfo{booktitle}{\emph{Proc. of the Ninth {IEEE}
  International Conference on Data Mining (ICDM)}}. \bibinfo{pages}{588--597}.
\newblock


\bibitem[\protect\citeauthoryear{Xu, Huang, Fox, Patterson, and Jordan}{Xu
  et~al\mbox{.}}{2009b}]%
        {xu2009detecting}
\bibfield{author}{\bibinfo{person}{Wei Xu}, \bibinfo{person}{Ling Huang},
  \bibinfo{person}{Armando Fox}, \bibinfo{person}{David Patterson}, {and}
  \bibinfo{person}{Michael~I Jordan}.} \bibinfo{year}{2009}\natexlab{b}.
\newblock \showarticletitle{Detecting large-scale system problems by mining
  console logs}. In \bibinfo{booktitle}{\emph{Proc. of the 22nd ACM Symposium
  on Operating Systems Principles (SOSP)}}.
\newblock


\bibitem[\protect\citeauthoryear{Yamanishi and Maruyama}{Yamanishi and
  Maruyama}{2005}]%
        {yamanishi2005dynamic}
\bibfield{author}{\bibinfo{person}{Kenji Yamanishi} {and} \bibinfo{person}{Yuko
  Maruyama}.} \bibinfo{year}{2005}\natexlab{}.
\newblock \showarticletitle{Dynamic syslog mining for network failure
  monitoring}. In \bibinfo{booktitle}{\emph{Proc. of the 11nd ACM SIGKDD
  International Conference on Knowledge Discovery and Data Mining (SIGKDD)}}.
  \bibinfo{pages}{499--508}.
\newblock


\bibitem[\protect\citeauthoryear{Yang, Park, and Ousterhout}{Yang
  et~al\mbox{.}}{2018}]%
        {DBLP:conf/usenix/YangPO18}
\bibfield{author}{\bibinfo{person}{Stephen Yang}, \bibinfo{person}{Seo~Jin
  Park}, {and} \bibinfo{person}{John~K. Ousterhout}.}
  \bibinfo{year}{2018}\natexlab{}.
\newblock \showarticletitle{NanoLog: {A} nanosecond scale logging system}. In
  \bibinfo{booktitle}{\emph{Proc. of the 2018 {USENIX} Annual Technical
  Conference (ATC)}}. \bibinfo{pages}{335--350}.
\newblock


\bibitem[\protect\citeauthoryear{Yao, de~P{\'{a}}dua, Shang, Sporea, Toma, and
  Sajedi}{Yao et~al\mbox{.}}{2020}]%
        {DBLP:journals/ese/YaoPSSTS20}
\bibfield{author}{\bibinfo{person}{Kundi Yao}, \bibinfo{person}{Guilherme~B. de
  P{\'{a}}dua}, \bibinfo{person}{Weiyi Shang}, \bibinfo{person}{Catalin
  Sporea}, \bibinfo{person}{Andrei Toma}, {and} \bibinfo{person}{Sarah
  Sajedi}.} \bibinfo{year}{2020}\natexlab{}.
\newblock \showarticletitle{Log4Perf: suggesting and updating logging locations
  for web-based systems' performance monitoring}.
\newblock \bibinfo{journal}{\emph{Empir. Softw. Eng.}} (\bibinfo{year}{2020}),
  \bibinfo{pages}{488--531}.
\newblock


\bibitem[\protect\citeauthoryear{Yao, Li, Shang, and Hassan}{Yao
  et~al\mbox{.}}{2019}]%
        {yaostudy}
\bibfield{author}{\bibinfo{person}{Kundi Yao}, \bibinfo{person}{Heng Li},
  \bibinfo{person}{Weiyi Shang}, {and} \bibinfo{person}{Ahmed~E Hassan}.}
  \bibinfo{year}{2019}\natexlab{}.
\newblock \showarticletitle{A study of the performance of general compressors
  on log files}.
\newblock \bibinfo{journal}{\emph{Empirical Software Engineering}}
  (\bibinfo{year}{2019}), \bibinfo{pages}{3043--3085}.
\newblock


\bibitem[\protect\citeauthoryear{Yu, Han, Zhang, and Xie}{Yu
  et~al\mbox{.}}{2014}]%
        {yu2014comprehending}
\bibfield{author}{\bibinfo{person}{Xiao Yu}, \bibinfo{person}{Shi Han},
  \bibinfo{person}{Dongmei Zhang}, {and} \bibinfo{person}{Tao Xie}.}
  \bibinfo{year}{2014}\natexlab{}.
\newblock \showarticletitle{Comprehending performance from real-world execution
  traces: A device-driver case}. In \bibinfo{booktitle}{\emph{Proc. of the 19th
  international conference on Architectural support for programming languages
  and operating systems (ASPLOS)}}. \bibinfo{pages}{193--206}.
\newblock


\bibitem[\protect\citeauthoryear{Yu, Joshi, Xu, Jin, Zhang, and Jiang}{Yu
  et~al\mbox{.}}{2016}]%
        {yu2016cloudseer}
\bibfield{author}{\bibinfo{person}{Xiao Yu}, \bibinfo{person}{Pallavi Joshi},
  \bibinfo{person}{Jianwu Xu}, \bibinfo{person}{Guoliang Jin},
  \bibinfo{person}{Hui Zhang}, {and} \bibinfo{person}{Guofei Jiang}.}
  \bibinfo{year}{2016}\natexlab{}.
\newblock \showarticletitle{Cloudseer: workflow monitoring of cloud
  infrastructures via interleaved logs}. In \bibinfo{booktitle}{\emph{Proc. of
  the 21st International Conference on Architectural Support for Programming
  Languages and Operating Systems (ASPLOS)}}. \bibinfo{pages}{489--502}.
\newblock


\bibitem[\protect\citeauthoryear{Yuan, Mai, Xiong, Tan, Zhou, and
  Pasupathy}{Yuan et~al\mbox{.}}{2010}]%
        {yuan2010sherlog}
\bibfield{author}{\bibinfo{person}{Ding Yuan}, \bibinfo{person}{Haohui Mai},
  \bibinfo{person}{Weiwei Xiong}, \bibinfo{person}{Lin Tan},
  \bibinfo{person}{Yuanyuan Zhou}, {and} \bibinfo{person}{Shankar Pasupathy}.}
  \bibinfo{year}{2010}\natexlab{}.
\newblock \showarticletitle{SherLog: error diagnosis by connecting clues from
  run-time logs}. In \bibinfo{booktitle}{\emph{Proc. of the fifteenth
  International Conference on Architectural support for programming languages
  and operating systems (ASPLOS)}}. \bibinfo{pages}{143--154}.
\newblock


\bibitem[\protect\citeauthoryear{Yuan, Park, Huang, Liu, Lee, Tang, Zhou, and
  Savage}{Yuan et~al\mbox{.}}{2012b}]%
        {DBLP:conf/osdi/YuanPHLLTZS12}
\bibfield{author}{\bibinfo{person}{Ding Yuan}, \bibinfo{person}{Soyeon Park},
  \bibinfo{person}{Peng Huang}, \bibinfo{person}{Yang Liu},
  \bibinfo{person}{Michael~Mihn{-}Jong Lee}, \bibinfo{person}{Xiaoming Tang},
  \bibinfo{person}{Yuanyuan Zhou}, {and} \bibinfo{person}{Stefan Savage}.}
  \bibinfo{year}{2012}\natexlab{b}.
\newblock \showarticletitle{Be conservative: enhancing failure diagnosis with
  proactive logging}. In \bibinfo{booktitle}{\emph{Proc. of the 10th {USENIX}
  Symposium on Operating Systems Design and Implementation (OSDI)}}.
  \bibinfo{pages}{293--306}.
\newblock


\bibitem[\protect\citeauthoryear{Yuan, Park, and Zhou}{Yuan
  et~al\mbox{.}}{2012a}]%
        {DBLP:conf/icse/YuanPZ12}
\bibfield{author}{\bibinfo{person}{Ding Yuan}, \bibinfo{person}{Soyeon Park},
  {and} \bibinfo{person}{Yuanyuan Zhou}.} \bibinfo{year}{2012}\natexlab{a}.
\newblock \showarticletitle{Characterizing logging practices in open-source
  software}. In \bibinfo{booktitle}{\emph{Proc. of 34th International
  Conference on Software Engineering (ICSE)}}. \bibinfo{pages}{102--112}.
\newblock


\bibitem[\protect\citeauthoryear{Yuan, Zheng, Park, Zhou, and Savage}{Yuan
  et~al\mbox{.}}{2011}]%
        {DBLP:conf/asplos/YuanZPZS11}
\bibfield{author}{\bibinfo{person}{Ding Yuan}, \bibinfo{person}{Jing Zheng},
  \bibinfo{person}{Soyeon Park}, \bibinfo{person}{Yuanyuan Zhou}, {and}
  \bibinfo{person}{Stefan Savage}.} \bibinfo{year}{2011}\natexlab{}.
\newblock \showarticletitle{Improving software diagnosability via log
  enhancement}. In \bibinfo{booktitle}{\emph{Proc. of the 16th International
  Conference on Architectural Support for Programming Languages and Operating
  Systems (ASPLOS)}}. \bibinfo{pages}{3--14}.
\newblock


\bibitem[\protect\citeauthoryear{Yuan, Shi, Liang, and Qin}{Yuan
  et~al\mbox{.}}{2019}]%
        {yuan2019approach}
\bibfield{author}{\bibinfo{person}{Yue Yuan}, \bibinfo{person}{Wenchang Shi},
  \bibinfo{person}{Bin Liang}, {and} \bibinfo{person}{Bo Qin}.}
  \bibinfo{year}{2019}\natexlab{}.
\newblock \showarticletitle{An approach to cloud execution failure diagnosis
  based on exception logs in openstack}. In \bibinfo{booktitle}{\emph{Proc. of
  the IEEE 12th International Conference on Cloud Computing (CLOUD)}}.
  \bibinfo{pages}{124--131}.
\newblock


\bibitem[\protect\citeauthoryear{Zaman, Han, and Yu}{Zaman
  et~al\mbox{.}}{2019}]%
        {zaman2019scminer}
\bibfield{author}{\bibinfo{person}{Tarannum~Shaila Zaman}, \bibinfo{person}{Xue
  Han}, {and} \bibinfo{person}{Tingting Yu}.} \bibinfo{year}{2019}\natexlab{}.
\newblock \showarticletitle{SCMiner: localizing system-level concurrency faults
  from large system call traces}. In \bibinfo{booktitle}{\emph{Proc. of 34th
  IEEE/ACM International Conference on Automated Software Engineering (ASE)}}.
\newblock


\bibitem[\protect\citeauthoryear{Zeng, Xiao, Chen, Sun, and Han}{Zeng
  et~al\mbox{.}}{2016}]%
        {zeng2016computer}
\bibfield{author}{\bibinfo{person}{Lei Zeng}, \bibinfo{person}{Yang Xiao},
  \bibinfo{person}{Hui Chen}, \bibinfo{person}{Bo Sun}, {and}
  \bibinfo{person}{Wenlin Han}.} \bibinfo{year}{2016}\natexlab{}.
\newblock \showarticletitle{Computer operating system logging and security
  issues: a survey}.
\newblock \bibinfo{journal}{\emph{Security and Communication Networks}}
  \bibinfo{volume}{9}, \bibinfo{number}{17} (\bibinfo{year}{2016}),
  \bibinfo{pages}{4804--4821}.
\newblock


\bibitem[\protect\citeauthoryear{Zeng, Chen, Shang, and Chen}{Zeng
  et~al\mbox{.}}{2019}]%
        {DBLP:journals/ese/ZengCSC19}
\bibfield{author}{\bibinfo{person}{Yi Zeng}, \bibinfo{person}{Jinfu Chen},
  \bibinfo{person}{Weiyi Shang}, {and} \bibinfo{person}{Tse{-}Hsun~(Peter)
  Chen}.} \bibinfo{year}{2019}\natexlab{}.
\newblock \showarticletitle{Studying the characteristics of logging practices
  in mobile apps: a case study on F-Droid}.
\newblock \bibinfo{journal}{\emph{Empirical Software Engineering}}
  (\bibinfo{year}{2019}), \bibinfo{pages}{3394--3434}.
\newblock


\bibitem[\protect\citeauthoryear{Zhang, Liu, Meng, Luo, Bu, Yang, Liang, Pei,
  Xu, Zhang, Chen, Dong, Qu, and Song}{Zhang et~al\mbox{.}}{2018}]%
        {zhang2018prefix}
\bibfield{author}{\bibinfo{person}{Shenglin Zhang}, \bibinfo{person}{Ying Liu},
  \bibinfo{person}{Weibin Meng}, \bibinfo{person}{Zhiling Luo},
  \bibinfo{person}{Jiahao Bu}, \bibinfo{person}{Sen Yang},
  \bibinfo{person}{Peixian Liang}, \bibinfo{person}{Dan Pei},
  \bibinfo{person}{Jun Xu}, \bibinfo{person}{Yuzhi Zhang}, \bibinfo{person}{Yu
  Chen}, \bibinfo{person}{Hui Dong}, \bibinfo{person}{Xianping Qu}, {and}
  \bibinfo{person}{Lei Song}.} \bibinfo{year}{2018}\natexlab{}.
\newblock \showarticletitle{PreFix: switch failure prediction in datacenter
  networks}.
\newblock \bibinfo{journal}{\emph{Proc. ACM Meas. Anal. Comput. Syst.}}
  \bibinfo{volume}{2} (\bibinfo{year}{2018}), \bibinfo{pages}{2:1--2:29}.
\newblock


\bibitem[\protect\citeauthoryear{Zhang, Xu, Lin, Qiao, Zhang, Dang, Xie, Yang,
  Cheng, Li, et~al\mbox{.}}{Zhang et~al\mbox{.}}{2019b}]%
        {zhang2019robust}
\bibfield{author}{\bibinfo{person}{Xu Zhang}, \bibinfo{person}{Yong Xu},
  \bibinfo{person}{Qingwei Lin}, \bibinfo{person}{Bo Qiao},
  \bibinfo{person}{Hongyu Zhang}, \bibinfo{person}{Yingnong Dang},
  \bibinfo{person}{Chunyu Xie}, \bibinfo{person}{Xinsheng Yang},
  \bibinfo{person}{Qian Cheng}, \bibinfo{person}{Ze Li}, {et~al\mbox{.}}}
  \bibinfo{year}{2019}\natexlab{b}.
\newblock \showarticletitle{Robust log-based anomaly detection on unstable log
  data}. In \bibinfo{booktitle}{\emph{Proc. of the 27th ACM Joint European
  Software Engineering Conference and Symposium on the Foundations of Software
  Engineering (ESEC/FSE)}}.
\newblock


\bibitem[\protect\citeauthoryear{Zhang, Rodrigues, Luo, Stumm, and Yuan}{Zhang
  et~al\mbox{.}}{2019a}]%
        {zhang2019inflection}
\bibfield{author}{\bibinfo{person}{Yongle Zhang}, \bibinfo{person}{Kirk
  Rodrigues}, \bibinfo{person}{Yu Luo}, \bibinfo{person}{Michael Stumm}, {and}
  \bibinfo{person}{Ding Yuan}.} \bibinfo{year}{2019}\natexlab{a}.
\newblock \showarticletitle{The inflection point hypothesis: a principled
  debugging approach for locating the root cause of a failure}. In
  \bibinfo{booktitle}{\emph{Proc. of the 27th ACM Symposium on Operating
  Systems Principles (SOSP)}}. \bibinfo{pages}{131--146}.
\newblock


\bibitem[\protect\citeauthoryear{Zhao, Rodrigues, Luo, Stumm, Yuan, and
  Zhou}{Zhao et~al\mbox{.}}{2017a}]%
        {DBLP:conf/hotos/ZhaoRLSYZ17}
\bibfield{author}{\bibinfo{person}{Xu Zhao}, \bibinfo{person}{Kirk Rodrigues},
  \bibinfo{person}{Yu Luo}, \bibinfo{person}{Michael Stumm},
  \bibinfo{person}{Ding Yuan}, {and} \bibinfo{person}{Yuanyuan Zhou}.}
  \bibinfo{year}{2017}\natexlab{a}.
\newblock \showarticletitle{The game of twenty questions: do you know where to
  log?}. In \bibinfo{booktitle}{\emph{Proc. of the 16th Workshop on Hot Topics
  in Operating Systems (HotOS)}}.
\newblock


\bibitem[\protect\citeauthoryear{Zhao, Rodrigues, Luo, Stumm, Yuan, and
  Zhou}{Zhao et~al\mbox{.}}{2017b}]%
        {DBLP:conf/sosp/ZhaoRLSYZ17}
\bibfield{author}{\bibinfo{person}{Xu Zhao}, \bibinfo{person}{Kirk Rodrigues},
  \bibinfo{person}{Yu Luo}, \bibinfo{person}{Michael Stumm},
  \bibinfo{person}{Ding Yuan}, {and} \bibinfo{person}{Yuanyuan Zhou}.}
  \bibinfo{year}{2017}\natexlab{b}.
\newblock \showarticletitle{Log20: fully automated optimal placement of log
  printing statements under specified overhead threshold}. In
  \bibinfo{booktitle}{\emph{Proc. of the 26th Symposium on Operating Systems
  Principles (SOSP)}}. \bibinfo{pages}{565--581}.
\newblock


\bibitem[\protect\citeauthoryear{Zhao, Rodrigues, Luo, Yuan, and Stumm}{Zhao
  et~al\mbox{.}}{2016}]%
        {zhao2016non}
\bibfield{author}{\bibinfo{person}{Xu Zhao}, \bibinfo{person}{Kirk Rodrigues},
  \bibinfo{person}{Yu Luo}, \bibinfo{person}{Ding Yuan}, {and}
  \bibinfo{person}{Michael Stumm}.} \bibinfo{year}{2016}\natexlab{}.
\newblock \showarticletitle{Non-intrusive performance profiling for entire
  software stacks based on the flow reconstruction principle}. In
  \bibinfo{booktitle}{\emph{Proc. of the 12th Symposium on Operating Systems
  Design and Implementation (OSDI)}}. \bibinfo{pages}{603--618}.
\newblock


\bibitem[\protect\citeauthoryear{Zhou, Peng, Xie, Sun, Ji, Li, and Ding}{Zhou
  et~al\mbox{.}}{2018}]%
        {zhou2018fault}
\bibfield{author}{\bibinfo{person}{Xiang Zhou}, \bibinfo{person}{Xin Peng},
  \bibinfo{person}{Tao Xie}, \bibinfo{person}{Jun Sun}, \bibinfo{person}{Chao
  Ji}, \bibinfo{person}{Wenhai Li}, {and} \bibinfo{person}{Dan Ding}.}
  \bibinfo{year}{2018}\natexlab{}.
\newblock \showarticletitle{Fault analysis and debugging of microservice
  systems: Industrial survey, benchmark system, and empirical study}.
\newblock \bibinfo{journal}{\emph{IEEE Transactions on Software Engineering}}
  (\bibinfo{year}{2018}).
\newblock


\bibitem[\protect\citeauthoryear{Zhou, Peng, Xie, Sun, Ji, Liu, Xiang, and
  He}{Zhou et~al\mbox{.}}{2019}]%
        {zhou2019latent}
\bibfield{author}{\bibinfo{person}{Xiang Zhou}, \bibinfo{person}{Xin Peng},
  \bibinfo{person}{Tao Xie}, \bibinfo{person}{Jun Sun}, \bibinfo{person}{Chao
  Ji}, \bibinfo{person}{Dewei Liu}, \bibinfo{person}{Qilin Xiang}, {and}
  \bibinfo{person}{Chuan He}.} \bibinfo{year}{2019}\natexlab{}.
\newblock \showarticletitle{Latent error prediction and fault localization for
  microservice applications by learning from system trace logs}. In
  \bibinfo{booktitle}{\emph{Proc. of the 2019 27th ACM Joint Meeting on
  European Software Engineering Conference and Symposium on the Foundations of
  Software Engineering (ESEC/FSE)}}. \bibinfo{pages}{683--694}.
\newblock


\bibitem[\protect\citeauthoryear{Zhu, He, Fu, Zhang, Lyu, and Zhang}{Zhu
  et~al\mbox{.}}{2015}]%
        {DBLP:conf/icse/ZhuHFZLZ15}
\bibfield{author}{\bibinfo{person}{Jieming Zhu}, \bibinfo{person}{Pinjia He},
  \bibinfo{person}{Qiang Fu}, \bibinfo{person}{Hongyu Zhang},
  \bibinfo{person}{Michael~R. Lyu}, {and} \bibinfo{person}{Dongmei Zhang}.}
  \bibinfo{year}{2015}\natexlab{}.
\newblock \showarticletitle{Learning to log: helping developers make informed
  logging decisions}. In \bibinfo{booktitle}{\emph{Proc. of the 37th {IEEE/ACM}
  International Conference on Software Engineering (ICSE)}}.
  \bibinfo{pages}{415--425}.
\newblock


\bibitem[\protect\citeauthoryear{Zhu, He, Liu, He, Xie, Zheng, and Lyu}{Zhu
  et~al\mbox{.}}{2019}]%
        {DBLP:conf/icse/ZhuHLHXZL19}
\bibfield{author}{\bibinfo{person}{Jieming Zhu}, \bibinfo{person}{Shilin He},
  \bibinfo{person}{Jinyang Liu}, \bibinfo{person}{Pinjia He},
  \bibinfo{person}{Qi Xie}, \bibinfo{person}{Zibin Zheng}, {and}
  \bibinfo{person}{Michael~R. Lyu}.} \bibinfo{year}{2019}\natexlab{}.
\newblock \showarticletitle{Tools and benchmarks for automated log parsing}. In
  \bibinfo{booktitle}{\emph{Proc. of the 41st International Conference on
  Software Engineering: Software Engineering in Practice (ICSE-SEIP)}}.
  \bibinfo{pages}{121--130}.
\newblock


\bibitem[\protect\citeauthoryear{Zuo, Wu, Min, Huang, and Pei}{Zuo
  et~al\mbox{.}}{2020}]%
        {zuo2020intelligent}
\bibfield{author}{\bibinfo{person}{Yuan Zuo}, \bibinfo{person}{Yulei Wu},
  \bibinfo{person}{Geyong Min}, \bibinfo{person}{Chengqiang Huang}, {and}
  \bibinfo{person}{Ke Pei}.} \bibinfo{year}{2020}\natexlab{}.
\newblock \showarticletitle{An intelligent anomaly detection scheme for
  micro-services architectures with temporal and spatial data analysis}.
\newblock \bibinfo{journal}{\emph{IEEE Transactions on Cognitive Communications
  and Networking}} (\bibinfo{year}{2020}).
\newblock


\end{thebibliography}

\end{document}


\title{Supplementary Material to: A Survey on Automated Log Analysis for Reliability Engineering}

\author{Shilin He}
\affiliation{%
  \institution{Microsoft Research}
}

\author{Pinjia He}
\affiliation{%
  \institution{School of Data Science, The Chinese University of Hong Kong, Shenzhen}
}

\author{Zhuangbin Chen}
\author{Tianyi Yang}
\author{Yuxin Su}
\author{Michael R. Lyu}
\affiliation{%
  \institution{Department of Computer Science and Engineering, The Chinese University of Hong Kong}
}

\renewcommand{\shortauthors}{S. He et al.}

\maketitle

\appendix

\section{Best Current Practices}
\label{sec:best_practice}
Log plays an essential role in industrial companies for its carried information, e.g., log management tools have been integrated in various cloud platforms. Although this survey is mainly research-oriented, in this section, we intend to discuss the current industrial best practices based on our experience and surveyed papers and articles. Note that the practices do not indicate the underlying technologies used by specific companies and we do not aim to include all the best practices.
Most existing practices concentrate on the logging, log collection, and monitoring aspects, which are summarized into six best practices as follows. 



\paragraph{Practice 1: Always follow the logging standards} It is crucial to follow the standards of logging during development, otherwise the produced logs would be hard to maintain, search and analyze. For example, the following logging standards are shared by various systems: (1) Timestamp: timestamp helps developers understand the sequential relationship among log events. Using correct timestamps (UTC/timezone adjusted) is necessary for debugging and analytic purposes. (2) Verbosity levels: proper verbosity levels ease log parsing and searching. In addition, aggregating logs by verbosity level is beneficial. (3) Format: log format is highly correlated with the parse and search procedures while most people might ignore. It is recommended to structure logs following an agreed standard (e.g., in JSON format) within the same project group. (4) Log message:  meaningful log messages facilitates the identification of the correct root cause for a failure. To construct meaning log messages, it is suggested to avoid duplicate logging descriptions (e.g., by assignment a unique ID to each logging statement).

\paragraph{Practice 2: Keep proper quantity of log messages}
Controlling the number of logging statements in the source code is very tricky. If logging too little, we may not have adequate information for problem diagnosis. On the contrary, if logging too much, we can easily get overwhelmed by the huge volume of logs and problem diagnosis is like looking for needles in haystacks. Moreover, too many logging statements could lead to unnecessary performance overhead~\cite{DBLP:conf/osdi/YuanPHLLTZS12, DBLP:conf/usenix/DingZLZLFZX15}. Hence, it is crucial to keep proper quantity of log messages.


\paragraph{Practice 3: Trace log cycle across services} 
The widely-employed micro-services and components in industry bring great challenges to the tracing of a request life cycle. Engineers may get overwhelmed by interleaving logs generated by different requests, hindering problem diagnosis in practice. Therefore, it is necessary to record the event ID information during log collection. A recent study~\cite{lou2017experience} revealed that many difﬁculties in automated log analysis are caused by logs of low quality, and a typical example is the missing of event IDs. Without event IDs, we cannot link log messages related to the operations of a single request together. Event ID is also widely leveraged in recent research work~\cite{xu2009detecting, he2018identifying, logcluster15}. Typically, developers add a unique event ID to the head of each log message, and use it during the inter-service communications.

\paragraph{Practice 4: Aggregate logs to a centralized location}
Since logs are often generated in different services and components separately, it is important to aggregate the logs to a centralized location for convenient log search and analysis. Moreover, there are many practical challenges to consider after aggregating logs, including log file name configuration, retention policy, storage size, backup strategy, etc. We take retention policy as an example. The increasing volume of log data boost the storage cost and query time, thereby an appropriate retention policy is highly in demand. In addition, logs of different types might have different retention requirements. For example, application logs for troubleshooting can be kept for only a few weeks, while audit logs or transaction logs require much longer retention times. Therefore, the retention policy should be flexible.

\paragraph{Practice 5: Safeguard the collected logs}
Due to the abundant information (e.g., database address) stored in logs, log data is often under a high risk of attacks. Adversaries tend to remove their trails of action, and an intuitive way is to remove any logs that might reveal their actions. Therefore, safeguarding the log files in case of an attack is a must. Because of similar reasons, industrial companies are usually not willing to share their log data to the public. Apart from setting up proper safeguarding strategies for the log storage, it is a highly-desirable practice to avoid logging any confidential information (e.g., password). Recording suspicious human activities (e.g., failed authentication) and system behaviors (e.g., spikes in resource consumption) are also crucial to log safeguarding.

\paragraph{Practice 6: Integrate logs with monitoring metrics}
In a large log dataset, manually inspecting every log message brings heavy workload to the developers. The log information is useful only when an issue is identified. In practice, logs are often used together with other signals such as monitors. A monitor  aggregates information from multiple sources and often serves as the first step towards troubleshooting. The monitoring metric could be a status count extracted from raw logs (e.g., the number of error events), a health indicator of service performance (e.g., the number of successful requests) or a resource monitor (e.g., CPU and memory consumption). Generally, monitors are embedded in a monitoring dashboard for real-time monitoring and alerting. Because of these monitors, troubleshooting by searching and analyzing logs becomes more effective.

\section{Future Directions}\label{sec:future}
In this section, we would like to discuss promising future directions.
Particularly, we start with concluding future directions for each topic reviewed in this survey: logging, log compression, log parsing, and log mining. Then we share our thoughts on the next-generation log analysis framework, which might shed light on future studies. In particular, we believe the next-generation log analysis should be ``autopilot" via better human-computer interaction and an end-to-end analytic structure. Note that these directions are mainly recommended by the authors or reviewers of this paper and we believe there are much more interesting topics beyond what we have discussed in the following.



\subsection*{Logging}

Although logging has been widely studied in recent years, we believe there is still much room for further improvement. In the following, we introduce two future directions: analysis-oriented logging and logging convention enforcement.

\label{sec:logging_future}
\subsubsection*{Analysis-Oriented Logging.} As the foundation of automated log analysis, current logging practices mostly focus on characterization and recommendation of logging statements (\textit{i.e.}, where, what, and how to log). Meanwhile, a recent industrial study~\cite{DBLP:conf/icse/BarikDDF16} pointed out that the inconsistent presentation format with-in and among teams poses a significant challenge in developing automated log analysis tools. In addition, even an effective log mining algorithm could generate meaningless results if the software runtime information is not well-documented in the collected logs. Thus, a crucial direction is designing advanced logging mechanisms that can coordinate with the subsequent log analysis steps during the development.

\subsubsection*{Logging Convention Enforcement} Although there are no golden rules of logging, developers typically intend to have consistent logging practice in the same project. For example, senior developers in a project may define a set of rules with examples and ask other developers to follow. However, it is difficult for junior developers to get familiar with all the suggested rules, which often change when they are involved in new projects. Thus, it is promising to design automated tools (e.g., lightweight static analysis tools) to recommend logging practice and report potential violations of the "missing"/"incorrect" logging code.

\subsection*{Log Compression}

\subsubsection*{Query-efficient Log Compression.} Different logs may play different roles and have different storage requirements. For example, failure logs could be frequently queried by developers and the corresponding log compressor should promptly respond to the query request. Therefore, how to incorporate the query demands into log compression becomes vital to log service providers. Particularly, several potential sub-problems could be: how to predict the user query request, how to store logs in a systematically modularized way for different demands, etc.

\subsection*{Log Parsing}


\subsubsection*{Distributed Online Log Parsing.} Existing approaches~\cite{He2017TDSC, DuTKDE18} provide offline parallel mechanisms to accelerate the log parsing process. However, there is a lack of distributed online parsing solution in large-scale distributed systems. The desired log parser should have multiple running threads in both the master and the working nodes and parse the log messages in a streaming manner. A trade-off will be considered between parsing accuracy and efficiency via an event template synchronization mechanism. With a distributed online parser, we could generate structured log messages on the fly, which facilities both log compression and log mining tasks.


\subsection*{Log Mining}
Although being extensively studied, log mining is still very challenging due to complex problem definition and poor data quality in different scenarios.  
In the following, we summarize insights inspired by the surveyed papers and propose some promising future directions.
Unlike some other research topics (e.g., log parsing) that are relatively mature, log mining still leaves a lot of space to explore.

\subsubsection*{Concept Drift-adapted Log Mining.} Most of the existing approaches train models with historical log data offline. However, modern software systems continuously undergo feature upgrades and system renewal; hence, the patterns of logs may drift accordingly~\cite{zhang2019robust}. Common strategies of online updating, \textit{e.g.}, periodically retraining the model, suffer from high false positive rates, and strategies like zero-positive learning~\cite{du2019lifelong} have limited generalization ability. Online learning and incremental learning that fit the log mining scenario are appealing research directions.


\subsubsection*{Interpretable Log Mining.} Whether a log analysis algorithm can provide interpretable results is vital for administrator and analysts to trust and act on the automated analysis. Some traditional machine learning algorithms possess the merit of interpretability~\cite{xu2009online,shang2013assisting,lou2010mining_a,nandi2016anomaly}. However, deep learning models, though having achieved impressive performance, are criticized as black-box oracles. Therefore, the community of log analysis should continue pursuing more explainable and trustworthy algorithms, for example, by utilizing attention mechanisms~\cite{brown2018recurrent}. 

\subsubsection*{Log Completeness} Log completeness~\cite{cohen2015have} measures whether developers have seen enough logs for different log mining tasks. Log completeness is important because (1) it facilitates measuring the reliability of the log mining results and (2) it is strongly related to the heuristics that may address the scalability issues of log mining~\cite{busany2016behavioral}. However, little work has been done on log completeness, which makes it a promising and important future direction.
 
\subsection*{Next-generation Log Analysis Framework}
Log analysis, as a core component in Artificial Intelligence for IT Operations (AIOps), is crucial to the reliability engineering of software systems. Particularly, the demand for automated log analysis draws dramatic attention as the increasing needs of cloud computing. We believe that the next-generation log analysis would be the "autopilot" in cloud environment. We list as below two major directions that the next-generation log analysis framework 
may contain.

\subsubsection*{Human-computer Interactive Log Analysis.}
Current automated log analysis research facilitates the failure diagnosis process by providing insightful decisions and clues for root cause analysis. In essence, developers are still at the core of failure diagnosis. For example, the diagnosis results should be confirmed and then debugged by developers. However, there is a lack of interactions between the advanced analysis techniques and developers-in-charge. Besides, the diagnosis process and results are seldom reused in the future log analysis. Therefore, it is interesting and promising to involve developers into the loop of automated log analysis. It will be promising to build an interactive dialogue system with automated analysis tools. The automated log analysis tools first provide suggestions to the developers for decision making. Then the developers can request for additional information of failures, and the diagnosis process with human involvement can be recorded and reused to improve the automated log analysis tools.

\subsubsection*{Intelligent and End-to-end Log Analysis.}
Current automated log analysis contains several consecutive steps, but they are not working seamlessly together. In particular, they do not form an end-to-end framework, indicating the possible loss of some critical information in the middle. In our vision, the ultimate goal of the next generation log analysis is an intelligent and end-to-end framework. In this framework, logging statements are automatically suggested by considering the failure diagnosis demand and the potential machine learning models used in log mining. Then, the framework intelligently stores raw logs following practical requirements. Logs are processed without information loss and are correlated with human-observable failures for diagnosis. Moreover, the framework would automatically generate failure reports by aggregating essential information 
from all analyzed system logs. It should also auto-repair the detected failures by observing experiences from human experts.

\bibliographystyle{ACM-Reference-Format}
\bibliography{bibliography}